\documentclass[aps,preprint,noshowpacs,nofootinbib,eqsecnum,floatfix,longbibliography,11pt]{revtex4-1}
\usepackage{amssymb,amsmath,graphicx}
\usepackage{bm}
\usepackage{hyperref}
\usepackage{color}
\hypersetup{colorlinks=true,linkcolor=blue,citecolor=red,urlcolor=blue}
\usepackage[normalem]{ulem}
\usepackage{tikz}
 \usetikzlibrary{decorations.pathmorphing,decorations.markings,trees,positioning,arrows}
 \usetikzlibrary{calc,patterns}
 \usetikzlibrary{shapes,backgrounds}

\newcommand{\non}{\nonumber\\}

\newcommand{\eps}{\epsilon}

\newcommand{\rar}{\rightarrow}
\newcommand{\cG}{{\cal G}}
\newcommand{\stout}[1]{{\color{blue}\sout{#1}}}
\renewcommand{\stout}[1]{}

\definecolor{fuchsia}{rgb}{0.55, 0.0, 0.55}

\definecolor{forest}{rgb}{0.13, 0.55, 0.13}

\newcommand{\dndT}{{\left(\frac{\partial \bar{n}}{\partial \bar T} \right)_{\bar \mu}}}
\newcommand{\dndmu}{{\left(\frac{\partial \bar{n}}{\partial \bar \mu} \right)_{\bar T}}}
\newcommand{\depsdT}{{\left(\frac{\partial \bar{\eps}}{\partial \bar T} \right)_{\bar \mu}}}
\newcommand{\depsdmu}{{\left(\frac{\partial \bar{\eps}}{\partial \bar \mu} \right)_{\bar T}}}
\begin{document}

\title{
\texorpdfstring{$T$}{T}-linear resistivity, optical conductivity and
	Planckian transport
 for a
	holographic local quantum critical metal in a
	periodic potential
}

\author{F. Balm, N. Chagnet$^{\ast}$, S. Arend, J. Aretz,
	 K. Grosvenor, M. Janse, O. Moors,
	\mbox{J. Post}, V. Ohanesjan, D. Rodriguez-Fernandez, K. Schalm and J. Zaanen.}
\affiliation{Instituut-Lorentz for Theoretical Physics, $\Delta$-ITP, Leiden University, The Netherlands.}
\email{chagnet@lorentz.leidenuniv.nl}

\begin{abstract}
High $T_c$ cuprate strange metals are characterized by a DC-resistivity that scales linearly with $T$ from the onset of superconductivity to the crystal melting temperature, characterized  by a current life time $ \tau_{\hbar} \simeq \hbar / (k_B T)$, the ``Planckian dissipation". At the same time, the optical conductivity ceases to be of the Drude form at high temperatures, suggesting a change of the underlying dynamics that surprisingly leaves the $T$-linear DC-resistivity unaffected. We use the AdS/CFT correspondence that describes strongly coupled, densely many body entangled metallic states of matter to study the DC thermo-electrical transport  properties and the optical conductivities of the local quantum critical Gubser-Rocha holographic strange metal in 2+1 dimensions in the presence of a lattice potential,  a prime candidate to compare with experiment. We find that the electrical DC-resistivity is linear in $T$ at low temperatures for a large range of potential strengths and wavevectors, even as it transitions between different dissipative regimes.  At weak lattice potential the optical conductivity evolves as a function of increasing temperature from a Drude form to a ``bad metal"  characterized by a mid-IR resonance without changing the DC transport, similar to that seen in cuprate strange metals. This mid-IR peak and notably its temperature evolution can be fully understood as a consequence of
Umklapp hydrodynamics: i.e. hydrodynamic perturbations are Bloch modes
in the presence of a lattice. At strong lattice potential an ``incoherent metal" is realized instead where momentum conservation no longer plays a role in the transport. We confirm that in this regime the thermal diffusivity  appears to be insensitive to the breaking of translations and can be explained by Planckian dissipation originating in universal microscopic chaos. A similar behavior has been found for holographic metals with strong homogeneous momentum relaxation.
The charge diffusivity does not submit to this chaos explanation, even though the continuing linear-in-T DC resistivity saturates to an apparent universal slope, numerically equal to a Planckian rate.
\end{abstract}
\maketitle

\tableofcontents

\section{The Planckian dissipation mystery versus computational holography.}

Are there states of matter that are governed by physical principles of a different kind from those  identified in the 20th century? This question arose in the study of strongly interacting electron systems realized in condensed matter, starting with the discovery of superconductivity at a high temperature in copper oxides. Their metallic states exhibit properties that appear to be  impossible to explain with the established paradigm explaining normal metals -- the Fermi-liquid theory  -- and these were accordingly called ``strange metals'' \cite{keimerQuantumMatterHightemperature2015, phillipsStrangerMetals2022}. 

An iconic signature is the linear-in-temperature electrical  resistivity \cite{husseyUniversalityMottIoffe2004}, an exceedingly simple behavior that is at odds with transport due to the quasiparticle physics of normal metals. A linear temperature dependence of the resistivity does occur naturally in conventional metals due to scattering of the quasiparticles against thermal disorder of the lattice above the Debye temperature. The problem in the cuprates and related systems is that the resistivity is linear all the way from the lowest to the highest temperatures where it has been measured. One anticipates some powerful principle of a new kind to be at work protecting this unreasonable simplicity. 

The measured optical conductivities reveal at lower temperatures a Drude response \cite{collinsReflectivityConductivityMathrmYBa1989,orensteinFrequencyTemperaturedependentConductivity1990,marelQuantumCriticalBehaviour2003,vanheumenStrangeMetalElectrodynamics2022}, signaling that the electrical conduction is controlled by a current relaxation time. Intriguingly, this time is very close to the ``Planckian dissipation'' time scale  $\tau_{\hbar} = \hbar / (k_B T)$. 
Planck's constant $\hbar$ plays a special role in dimensional analysis, as for instance the Planck scale of quantum gravity. Since $\hbar$ carries the dimension of action, $\tau_{\hbar}$ is a time scale associated with the thermal physics property of dissipation, the conversion of work into heat \cite{zaanenWhyTemperatureHigh2004,hartnollTheoryUniversalIncoherent2015}.  The case was made based on DC data that this Planckian time is remarkably universal also involving a variety of non-cuprate  unconventional metals exhibiting the linear resistivity \cite{bruinSimilarityScatteringRates2013,legrosUniversalTlinearResistivity2019,hartnollHolographicQuantumMatter2018}.

However, upon raising temperature further,  in the ``bad metal'' regime above the Mott-Ioffe-Regel bound optical conductivity studies show that the dynamical response changes drastically. Instead of a Drude response, a mid-infrared resonance develops with a characteristic energy that appears to increase with temperature, leaving a rather incoherent response at low energy \cite{delacretazBadMetalsFluctuating2017}. Remarkably, there is no sign of this radical reconfiguration of the dynamical response in the DC resistivity that continues to be a perfectly straight line, seemingly controlled by $\tau_{\hbar}$.

The occurrence of this universality of electrical conduction poses quite a problem of principle. On the one hand, considerable progress has been made in the understanding of dissipative phenomena in terms of  quantum thermalization,  explaining it in terms  of unitary time evolution and the collapse of the wave function (e.g. \cite{dalessioQuantumChaosEigenstate2016}). An early result is the identification of $\tau_{\hbar}$ as the characteristic universal dimension for the dissipation time of non-conserved quantities associated with  densely many-body entangled quantum critical states  \cite{zaanenLecturesQuantumSupreme2021a} realized at strongly interacting bosonic quantum phase transitions \cite{chakravartyTwodimensionalQuantumHeisenberg1989,sachdevQuantumPhaseTransitions2011}. 

This was very recently further clarified using both holographic duality (AdS/CFT correspondence) as well as studies in the closely related SYK models that connect macroscopic transport in such strange metals to microscopic quantum chaos. The central issue is that thermalization leading to local equilibrium may proceed very rapidly in densely entangled systems compared to  quasiparticle systems. Using out-of-time-order correlators (OTOC's) one can identify a quantum Lyapunov time $\tau_{\lambda}$ characterizing the microscopic time associated with the onset of quantum chaos that turns out to be bounded from below by  $\tau_{\hbar}$. In strongly correlated strange metals this microscopic time scale together with the chaos propagation ``butterfly" velocity $v_B$ can set the natural scale for the charge/heat and momentum diffusivities controlling the dissipative properties of the macroscopic finite temperature hydrodynamical fluid \cite{blakeUniversalChargeDiffusion2016,blakeUniversalDiffusionIncoherent2016,blakeThermalDiffusivityChaos2017}. 

However, in ordinary metals electrical conduction is controlled by total momentum conservation, as a ramification of translational invariance: any finite density system in the Galilean continuum {\em has} to be a perfect conductor. A finite resistivity is therefore rooted in the breaking of translation invariance. But how can this ever give rise to a universal resistivity controlled by  $\tau_{\hbar}$? This is the core of the mystery -- all explanations we are aware off rely on accidental, fine tuning circumstances, e.g. \cite{zaanenPlanckianDissipationMinimal2019,hartnollHolographicQuantumMatter2018,murthyStabilityBoundTlinear2021}. 

Holographic duality is now widely appreciated as a mathematical machinery that has a remarkable capacity to shed light on general principles associated with densely entangled matter \cite{ammonGaugeGravityDuality2015,zaanenHolographicDualityCondensed2015,hartnollHolographicQuantumMatter2018,zaanenLecturesQuantumSupreme2021a}, the ``scrambling" that we just discussed being a case in point. It achieves this by dualizing the densely entangled quantum physics into a gravitational problem in one higher dimension that is computable with (semi-)classical General Relativity. However, this is only a relatively easy mathematical affair  for a homogeneous translationally invariant space. When one breaks the spatial translation symmetry the Einstein equations become  a system of highly non-linear partial differential equations. If one wishes to have a full view on what holography has to say about transport in the laboratory systems one has to confront this challenge. Invariably a very strong effective potential  due to the background of ions is present in the laboratory strange metals, and it is even believed to be a {\em necessary} condition to obtain strongly correlated electron behavior
\cite{zaanenBandGapsElectronic1985,leeDopingMottInsulator2006,phillipsMottness2007}. But what has holography to tell about the effects of strong lattice potentials on strange metal transport? 

This can only be accomplished numerically. Although relatively efficient numerical relativity algorithms are available, the computations are demanding. Proof of principle was delivered that it can be done \cite{horowitzOpticalConductivityHolographic2012,horowitzFurtherEvidenceLatticeInduced2012,gauntlettQuantumCriticalityHolographic2010,donosThermoelectricPropertiesInhomogeneous2015,withersHolographicCheckerboards2014} and we set out to explore this more systematically. We focused specifically on the so-called Gubser-Rocha (GR) holographic strange metal \cite{gubserPeculiarPropertiesCharged2010}. This is unique in the regard that it is characterized by ``local quantum criticality" (a dynamical critical exponent $ z \rightarrow \infty$) as well as a Sommerfeld  entropy $s \sim T$ in the regime $T \ll \mu$, generic properties that appear to be realized by the cuprate strange metals \cite{zaanenLecturesQuantumSupreme2021a}. In such strongly coupled systems this then also predicts a linear-in-$T$ resistivity \cite{davisonHolographicDualityResistivity2014}. For comparison we also include results for the elementary Reissner-Nordström holographic strange metal. This also exhibits local quantum criticality, but it has a (pathological)
finite zero temperature entropy.
 
 \subsection{Main observations and summary of the results.}
 
 We consider a 2+1 dimensional 
strongly interacting strange metal holographically dual to the Gubser-Rocha model in the presence of a harmonic square ionic lattice  background encoded in the chemical potential
 \begin{align}
\mu(x,y) &= \bar{\mu}  \left( 1 + \frac{A}{2}  ( \cos ( G x) + \cos ( G y) )  \right)
\end{align}
We numerically compute the full set of DC thermo-electrical transport coefficients --- electrical conductivity $\sigma$, thermal conductivity $\bar{\kappa}$, the thermo-electrical coefficient $\alpha$ --- up to very large potentials ($A \simeq 8$) and  temperatures as low as  $T\simeq 0.005 \mu$. For stronger potentials
we sometimes resort to uni-directional 1D potentials to maintain numerical control. In addition, we also compute the optical conductivities. Because of numerical difficulties we encountered this is limited to intermediate potential strength ($A \lesssim 1-2$) and 1D lattices. 
 
From this computational experiment we make three remarkable observations:
\begin{figure}[t!]
		\includegraphics[width=0.8\textwidth]{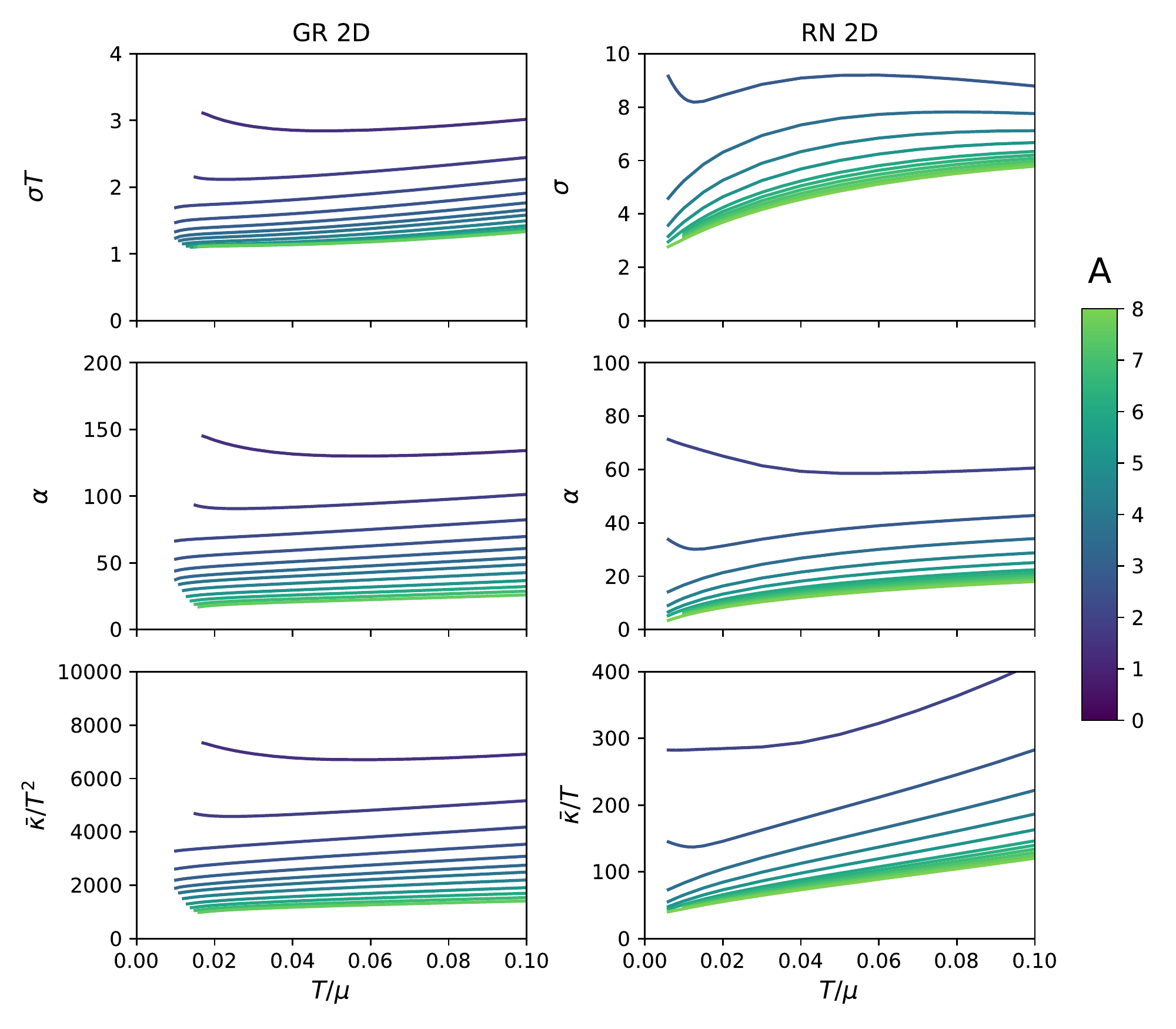}
		\caption{The thermo-electrical DC transport coefficients as functions of temperature $T$ in units of the chemical potential $\mu$ for the Gubser-Rocha (GR, left column) and Reissner-Nordström (RN, right column) metals in a 2D square lattice harmonic background potential with wave vector $G = 0.1 \sqrt{2}\, \mu$ and a strength $0 < A < 8$. $\sigma$, $\alpha$ and $\bar{\kappa}$ are the electrical conductivity, thermo-electrical cross conductivity and the overall thermal conductivity respectively. The electrical conductivity of the GR metal (top-left panel) shows for all potentials  a nearly  linear in temperature resistivity ($\rho = 1/\sigma \sim T$) with a slope that shows saturating behavior for large potentials.}  \label{fig:lin-in-temp-res-plus-saturation-part2}
\end{figure}

\begin{enumerate}

\item 
	The DC electrical resistivity of the Gubser-Rocha metal becomes to good approximation linear in temperature at low temperatures, see the upper left panel in Fig.~\ref{fig:lin-in-temp-res-plus-saturation-part2}. Strikingly, we find the slope of this linear resistivity  to saturate for an increasing potential strength after correcting for a spectral weight shift. 
 This suggests a connection with the universal Planckian dissipation bound: using the optical conductivity to deconvolve this in a total spectral weight and a current life time,  the saturation value for the latter is close to $\tau_{GR} = \frac{1}{2\pi} \hbar /(k_B T)$ (see Fig.~\ref{fig:saturation-corrected}).

The electrical conductivity of the Reissner-Nordstr\"{o}m (RN) metal also saturates for large potential strength at a roughly temperature independent value, although less perfect. The gross differences in temperature dependencies of the GR and RN metals between the electrical conductivity appear to reflect the different temperature dependencies of the entropies. We will discuss below why this is not so. Despite first appearances, the thermo-electric ($\alpha$) and heat ($\bar{\kappa}$) conductivities do not saturate at larger lattice potentials, but vanish as $1/A$ (see Fig.~\ref{fig:saturation-all}).

\begin{figure}[t]
\includegraphics[width=0.9\textwidth]{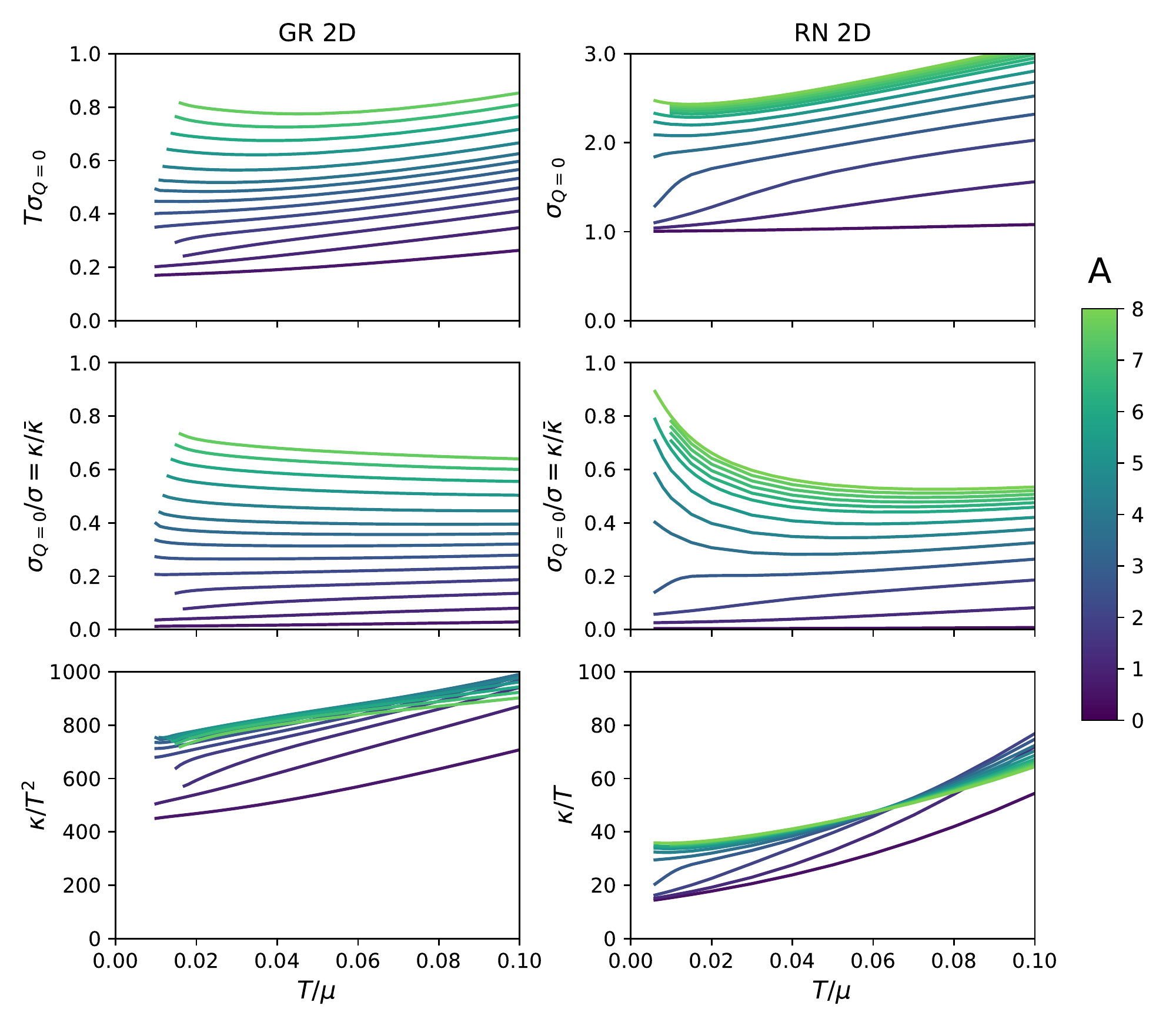}
\caption{ The electrical conductivity at zero heat current $\sigma_{Q=0}$ shows that as the lattice strength is increased the non-convective current anchored in charge diffusion becomes the dominating conduction channel. At the largest lattice strength where $A = 8$, the ratio of non-convective to convective transport $\sigma_{Q=0}/\sigma$ reaches up to 80\%,  signalling that momentum conservation is nearly completely destroyed. By definition, the fraction $\sigma_{Q=0}/\sigma$ is equal to the ratio $\kappa/\bar{\kappa}$. The open boundary thermal conductivity $\kappa$ anchored in thermal diffusion is rather independent of the lattice strength, barely changing after a moderate value of $A=1$ has been reached. Parameters are the same as in Fig.~\ref{fig:lin-in-temp-res-plus-saturation-part2}.
} 
\label{fig:incoherentransportcoef}
\end{figure}

\item
We can separate out the convective overall transport from more microscopic diffusive transport by considering the heat conductivity with zero electrical current \mbox{$\kappa=\bar{\kappa}-T\alpha^2/\sigma$}, also known as the open boundary heat conductivity.
Similarly, one can define an electrical conductivity without heat transport \mbox{$\sigma_{Q=0}=\sigma-T\alpha^2/\bar{\kappa}$} that is a (non-perfect) proxy for transport anchored in charge diffusion --- it is proportional to charge diffusion, but its thermodynamic scaling is also determined by cross-terms with the convective part. These 
are shown in Fig.~\ref{fig:incoherentransportcoef}. The $\sigma_{Q=0}$ is also (nearly) inversely proportional to temperature up to the largest potentials, similar to the overall $\sigma$. Most importantly, however,  we see that for large potentials this
diffusion-anchored contribution to the conductivity dominates the transport (middle panels): up to $\sim 80 \%$ of the electrical currents is anchored in the diffusive sector. Similarly, the diffusion-anchored open boundary thermal conductivity ($\kappa$, lowest panels) accounts for almost the full heat conductivity $\bar{\kappa}$ of Fig.~\ref{fig:lin-in-temp-res-plus-saturation-part2} 
in the large potential regime. This signals that for the strongest potentials the system approaches closely the incoherent metal regime addressed by Hartnoll \cite{hartnollTheoryUniversalIncoherent2015} where there is no longer a sense of momentum conservation;
It is  governed  instead by a ``hydrodynamics" that only relies on energy- and charge conservation. 
A key observation is that this is the regime which displays the ``Planckian saturation" of the electrical resistivity highlighted above in Fig.~\ref{fig:lin-in-temp-res-plus-saturation-part2}. In other words, this is the regime that should contain the clue behind the saturation phenomenon. 

\begin{figure}[t!]
		\centering
		\includegraphics[width=0.49\textwidth]{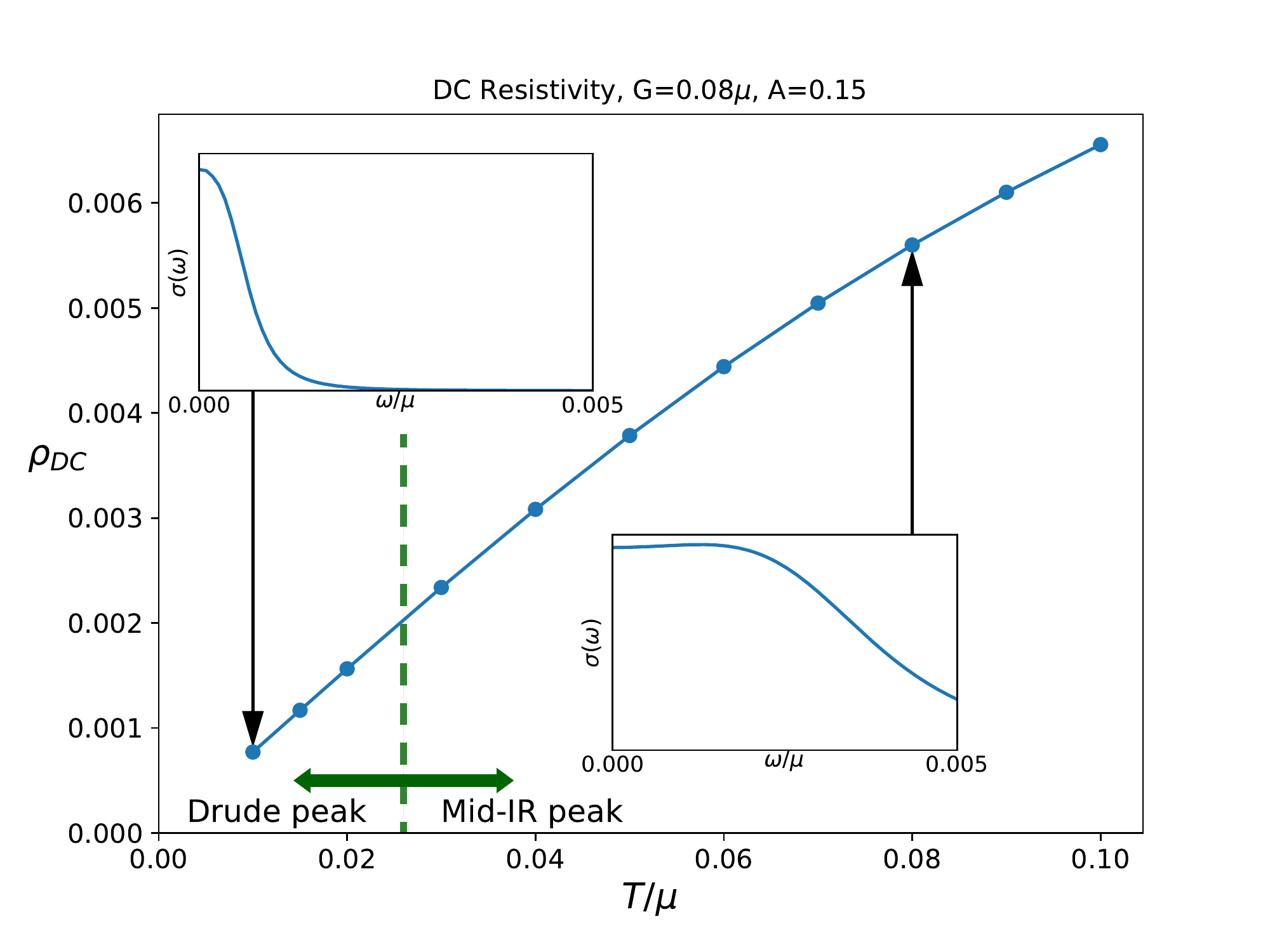}
		\includegraphics[width=0.49\textwidth]{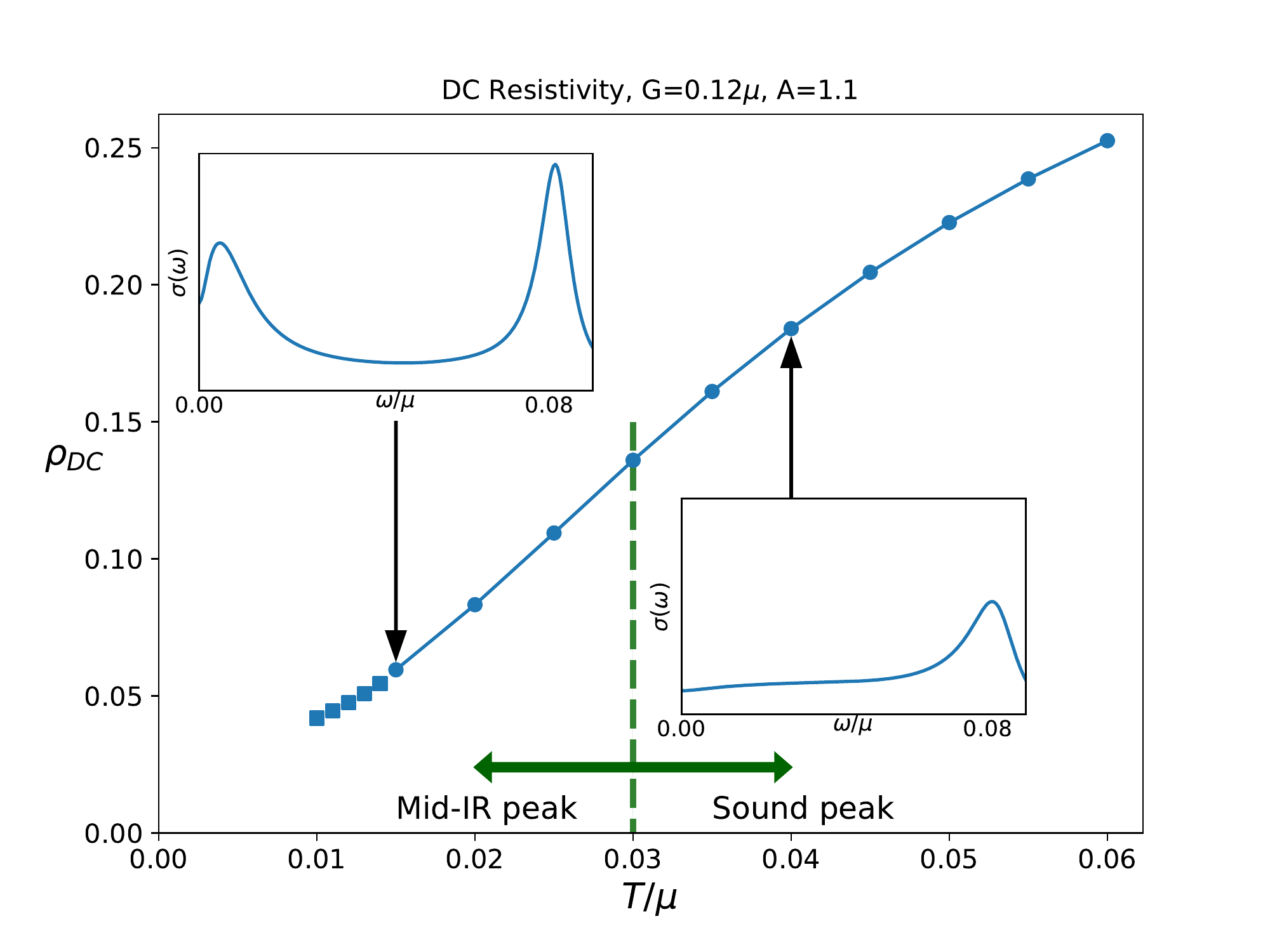}
		\caption{The DC resistivities for the small- ($A =0.15$, left panel) and intermediate ($A=1.1$, right panel) lattice potential  of the Gubser-Rocha metal are in both cases (nearly) linear in temperature. However, in both cases the optical conductivity (insets) undergoes radical changes when temperature increases. At the lowest temperatures in the small potential case (left panel)  this consists of a simple Drude peak that gradually turns into an incoherent ``flat top'' low frequency response terminating at a developing ``mid IR peak". The characteristic temperature where this happens decreases for increasing potential strength. In the right panel, a full fledged mid IR peak has already developed at a low temperature $T \sim 0.015 \mu$ (left inset), while it is accompanied by a high energy peak at $\omega =c_sG = \frac{1}{\sqrt{2}}G$ that is identified to be the ``Umklapped sound peak". Upon further raising temperature, the mid IR peak moves up in energy to eventually merge with the sound peak (right inset).}
		\label{fig:lin-in-temp-res-plus-saturation}
\end{figure}

\item 
Computing the optical conductivities, we find
	for small lattice potential
	at the lowest temperature a perfect Drude peak (left panel  Fig.~\ref{fig:lin-in-temp-res-plus-saturation}). 
	Strikingly, upon raising temperature this evolves into a  mid IR peak,  reminiscent of what is seen in experiment. Although the dynamical response shows such drastic changes, these do not imprint {\em at all} on the linearity in temperature of the DC resistivity remarkably. This finding is repeated in the intermediate potential case.
There, the electrical DC resistivity can even stay linear-in-T through a second change in relaxational dynamics from the mid-IR-peak regime to a fully incoherent metal.  
Just within reach of our numerics, the spectrum at the lowest temperature (left inset)  
now already displays the mid-IR peak, and we have good reasons to expect that at even lower temperatures, outside of our numerical reach, a Drude response should still be present.  There is also a second peak at higher frequencies that can be identified with the ``Umklapp copy" of the sound mode
at an energy  $\omega=c_s G$  where $c_s$ is the speed of sound and $G$ the lattice wavevector (Section \ref{sec:weaklattice-conductivities}). Upon raising temperature the mid-IR peak moves to higher frequency to eventually merge with the ``Umklapped sound" peak, transitioning to a fully bad incoherent metal regime (right inset), while the DC resistivity stays essentially linear-in-T throughout.  

These observations are reminiscent of the experimental observation that the linear-in-T DC resistivity appears to be completely insensitive to the change from ``good metal'' to ``bad metal'' behavior when temperature increases. This transition can be defined using the absolute value of the resistivity crossing the Mott-Ioffe-Regel limit  but perhaps a better way is to identify it through the dynamical response, associating the good metal regime with a Drude response while the bad metal has the incoherent ``mid IR peak" type of behavior as in our computations.

\end{enumerate}

To dissect these numerical results is an intensive exercise. We therefore provide an executive summary of the paper here. The reader interested in the details may proceed directly to Section \ref{homogeneousprop} and skip the remainder of this Introduction.

\subsubsection{The local quantum critical strange metals of holography and hydrodynamical transport} 

	Transport in holographic strange metals is governed by hydrodynamics (Section \ref{homogeneousprop}). Holographic strange metals originate in the quantum critical state of a non-trivial IR fixed point and the GR metal is singled out as the one with the right scaling properties to reproduce both the local quantum criticality and Sommerfeld entropy of the cuprate strange metals. The non-trivial fixed point is of a special kind in that it still has an intrinsic correlation length  $\xi \sim {1}/{\mu}$ (\cite{iqbalSemilocalQuantumLiquids2012} and Appendix \ref{app:causal}).
	Hydrodynamics has long been utilized to describe transport in such densely entangled critical states, and holography is no different; though it it is still an important open question whether transport in cuprate strange metals is hydrodynamical. In the Galilean continuum hydrodynamics is governed by (near) {\em momentum conservation} captured by the Navier-Stokes equations describing {\em convective} currents, also called ``coherent" in the condensed matter- and holographic communities. However, there are also transport channels that are controlled by only
	{\em diffusive} (or ``incoherent") transport. The overall electrical ($\sigma$) thermo-electric ($\alpha$) and thermal ($\bar{\kappa}$) transport coefficients are set by the sum of both convective and diffusive transport channels. 
	The open boundary thermal conductivity $\kappa=\bar{\kappa}-T\alpha^2/\sigma$ and the charge-without-heat transport $\sigma_{Q=0}=\sigma-T\alpha^2/\bar{\kappa}$ can be used to disentangle these.
	These zero out the dominant convective contribution. 
If Planckian dissipation occurs, the natural channel is this diffusive channel which can reflect universal microscopic dynamics. The convective channel is controlled by the way translational symmetry is broken and therefore unlikely to be universal. However, the convective channel dominates when translational symmetry is only broken weakly, and Planckian dissipation is therefore most natural in systems with strong translational symmetry breaking.

\subsubsection{Convective hydrodynamics in the presence of a weak lattice potential}

The presence of a lattice potential plays an important role in cuprate strange metals and this is the obvious way translational symmetry is broken. Placing the holographic strange metals in a background lattice with a perturbatively small potential strength the nature of the linear response of hydrodynamical transport is in fact familiar (Section \ref{sec:weakpotperttheory}).
Hydrodynamic fluctuations must be decomposed in Bloch modes that Umklapp at Brillouin zone boundaries. This holds for purely diffusive as well as propagating modes.
Well known is that the translational symmetry breaking by the lattice makes
momentum relax due to shear drag with
a
life time $\Gamma_{\text{shear}} = \eta G^2  /(\varepsilon + P) $ ($\varepsilon$ and $P$ being the energy density and pressure and $\eta$ the shear viscosity).
However,
a careful analysis reveals that
the Umklapp potential
gives rise to a mode coupling between this relaxational mode and the 
Umklapped
charge diffusion mode
characterized by a relaxation rate $\Gamma_{\text{charge}} = D_ c G^2$, where $D_c$ is the {charge} diffusivity.
For weak lattices $A\ll 1$ the result of this generic mode coupling problem 
is 
an optical conductivity  of the form
(cf. Eq.~\eqref{eq:12b} \& Eq.~\eqref{eq:two-pole-conductivity-formula}),
\begin{equation}
 \sigma (\omega) \sim \frac{\Omega - i \omega}{(\Gamma - i \omega) (\Omega - i \omega)  + \omega^2_0}
 \label{polecolsketch}
 \end{equation}      
 where $\omega_0$ is related to the strength of the mode coupling and $\Gamma$ and $\Omega$ are combinations of $\Gamma_{\text{shear}}$ and $\Gamma_{\text{charge}}$. Taking the DC limit
 gives an overall current relaxation rate $\Gamma_{\text{DC}} = \Gamma + \omega_0^2 / \Omega$ controlled by two separate dissipative channels.

The above hydrodynamic analysis is only valid for lattice sizes $a=G^{-1}$ greater than the earlier emphasized retained correlation length $\xi \simeq 1/\mu$ of the IR fixed point or equivalently $G\ll \mu$
(Section \ref{sec:appl-of-hydro}). 
This length $\xi\simeq 1/\mu$ 
 where hydrodynamics provides the better perspective on transport than the quantum critical
 power law response set by the near horizon geometry as elucidated by Hartnoll and Hofman \cite{hartnollLocallyCriticalResistivities2012}.  In a lattice background this reflects itself in a strong change in the transport properties when the lattice momentum $G$ crosses this scale. The results in the above are all associated with the hydrodynamical regime ($G \ll \mu$); for large lattice momenta ($G \gtrsim \mu$)
 the additional Umklapp contribution to the dissipation of the currents is strongly suppressed (Fig.~\ref{fig:largeG-ACconductivity}). 

This Umklapp hydrodynamics can explain our observations at weak lattice potential (Section \ref{sec:weaklattice-conductivities}). When $| \Gamma - \Omega |> 2 \omega_0$ the AC conductivity displays
 a single peak, explaining the low temperature Drude-like result of Fig.~\ref{fig:lin-in-temp-res-plus-saturation}. Only for the lowest temperatures is this a pure Drude peak controlled by a single pole, however. In detail it originates in two diffusive poles, the Drude $k=0$ sound pole and the Umklapped charge diffusion pole; for each we fully understand their temperature dynamics from the underlying  hydrodynamic computation and the thermodynamical properties of the holographic strange metal.

At higher temperatures (and/or at stronger lattices) generically $| \Gamma - \Omega | < 2 \omega_0$ and a {\em real}, propagating part develops in modes controlling the AC conductivity. This pole collision explains the emergence of the mid-IR-peak in the dynamical response -- the numerical results are perfectly fitted by this form.

The same two-relaxational-current response was identified in the context of
a hydrodynamical fluid coupled to the fluctuations of a damped pinned charge density wave \cite{delacretazDampingPseudoGoldstoneFields2022}. 
There the peak emerges as the temperature is lowered as it can be identified as a pseudo-Goldstone mode of spontaneous translational symmetry breaking, that is absent at high temperatures.\footnote{Because the lattice is ultimately irrelevant in the deep IR, at the lowest temperatures the pseudo-Goldstone boson mid IR peak from spontaneous translational symmetry breaking will move again to $\omega=0$ or equivalently disappear as the temperature is lowered; see e.g. \cite{amorettiUniversalRelaxationHolographic2019}.}
Our discovery is that Umklapp hydrodynamics gives the right temperature evolution necessary to have a mid-IR-peak appear as temperatures increase. As emphasized in the introduction, this same development of a mid-IR peak in the optical conductivity as temperatures increase is observed in the strange metal phase of the high $T_c$ cuprates.

As emphasized, the DC resistivity can remain linear throughout this transition.
This can be explained by the fact that the scaling properties of the hydrodynamic parameters are inherited from the underlying non-trivial quantum critical IR fixed point.
For the GR strange metal both relaxation rates scale as $T$, whereas for the RN metal one scales as $T^0$ and the other as $T^2$. This manifestation of the differing detailed expressions for both relaxation rates shows that a simple interpretation of the scaling of the resistivity in terms of the entropy fails. 
Instead their scaling is determined at a deeper level by the quantum critical IR fixed point. It behooves us to point out at this stage that we are considering a rigid lattice only. We are at this stage not taking lattice vibrations or phonons into account. The underlying assumption is that in these intrinsically densely entangled system the strongly self-interacting degrees of freedom dominate all the physics and any phonon contribution is negligible. We comment on this further in the conclusion.

At intermediate lattice strengths a similar scenario can take place. Now the transport response is determined by four modes, the two modes above and two Umklapped sound modes at Re $\omega =\pm c_sG$.
Upon raising temperature
the pole responsible for the mid-IR peak
moves up with temperature to approach close to the Umklapped sound pole, such that it gets obscured and only one peak remains in the AC conductivity (right inset of Fig.~\ref{fig:lin-in-temp-res-plus-saturation}). From this temperature onward the low frequency AC spectrum becomes roughly temperature independent. We can track this
in terms of the quasinormal modes  (Fig.~\ref{fig:evolution-intermediatelattice}) although we can no longer rely on the perturbative expansion to enumerate it.
For a large part of this intermediate lattice regime, the DC resistivity is still effectively captured by the expression $\sigma_{DC}=\omega_p^2/(\Gamma+\omega_0^2/\Omega)$, though one needs a careful AC-fit to extract the values. Again, its temperature scaling is set by the non-trivial IR fixed point and can remain unaffected by the change in dissipative dynamics in the AC conductivity. 

\subsubsection{The incoherent hydrodynamics at large lattice potential.}

At large lattice potentials momentum is strongly broken and we enter in a qualitatively different regime (Section \ref{sec:saturation}).
Observationally this is where the numerically extracted relaxation rate of the DC conductivity of the GR metal saturates at about the Planckian value $\Gamma_{\text{corrected}} \simeq 2\pi T$ (Fig.~\ref{fig:saturation-corrected}). Because momentum is strongly broken, the framework to understand whether this can be verified is the one where
 transport is governed by only two conserved quantities, energy and charge \cite{hartnollTheoryUniversalIncoherent2015}. Their fluctuations consist of two coupled diffusive modes
with diffusion constants that are not the same as they are in the homogeneous system. At strict $T=0$, charge and energy transport formally decouple and the electrical conductivity is governed by one of these modes $\sigma = \chi D_+$ with $\chi$ the charge susceptibility, while  the thermal conductivity $\kappa = c_n D_-$ is governed by the other with $c_n$ is the specific heat at constant charge density. At low but finite temperature they mix perturbatively, but are still dominated by their $T=0$ scaling. From our numerics we conclude that $D_+ \sim T^{-1}$ whereas $D_-\sim T$ for the GR metal; similar behavior has been established in homogeneous holographic strange metals with strong  momentum  relaxation (GR metal in a Q-lattice) where the homogeneous geometry allows analytical solutions \cite{niuDiffusionButterflyVelocity2017}.
It has been argued that the temperature dependence of the thermal diffusivity empirically defined as $D_T \equiv \kappa/c_n$ should be insensitive to the breaking of translations and reduces to one of the incoherent diffusivities $D_-$ at low temperature and strong lattices. Moreover, it can be related to microscopic chaos through a butterfly velocity $v_B^2$ times a maximal Lyapunov rate $\lambda= 2\pi T$ that embodies Planckian dissipation $D_T=\frac{1}{2}v_B^2/(2\pi T)$ \cite{blakeUniversalChargeDiffusion2016,blakeUniversalDiffusionIncoherent2016,blakeThermalDiffusivityChaos2017}. Provided we can extrapolate from the homogeneous result that in the non-trivial IR fixed point of the GR metal in a strong lattice the butterfly velocity still scales as $v_B^2 \sim T^2$, this is consistent with our findings. The puzzle is the DC-conductivity and charge response. We conjecture that the Planckian relaxation set by the maximal Lyapunov rate should still govern charge transport as well. Given that on dimensional grounds $D_+\sim (v^{\text{charged}})^2/(2\pi T)$, this can be only so if the velocity appearing in charge diffusion is not set by the universal butterfly velocity. In other words scrambling depends on the quantum numbers of the operators probing chaos; there are hints that this is true \cite{ageevWhenThingsStop2019,sorokhaibamPhaseTransitionChaos2020,colangeloChaosOverlineQSystem2020,chenManybodyQuantumDynamics2020}. If it can be shown that $v^{\text{charged}} \sim T^0$ this could explain not only the observed linear-in-T resistivity at strong lattice potentials in the GR metal, but also its saturation to the Planckian value.

We will end with a short discussion in Section \ref{outlook} of these results with a focus on the possible relevance to experiment. We also include a number of Appendices where we discuss various technical details.

\section{Holographic strange metals, transport and translational symmetry breaking.}
\label{homogeneousprop}

In the absence of a lattice, the homogeneous finite density strange metals \cite{ammonGaugeGravityDuality2015,zaanenHolographicDualityCondensed2015,hartnollHolographicQuantumMatter2018,zaanenLecturesQuantumSupreme2021a}  of holography are characterized by a non-trivial IR fixed point. These are specified by a handful of anomalous scaling dimensions: the dynamical critical exponent $z$, the hyperscaling violation dimension $\theta$ and the charge exponent $\zeta$, expressing the scaling of time with space, the scaling of the thermodynamically relevant degrees of freedom with volume, and the running of the charge, respectively. Experimental evidences suggest that the cuprates are ``local quantum critical'' \cite{varmaPhenomenologyNormalState1989,varmaColloquiumLinearTemperature2020,mitranoAnomalousDensityFluctuations2018}, referring to $z \rightarrow \infty$, while electronic specific heat measurements in the high temperature strange metal regime exhibit a Sommerfeld entropy, $s \simeq k_B T/\mu$ (see e.g. \cite{loramElectronicSpecificHeat1994}) where $\mu$ is the chemical potential taking the role of the Fermi energy. Though the notion that cuprate strange metals are explained by a non-trivial IR fixed point was put forth independently of holography, the fixed point that shares the rough qualitative characteristics was first discovered using AdS/CFT. Amongst the holographic strange metals this is the so-called Gubser-Rocha strange metal \cite{gubserPeculiarPropertiesCharged2010}, being the only holographic strange metal in the general classification that reconciles $z \rightarrow \infty$ with Sommerfeld entropy.  Within the larger class of holographic strange metals, the critical scaling at the IR fixed point insists that the entropy should scale as $s \sim T^{(d-\theta)/z}$. For $z \rightarrow \infty$ and $d - \theta$ finite the entropy should therefore be temperature independent, implying a zero temperature entropy. This is the case for the holographic strange metal dual to the Reissner-Nordstr\"{o}m black hole and the closely related SYK systems. The GR metal is characterized by a double scaling limit such that $z, -\theta \rightarrow \infty$ while $-\theta/z = 1$. This reconciles a low temperature Sommerfeld entropy $s \sim T+\ldots$ with local quantum criticality.  For comparison we will also present results for the Reissner-Nordstr\"{o}m strange metal \cite{zaanenHolographicDualityCondensed2015,hartnollHolographicQuantumMatter2018,faulknerHolographicNonFermiLiquid2011}.
For a qualitative understanding of our results nothing more than the thermodynamics of the fixed point are required (summarized in Table I). The precise details RN and GR holographic strange metal and the duality map are discussed in Appendix \ref{app:set-up}.
\begin{table}[!ht]
\centering
\begin{tabular}{r|c|c|c|}
&IR Scaling &\begin{tabular}{c}RN\\ $\theta = 0, z = \infty$\end{tabular}&\begin{tabular}{c}GR\\ $z, -\theta = \infty$\end{tabular}\\
\hline
&&&\\[-1em]
Entropy&~ $s/\mu^2 \sim (T/\mu)^{(d-\theta)/z}$ ~&~ $ s/\mu^2 \sim (T/\mu)^0$ ~&~ $s/\mu^2 \sim (T/\mu)^1$~\\
&&&\\[-1em]
Charge Density& * &$n/\mu^2 \sim (T/\mu)^0$ & $n/\mu^2 \sim (T/\mu)^0$\\
\hline
\end{tabular}
\caption{
IR scaling behavior in holographic strange metals of the entropy density $s$ and the charge density $n$ in terms of the chemical potential $\mu$ and the temperature $T$. The first column highlights the general formula of holographic scaling geometries. The last two columns focus on the two holographic models with local quantum criticality ($z\rightarrow \infty$) of interest in this paper: the Reissner-Nordstr\"{o}m and the Gubser-Rocha model. 
$(*)$: For a discussion on this, see section 4.2.4 of \cite{hartnollHolographicQuantumMatter2018}.
}
\end{table}

The motivation for this study is that all experimental strange metals are known to occur in the presence of an excessively strong effective ionic background potential felt by the electron system, the Mottness of the cuprates being case in point (see e.g., \cite{zaanenBandGapsElectronic1985,leeDopingMottInsulator2006,phillipsMottness2007}). The commonality of this lattice potential suggests an importance in observed systems of which the effects on the holographic strange metals have not yet been systematically investigated. We shall study the GR and the RN AdS black holes dual to 2+1 dimensional strange metals  where we break translations by either a one dimensional or two-dimensional explicit periodic square ionic lattice potential encoded in the local chemical potential
\begin{align}
\mu_{1D}(x,y) &= \bar{\mu}  \left( 1 + A \cos ( G x) \right)\non
\mu_{2D}(x,y) &= \bar{\mu}  \left( 1 + \frac{A}{2}  ( \cos ( G x) + \cos ( G y) )  \right)
\label{squarelatpot}
\end{align}
The parametrization is such that the maximal deviation from the average is $\pm A$  in both cases. 

The above explicit lattice condition appears as boundary conditions in the dual holographic gravitational description of the strange metal system in question. The difficulty is that studying such explicit translational symmetry breaking is only possible numerically outside perturbation theory. We solve the full set of spatially dependent Einstein-Maxwell-Dilaton equations of motion for the GR and RN strange metals using the DeTurck gauge in a Newton-Raphson scheme \cite{krikunNumericalSolutionBoundary2018,headrickNewApproachStatic2010,adamNumericalApproachFinding2012}. A summary is given in Appendix \ref{app:numerics}. DC transport is computed by numerically solving for the Stokes flow problem at the horizon \cite{donosThermoelectricDCConductivities2014,banksThermoelectricDCConductivities2015,donosDCConductivityMagnetised2016,donosHolographicDCConductivity2017}. All numerical computations employ a higher-order finite difference scheme where the radial coordinate is discretized on the Chebyshev-Lobatto nodes (Appendix \ref{app:numerics}).
 
We treat the numerical data obtained as the outcome of an experiment. However, the framework in which to analyze this data is known.
As we already emphasized, the dense entanglement of the quantum many body system described holographically by its dual gravity theory drives a very rapid quantum thermalization. This implies that local equilibrium sets in very rapidly, which in turn implies that, in the homogeneous background with no lattice, transport at macroscopic times and lengths is governed by hydrodynamics. Different from the quasiparticles in Fermi-liquid metals, a strange metal flows like water. 
It is a general hydrodynamical principle that it can be decomposed in convective- (also called ``coherent'') and diffusive (``incoherent'') flows. The former refers to the motion of the fluid as a whole as protected by the conservation of total momentum in the translationally invariant homogeneous background. When the translational symmetry is weakly broken, ---
introduced by hand through a momentum decay rate $\Gamma_{\text{mom.rel.}}=\tau_{\text{mom.rel.}}^{-1}$ as the largest relaxation time,--- a straightforward hydrodynamic analysis yields\footnote{see e.g. the review \cite{Baggioli:2022pyb}.} 
\begin{eqnarray}
\sigma (\omega) & = & \frac{n^2} {\chi_{\pi\pi}} \frac{1}{\Gamma_{\text{mom.rel.}}} + \sigma_{\mathrm{inc}}  \nonumber \\
\alpha (\omega ) & = &\frac{n s} {\chi_{\pi\pi}} \frac{1}{\Gamma_{\text{mom.rel.}}}  + \alpha_{\mathrm{inc}} \nonumber \\
\bar{\kappa} (\omega) & = & \frac{s^2 T} {\chi_{\pi\pi}} \frac{1}{\Gamma_{\text{mom.rel.}}} + \bar{\kappa}_{\mathrm{inc}} 
\label{thermoeltransport}
\end{eqnarray}
Here $n, s$ of the convective terms are the charge and entropy density respectively, and $\chi_{\pi\pi}$ is the momentum susceptibility.
For non-relativistic hydrodynamics $\chi_{\pi\pi}=nm$ with $m$ the constituent quasiparticle mass and one recognizes the Drude model. For relativistic hydrodynamics appropriate to strange metals where a linear dispersion relation of charged constituents induces an emergent Lorentz symmetry, and for holographic strange metals studied here the momentum susceptibility equals $\chi_{\pi\pi}=\eps+P$, the sum of the energy and pressure density respectively.
The Lorentz symmetry also demands that the incoherent contributions are related to each other by
$\sigma_{\mathrm{inc}}  = \sigma_Q,  \alpha_{\mathrm{inc}} = - \frac{\mu}{T} \sigma_Q$ and $\bar{\kappa}_{\mathrm{inc}}  = \frac{\mu^2 }{T} \sigma_Q$ in terms of a transport coefficient $\sigma_Q$.\footnote{There is one exception. If the translational symmetry breaking happens in only one of the spatial dimensions $\alpha_\text{inc}$ and $\kappa_{\text{inc}}$ vanish \cite{davisonDissectingHolographicConductivities2015}. In that particular case a subleading term in the numerator of the convective term precisely cancels the incoherent term in the thermo-electric and heat conductivity.} 

Writing $\sigma_{\text{inc}} = \frac{T}{\mu^2} \bar{\kappa}_{\text{inc}}$, $\alpha_{\text{inc}} = -\frac{1}{\mu}\bar{\kappa}_{\text{inc}}$ instead, this reveals that in a Galilean invariant system where both $c\rightarrow \infty $ and $\mu = m_ec^2+\ldots \rar \infty$, only the incoherent heat contribution survives.
It is a highlight of non-relativistic finite temperature Fermi-liquid theory that such a diffusive heat conduction is present even dealing with spin-less fermions,
mediated by the Lindhard continuum. This $\bar{\kappa}_{\text{inc}} = c_n D_{T}$, where the specific heat at constant density (equal to the specific heat at constant volume) $c_n \sim T$, while the thermal diffusivity $D_T \simeq v^2_F \tau_{\mathrm{col}}$ where  $\tau_{\mathrm{col}} \simeq \frac{E_F}{k_B T} \tau_{\hbar}$; therefore $\bar{\kappa}_{\text{inc}} \sim 1/T$ as verified e.g. in the $^3$He Fermi liquid. 
In contrast in the non relativistic limit $\epsilon + P \rightarrow \mu n\sim n m c^2$ the electrical conductivity becomes purely convective and  one recognizes the familiar Drude weight expressed in the plasma frequency as $\omega^2_p = ne^2/m$. 

The incoherent contributions to transport are in principle measurable in the laboratory by zeroing out the coherent part. This can be done by measuring heat transport in the absence of  charge transport (open circuit thermal
conductivity) $\kappa$ or  charge transport without heat,  $\sigma_{Q=0}$ equal to
\begin{eqnarray}
\kappa & = & \bar{\kappa} - \frac{T \alpha^2}{\sigma}~, \nonumber \\
\sigma_{Q=0} & = & \sigma - \frac{T \alpha^2}{\bar{\kappa}}~.
\label{diffusioncontr}
\end{eqnarray}
Note that in the Galilean limit when there is only an incoherent heat conductivity $\kappa = \bar{\kappa}_{\text{inc}}$; note therefore that in ordinary metals the thermal conductivity consists completely of the incoherent contribution in this language (see \cite{lucasElectronicHydrodynamicsBreakdown2018}).

These incoherent contributions are diffusive. The open boundary combinations Eq.\eqref{diffusioncontr} are therefore a mixture of diffusive and convective transport. Nevertheless, it is useful and conventional to define the charge and thermal diffusivities $D_c \equiv \sigma/\chi$ and $D_T \equiv \kappa/c_n$, where $\chi$ is the charge susceptibility, and $c_n$ the heat capacity. In the remainder of this text, we will see that when translational symmetry is strongly broken and the convective part is strongly suppressed, these diffusivities are directly related to diffusion constants in transport. These ``incoherent metal'' diffusivities and diffusion constants should not be confused with the well-known diffusion of charge $D_\rho$ and energy $D_\pi$ in weak or vanishing translational symmetry breaking. As we shall see in the Gubser-Rocha metal the latter are both linear-in-$T$ at low temperature while they are $T$-independent at low temperature in Reissner-Nordstr\"om. In the incoherent metal, in contrast, we will see that $D_T \sim T$ while $D_c \sim T^{-1}$.

\subsection*{Will the real Planckian dissipating channel make itself known?}

The point of this brief hydrodynamical exposition is to highlight the fundamental issue we address in this article. The above illustrates that even in the simplest Drude hydrodynamics there are two dissipative channels: the convective coherent Drude term encoding the way translational symmetry is broken, and the incoherent term related to a diffusion of microscopic origin. 
For weak lattice potentials, or more generally for weak translational symmetry breaking, the convective Drude term is much larger than the incoherent term. With the conjecture that in strongly correlated critical points the shear viscosity is bounded by the entropy $\eta \geq s/4\pi$, two of us, together with R. Davison, proposed that in disordered strange metals the usual shear viscosity based momentum relaxation rate $\Gamma_{\text{mom.rel.}} \sim \eta \sim s $ can explain a linear-in-T resistivity for a system with Sommerfeld entropy \cite{davisonHolographicDualityResistivity2014}. The connection between the resistivity and the entropy would explain the universality and the minimal viscosity would be the encoding of Planckian dissipation. Moreover, this argument is also consistent with a Drude response in the optical conductivity. The counterargument is that this only holds in detail for marginal disorder. Relevant or irrelevant disorder would significantly limit the regime of applicability of this argument \cite{hartnollTransportIsingnematicQuantum2014,lucasScaleinvariantHyperscalingviolatingHolographic2014}. 

Taking a step back, it actually is difficult to argue that a universal phenomenon such as Planckian dissipation should manifest itself through the convective channel, as this coherent channel will generically depend on the details of translational symmetry breaking \cite{blakeUniversalChargeDiffusion2016,erdmengerSwaveSuperconductivityAnisotropic2015}. The far more natural channel for Planckian dissipation would be the incoherent diffusive channel. But if one takes this point of view, one can no longer use it to explain the universal linear-in-T DC resistivity in strange metals. These all show strong Drude behavior in the optical conductivity, and the DC conductivity is therefore  set by the coherent response in the context of weak translational symmetry breaking. It appears to be a Catch-22.\footnote{A Catch-22 is a paradoxical situation which cannot be escaped by design. It originates from the eponymous novel written by Joseph Heller and published in 1961.} Either a Planckian dissipation can set the universally observed linear-in-T resistivity in strange metals, but then the AC conductivity ought to be Drude, or weak translational symmetry breaking sets the resistivity, but then it is hard to see how it can be universal.

We will resolve this conundrum by showing explicitly that in weak lattice near a non-trivial IR fixed point, the thermodynamics of the fixed point together with a fixed-point-controlled scaling of transport coefficients can set the DC resistivity in a universal sense, independent of the dissipative channel shown in the AC conductivity. Qualitatively this is an extension of the Davison-Schalm-Zaanen \mbox{$\Gamma_{\text{mom.rel.}} \sim \eta \sim s$} result. 
At the same time, for large lattice strengths the incoherent part becomes dominant and indeed shows universal Planckian dissipation as surmised by Blake and others \cite{blakeUniversalChargeDiffusion2016,blakeUniversalDiffusionIncoherent2016,blakeThermalDiffusivityChaos2017}. For good measure we state that there may still be a deeper way to also understand the weak lattice results in terms of Planckian dissipation. Even though they appear non-universal, the observed scaling, together with the way the Sommerfeld entropy is a natural bounding behavior at low temperatures, leaves this possibility open.

\section{Umklapp hydrodynamics for weak lattice  potentials.}
\label{sec:weakpotperttheory}

As we emphasized, in the low frequency limit at macroscopic long wavelengths holography reduces to hydrodynamics albeit with specific transport coefficients \cite{Bhattacharyya:2007vjd}. 
A fundamental principle behind the theory of hydrodynamics is local
equilibrium. The state of the fluid can be described by a slowly
spatially varying energy-momentum tensor $T_{\mu\nu}(x)$ and in the
presence of a $U(1)$ charge, a current $J^{\mu}(x)$. In turn the local equilibrium condition 
implies that one can also describe fluid behavior in the presence of a
slowly spatially varying external potential whether temperature
$T(x)$, pressure $P(x)$, or chemical potential $\mu(x)$ \cite{lucasConductivityStrangeMetal2015,lucasHydrodynamicTransportStrongly2015,lucasTransportInhomogeneousQuantum2016},
Suppose this background is periodic in the coordinate $x$.
The hydrodynamical problem of relevance is nothing else than that of a hydrodynamical fluid like water that is flowing through a periodic ``array" of obstacles weakly perturbing the flow, characterized by a microscopic ``lattice constant".
This is a rather unusual circumstance in standard hydrodynamics and 
we are not aware of any literature addressing the role of Umklapp in the AC structure of the correlators, though a beginning was made in \cite{donosDiffusionInhomogeneousMedia2017}.

But it represents an elementary exercise, and the answer is readily understood.
From elementary solid state physics it is well known that a quantum
mechanical wave function in a periodic background experiences
Umklapp. This is purely a wave phenomenon and the principle therefore
also applies to classical waves as described by hydrodynamics.
Both a quantum mechanical wave function and linearized hydrodynamic fluctuations
around equilibrium are described by a differential equation of the form
\begin{align}
	\label{eq:1}
	(\partial_t + M(x))\phi(x) =0
\end{align}
If $M(x)$ is periodic $M(x+\frac{2\pi n}{G})=M(x)$, then $\phi(x)$ can be decomposed
in Bloch waves $\phi(x) = \frac{1}{2\pi G}\int_{-G/2}^{G/2} dk\sum_{n}
\phi_n(k) e^{i(k+nG)x}$. Taking $M(x) = -M_0\partial_x^2+ A \cos(Gx)$
as canonical example, one can solve Eq.~\eqref{eq:1} perturbatively in
$A$. Defining $\phi_n(k)= \phi^{(0)}_n(k)+A\phi^{(1)}_n(k) +
\frac{A^2}{2} \phi^{(2)}_n(k) + \ldots$, the solution to first order $A$ is
\begin{align}
	\label{eq:3}
	\phi_n(k) =\phi_n^{(0)}(k)+
	\frac{A}{2G(G-2k)M_0}\phi_{n-1}^{(0)}(k)-\frac{A}{2G(G+2k)M_0}\phi_{n+1}^{(0)}(k) +\ldots
\end{align}
This mixing between the different Bloch waves is Umklapp. In hydrodynamics these Umklapped responses have 
already been observed several years ago in numerical computations of holographic metals in
explicit periodic lattices in \cite{horowitzFurtherEvidenceLatticeInduced2012,lingHolographicLatticeEinsteinMaxwellDilaton2013,donosThermoelectricPropertiesInhomogeneous2015}.
Fig.4 in
the article \cite{donosThermoelectricPropertiesInhomogeneous2015} shows an Umklapped sound mode at $\omega = v_s G$ in the
optical conductivity with $G$ the lattice momentum. However, a full treatment has been lacking.

For $U(1)$ charged relativistic hydrodynamics the fluctuation equations in the
longitudinal sector in a spatially constant background are the
coupled equations \cite{kovtunLecturesHydrodynamicFluctuations2012}
\begin{align}
	\label{eq:4}
\left(
\begin{matrix}
-i\omega  & i k & 0\\
i k \beta_1  & D_\pi k^2 - i \omega & i k \beta_2 \\
- D_{n_1} k^2  & i k \beta_3 & D_{n_2} k^2 - i \omega\\
\end{matrix}  
\right)
\left(\begin{matrix}
\delta \eps \\
\delta \pi_x \\
\delta n
\end{matrix}
\right)
= 0
\end{align}
Here $\delta \eps, \delta n, \delta \pi_x$ are the fluctuations in energy-,
charge-, and longitudinal momentum density respectively.
The upper two-by-two block is the sound sector with 
$\beta_1=
\left(\frac{\partial \bar{p}}{\partial \bar{\eps}}\right)_n$. At finite density this interacts with a charge diffusion mode
in the bottom one-by-one block through the interactions $\beta_2 =
\left(\frac{\partial \bar{p}}{\partial \bar{n}}\right)_\eps$, $\beta_3 = \frac{\bar{n}}{\bar{\eps}+\bar{p}}$ and the diffusion constant $D_{n_1}$. The diffusion constants equal
\begin{equation}
	\label{eq:5}
\begin{aligned}
 D_{\pi} & =\left({2\left(1-\frac{1}{d}\right)\eta +\zeta}\right)\left(\frac{\partial v_x}{\partial \pi_x}\right)_\eps =\frac{2\left(1-\frac{1}{d}\right)\eta +\zeta}{\bar{\eps}+\bar{p}} ~,\\
	D_{n_1} & = \sigma_Q \bar{T} \left(\frac{\partial (\bar \mu/\bar T)}{\partial \bar \eps}\right)_{\bar n} =- \sigma_Q \dfrac{ \dndT + \frac{\bar \mu}{\bar T} \dndmu }{ \dndmu \depsdT -\dndT \depsdmu}~,\\
	D_{n_2} & = \sigma_Q \bar{T} \left(\frac{\partial (\bar \mu/\bar T)}{\partial \bar n}\right)_{\bar \eps}= \sigma_Q \dfrac{\depsdT + \frac{\bar \mu}{\bar T} \depsdmu }{\dndmu \depsdT -\dndT \depsdmu}~.
\end{aligned}
\end{equation}

In the last two equations, the last equality leads to a seemingly more complicated form, but each of these derivatives is much simpler to compute. Barred quantities denote the (spatially constant) equilibrium background, and  $\eta, \zeta,
\sigma_Q$ are the microscopic transport coefficients: the shear- and bulk-viscosity and the momentum-independent contribution
to the conductivity. 
As discussed, the holographic models we consider have $d = 2$ with an underlying conformal symmetry for which the equation of state $\bar{\epsilon} = 2 \bar{p}$ implies that $\zeta = 0$, $\beta_2 = 0$ and $\beta_1=c_s^2 = 1/2$; we will limit our focus to conformal hydrodynamics in the remainder.

Placing such a system in a spatially varying
chemical potential  $\mu(x) = \bar{\mu} \left(1 + A\cos(Gx)\right)$ the Umklapp interactions follow from a re-derivation of the fluctuation equations in this background.
A detailed derivation for both conformal and non-conformal hydrodynamics and discussion with a natural generalization to a two-dimensional lattice
$\mu(x) = \bar{\mu} \left(1 + \frac{A}{2}\cos(Gx)+\frac{A}{2}\cos(Gy)\right)$ is given in a companion
article \cite{chagnetUmklappHydrodynamics2022}.
In summary, to maintain
equilibrium with spatially constant temperature also requires a spatially varying
charge density $n(x)=\bar{n} + \bar{\mu} A \dndmu  \cos(Gx)+\ldots$ and pressure
$p(x) = \bar{p} + A \bar{\mu} \bar{n} \cos(G x)+\ldots$ to leading order in $A$. The exact equation of state $\bar{\epsilon} = 2 \bar{p}$ in a conformal fluid means the energy density follows the pressure.
By viewing the lattice as a small perturbation on the thermal equilibrium, we can express the perturbations in terms of the chemical potential modulation and the thermodynamic susceptibilities of the background.
These corrections to the background are responsible for the Bloch decomposition and Umklapp interactions mixing them. To first order in
the lattice strength $A$ the three modes of the longitudinal sector\footnote{Substituting this spatially varying background into the defining conservation equations of hydrodynamics and expanding in
fluctuations, they no longer decompose in a longitudinal and transverse sector. It can be shown, however, that in the presence of a orthogonal lattice the naively longitudinal sector along one of the lattice directions is self-contained.
} 
mix with their six Umklapp copies. 
Our interest in this article is how this Umklapp affects the response at low frequencies $\omega \ll G$ and zero momentum $k=0$. At $k=0$ the un-Umklapped charge diffusion mode  decouples, and the remaining eight modes decompose into four parity-odd-in-$G$ ones and four parity-even modes.
The latter include the $k=0$ sound mode $\delta \pi^{(0)}$, two Umklapped sound modes built on $\delta \eps^{(S)} =\int\!\text{d}x\sin(Gx) \delta \eps(x)$,
$\delta \pi^{(C)} =\int\!\text{d}x \cos(Gx)\delta \pi(x)$;
and 
one Umklapped charge diffusion mode built on $\delta n^{(S)} = \int\!\text{d}x \sin(Gx)\delta n(x)$ that interact as
\begin{equation}
		\label{eq:hydroEq}
	(\partial_t+  M) \cdot \delta \phi = R
\end{equation}
with
\begin{equation}
\label{eq:dynMatrixHydro}
M =
\begin{pmatrix}
0 & \frac{1}{2}AG\bar{\mu} & \frac{1}{2}AG\bar{\mu} & 
-\frac{3}{2}i \omega A\bar{\mu} \beta_3 \\
-\frac{AG\bar{\mu}}{(\bar{\eps} + \bar{p}) \alpha_n} & D_{\rho}G^2 & 0 & 0\\
-2 AG\bar{\mu} \beta_3^2 &0 &0 &- G \beta_3 \\
 -3 i \omega A\bar{\mu} \beta_3 & 0 &  \frac{G}{2 \beta_3} & D_{\pi} G^2
\end{pmatrix}
\end{equation}
and
\begin{equation}
\label{eq:dynMatrixHydroSupp}
\delta\phi  =\begin{pmatrix}
\delta \pi^{(0)}_x\\
\delta n^{(S)}- \beta_3\delta \eps^{(S)}\\
\beta_3\delta \eps^{(S)}\\
\delta \pi_x^{(C)}\\
\end{pmatrix}~, ~~
R  = \begin{pmatrix}
		\bar{n}\\
		\bar{\mu}A \beta_3 \frac{D_\rho G}{\alpha_n}\\
		-\bar{\mu}A \beta_3 \frac{D_\rho G}{\alpha_n}\\
		\bar{\mu} A \left( \alpha_n^{-1} + \frac{\bar{n}^2}{(\bar{\eps} + \bar{p}) c_s^2} \right)
\end{pmatrix} \delta \bar E_x
\end{equation}
where we have defined $D_\rho = D_{n_2}$ the charge diffusion constant and where we used the coefficient $\alpha_{n} \equiv \bar T \left( \frac{\partial (\bar \mu/\bar T)}{\partial \bar n} \right)_{\bar \eps}$ which entered the definition of $D_{n_2}$. It is purely thermodynamic and has a universal scaling behavior determined by the scaling of entropy, as we will later highlight. We have added to our system a perturbatively small time-varying electric field $\delta E_x(t) = -\delta \bar E_x e^{-i \omega t}$ which will externally source a longitudinal current $\delta J^x$. This term will also enter the hydrodynamic system as an extra term in the current constitutive relation through $\partial_x \mu \to \partial_x \mu + \delta E_x(t)$.

We can now therefore linearize the constitutive relation $\delta J^x=n \delta v^x -\sigma_Q\left[\partial^x(\delta \mu - \frac{\mu}{T} \delta T) + \delta E_x \right]$ 
for the current density defined as 
\begin{equation}
		\begin{aligned}
		\delta J^x(t) & = \left(\dfrac{\partial J^x}{\partial \phi}\right)^\intercal \cdot \delta \phi(t) + \sigma_Q \delta \bar E_x e^{-i \omega t}\\
		\text{ with } \left(\dfrac{\partial J^x}{\partial \phi}\right)^\intercal &  = \begin{pmatrix} \beta_3, & -\bar{\mu} A \dfrac{D_\rho G}{2} \beta_3, & \bar{\mu} A \dfrac{ D_\rho G}{4 \bar{n} \alpha_n}, & \bar{\mu} A \left(\beta_3 + \frac{1}{2 \bar{n} \alpha_n} \right) \end{pmatrix}~.
		\end{aligned}
\end{equation}
We make use of the dynamical system \eqref{eq:hydroEq},  
to obtain the time-evolution of the dynamical fields $\delta \phi(\omega) = \left(-i \omega I_4 + M(\omega) \right)^{-1} \cdot \left( \delta \phi(t = 0) + R \delta \bar E_x \right)$. Since we have turned on the external electric field, we are not interested in explicitly sourcing any of the hydrodynamical variables and therefore we set $\delta \phi(t = 0)$ as an initial condition such that $\delta \phi(\omega) \propto \delta \bar E_x$ and by extension so will be $\delta J^x$. Finally, the optical conductivity can be computed as \cite{chagnetUmklappHydrodynamics2022}
\begin{align}
\label{eq:cond}
		\sigma(\omega) = \dfrac{\delta J^x}{\delta \bar E_x}.
\end{align}
The inverse $(-i\omega I_4 +M)^{-1}$ is dominated by the vanishing of its determinant. These zeroes show up as poles in the conductivity. Expanding the determinant\footnote{
Strictly speaking, there are terms at order $\mathcal O(A^2)$ in the lower-right $3 \times 3$ sub-block of $M$ 
that are ignored in Eq.\eqref{eq:dynMatrixHydro} but will contribute to the eigenvalues at that order. 
However, as we show in detail in \cite{chagnetUmklappHydrodynamics2022}, these contributions to the poles \eqref{eq:polesHydro} are also higher order in $G$. They contribute at order $\mathcal O(A^2 G^2)$. 
Crucially, moreover, these corrections will not affect $\omega_1$ but only correct the diffusion constants and sound velocities of $\omega_2$ and $\omega_\pm$. They will not qualitatively change the pole structure \eqref{eq:polesHydro} therefore, nor the decomposition \eqref{eq:hydroToCDW}. Since $A^2 G^2$ will typically be very small compared to the finite position of these poles, we ignore these corrections here. Figure~\ref{fig:5} illustrates that this assumption is justified in the numerical range we consider.} to order $A^2$, there are four poles at
\begin{equation}
		\label{eq:polesHydro}
		\begin{aligned}
		\omega_1 &= -i(\Gamma_{\eta}+\Gamma_d)+ \mathcal O(A^4)~,\\
		\omega_2 &= -i(D_{\rho}G^2 -\Gamma_d)+ \mathcal O(A^4)~,\\
		\omega_\pm &= \pm \frac{G}{\sqrt{2}} \left[1-\bar{\mu}^2 A^2 \beta_3^2 + \mathcal O(A^4) \right] -i\frac{1}{2} \left[ D_\pi G^2-\Gamma_\eta + \mathcal O(A^4) \right] + \mathcal O(G^3)
		\end{aligned}
\end{equation}
with 
\begin{align}
\label{eq:gamma-definitions}
		\Gamma_d &\equiv A^2 \dfrac{\bar{\mu}^2}{2 (\bar{\eps}+\bar{p}) D_\rho \alpha_n}~, \non 
		\Gamma_\eta &\equiv 2 \bar{\mu}^2 A^2 \beta_3^2  D_\pi G^2 = 2 A^2\frac{\bar{\mu}^2 \bar{n}^2}{(\bar{\eps}+\bar{p})^2}  D_\pi G^2
\end{align}
At low frequency $\omega \ll c_s G$, the contribution from the two sound poles $\omega_\pm$ should be negligible in the conductivity. By expanding the expression \eqref{eq:cond} as a quadruple Laurent series
\begin{align}
		\label{eq:LaurentSeries}
		\sigma({\omega}) = \sigma_0 + \sum_{i=1,2,\pm} \frac{Z_i}{\omega-\omega_i}
\end{align}
and truncating the two sound modes,
one finds that it takes the form\footnote{An attempt to formally decouple the sound modes by taking the limit $c_s^2\rightarrow \infty$ requires that $\Gamma_\eta\sim \frac{1}{c_s^2}$ and will therefore shift the poles. The truncated Laurent expansion keeps the poles in the right location.}
\begin{align}
	\label{eq:12b}
	\sigma_{\text{no sound}}(\omega) = 
	\sigma_0 +
	\dfrac{Z_1}{\omega-\omega_1}+
	\dfrac{Z_2}{\omega-\omega_2}
	= \sigma_0 +
	{Z_{\text{eff}}} \frac{\Omega-i\omega}{(\Gamma-i\omega)(\Omega-i\omega) + \omega_0^2}
\end{align}
with

\begin{equation}
\label{eq:hydroToCDW}
\begin{aligned}
		\Omega       & = \overbrace{D_\rho G^2}^{\mathcal{O}(1)}          && - && \overbrace{ 2 D_\rho^2 G^2\Gamma_d+ \mathcal O(G^3)}^{\mathcal{O}(A^2)} && + \mathcal O(A^4)~,\\
		\Gamma         & = \hphantom{D_\rho G^2}  && \hphantom{+} && 2 D_\rho^2 G^2 \Gamma_d + \Gamma_\eta + \mathcal O(G^3) && + \mathcal O(A^4)~,\\
		\omega_0^2     & = \phantom{D_\rho G^2}  &&\hphantom{-} && D_\rho G^2 \Gamma_d \left[ 1 - 2 D_\rho^2 G^2 + \mathcal O(G^3) \right] && + \mathcal O(A^4)~,\\
		Z_{\text{eff}}/\omega_p^2 & = 1 && + && 4 \bar{\mu}^2 A^2 \beta_3^2 D_\pi^2 G^2 - D_\rho \Gamma_d \left[4 + D_\rho (D_\rho - 4 D_\pi) G^2\right] + \mathcal O(G^3) && + \mathcal O(A^4)~,\\
		\sigma_0       & =\sigma_Q                 &&  && && + \mathcal O(A^4)~,
\end{aligned}
\end{equation}
where the plasmon frequency is $\omega_p^2 = \frac{\bar{n}^2}{\bar{\eps}+\bar{p}}~$.

The form Eq.~\eqref{eq:12b} is well known from studying the hydrodynamics of
decaying charge density waves or other 
pseudo-spontaneously broken $U(1)$
superfluids
\cite{delacretazBadMetalsFluctuating2017,delacretazTheoryHydrodynamicTransport2017,amorettiDcResistivityQuantum2018,amorettiUniversalRelaxationHolographic2019, 
armasHydrodynamicsChargeDensity2020,ammonPseudospontaneousSymmetryBreaking2022,armasApproximateSymmetriesPseudoGoldstones2022}. This
is not surprising as the underlying physics is that of two damped
currents cross-coupled with an interaction $\omega_0$ (see Appendix \ref{app:lorentz}). Both a decaying (i.e. damped) pseudo-Goldstone boson, as
well as an Umklapp hydrodynamics interaction belong to this class.

Given an appropriate temperature scaling of $\Gamma,\Omega, \omega_0$
or equivalently $Z_{1,2}, \omega_{1,2}$ it was already proposed that such a
conductivity could explain the emerging mid-IR peak at high
temperature in the cuprates. We will argue below that this Umklapp
hydrodynamics in an holographic AdS$_2$ metal with Sommerfeld specific
heat provides precisely the right scaling.

\section{The applicability of hydrodynamics and the imprint of local quantum criticality}
\label{sec:appl-of-hydro}

Despite the fact that the interplay between holography and hydrodynamics has been
formidable, it is not a given that a hydrodynamical understanding as
given above applies directly to holographic strange AdS$_2$ metals in explicit
lattices. Even though holography describes strongly coupled systems
which implies a large hydrodynamical regime, this regime is finite as has been emphasized in several recent articles
\cite{areanHydrodynamicDiffusionIts2021,hellerConvergenceHydrodynamicModes2021,wuUniversalityAdS2Diffusion2021,jeongBreakdownMagnetohydrodynamicsAdS22022,liuBreakdownHydrodynamicsHolographic2022}, and bounded by $\omega = 2\pi \Delta T$ where
$\Delta$ is the scaling dimension of the lowest irrelevant operator
from the strange metal fixed point. This argument against
hydrodynamics can be sharpened by the fact that momentum dependent longitudinal 
DC-conductivities
at zero frequency $\sigma(\omega=0,k\neq 0)$ vanish
\cite{anantuaPauliExclusionPrinciple2013}.\footnote{Recall that momentum-dependent conductivities
	at finite momentum need not be in the hydrodynamic regime. Within hydrodynamics, longitudinal diffusive conductivities obeying $\sigma(\omega,k) = \dfrac{i\omega D\chi}{i\omega - Dk^2}$ give an exactly vanishing DC conductivity at finite momentum, but a finite DC conductivity at zero momentum obeying Einstein's relation $\sigma =D \chi$.} This is an unavoidable
consequence of $U(1)$ current conservation:  $\omega\rightarrow 0$ implies $G\cdot J =0$.  Naively considering Umklapp as the mixing of the $\sigma(\omega,0)$ and $\sigma(\omega,k=G)$, would argue that the amplitude of the mixed-in
Umklapp wave is thus very small. This is illustrated by a memory
matrix computation \cite{hartnollLocallyCriticalResistivities2012,anantuaPauliExclusionPrinciple2013}.
The
momentum-dependent 
density correlation function $G_{J^t J^t}$ in a homogeneous AdS$_2$
metal, which is the operator to consider for our choice of lattice, scales as a function of the temperature as 
\begin{equation}
	\label{eq:8}
	\begin{aligned}
	\operatorname{Im} G_{J^t J^t}^{\mathrm{homogeneous}}(\omega=0,k) &\sim T^{2\nu_k} + \ldots~,\\
	\nu_k & = \frac{1 + \hat \eta}{2 \sqrt{2 + \hat \eta}}\sqrt{10 + \hat{\eta}  + 4 (2 + \hat{\eta}) \bar{k}^2 - 8 \sqrt{1 + (2+\hat{\eta})\bar{k}^2}}
	\end{aligned}
\end{equation}
where $\hat{\eta} \equiv -\theta/z$ characterizes the near-AdS${}_2$ region and 
$\bar k$ is a wavevector renormalization that correctly rescales to the emergent near horizon AdS${}_2$ geometry in a lattice \cite{anantuaPauliExclusionPrinciple2013,donosThermoelectricPropertiesInhomogeneous2015}.
For GR $\hat{\eta} = 1$ and for RN, $\hat{\eta} = 0$ while in both cases, 
$\bar k = \frac{k}{\mu}$.
This scaling of $G_{J^t J^t}$ follows from a near-far matching method in the AdS$_2$ bulk which
shows that a generic Green's function takes the form
\begin{align}
	\label{eq:10}
	G = \frac{ A+B\cG}{C+D\cG}
\end{align}
with $A,B,C,D$ purely real and $\cG$ the AdS$_2$ Green's function \cite{faulknerEmergentQuantumCriticality2011}
\begin{align}
	\label{eq:11}
	\cG(\omega, k) \propto T^{2\nu_k} \frac{\Gamma\left(1-\nu_k\right)\Gamma\left(\frac{1}{2} + \nu_k - \frac{i\omega}{2\pi T}\right)}{\Gamma\left(1 + \nu_k\right)\Gamma\left(\frac{1}{2}-\nu_k -\frac{i\omega}{2\pi T}\right)}
\end{align}
The imaginary part of the density correlator is proportional to the imaginary part of the AdS${}_2$ correlator as $\operatorname{Im} G_{J^t J^t} \sim \operatorname{Im} \mathcal G$.
Though this scaling as a function of the temperature is exact, it
ignores the possibility that there can still be a large amplitude as a
function of the other parameters. This is in fact what
happens when one extrapolates the exact answer for the momentum-dependent transverse conductivity $\sigma_\perp(\omega,k)$
to the hydrodynamic regime $k\ll \mu$ \cite{davisonHydrodynamicsColdHolographic2013}. The
momentum dependent
current-current correlation function in an AdS$_2$ metal behaves as
\begin{align}
	\label{eq:9}
	G_{J_\perp J_\perp} (\omega, k) = - \frac{ k^2 \cG(\omega)}{\omega^2 + \frac{k^2}{2r_0}\cG(\omega)}
\end{align}
Although the scaling is indeed captured by the Hartnoll-Hofman result
Eq.~\eqref{eq:8} one also sees that for small $k$ the hydrodynamic pole
at $\omega = -iDk^2$ becomes far more important than the
$(\omega/T)^{2\nu}$-suppression. For $k\leq \mu$ the
hydrodynamic pole captures the physics far better than the AdS$_2$
power-law.

\begin{figure}[t!]
	\centering
	\includegraphics[width=0.49\textwidth]{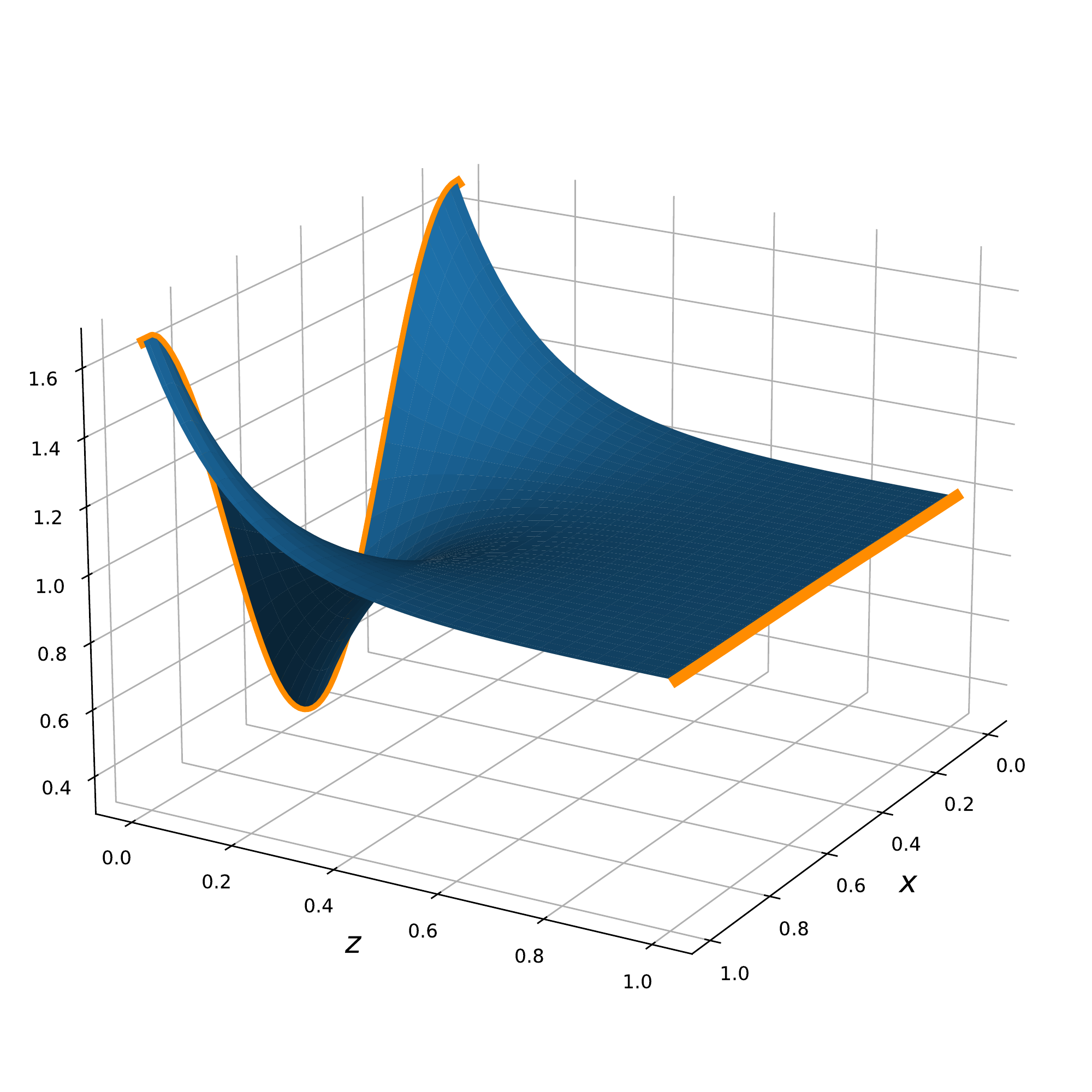}
	\includegraphics[width=0.49\textwidth]{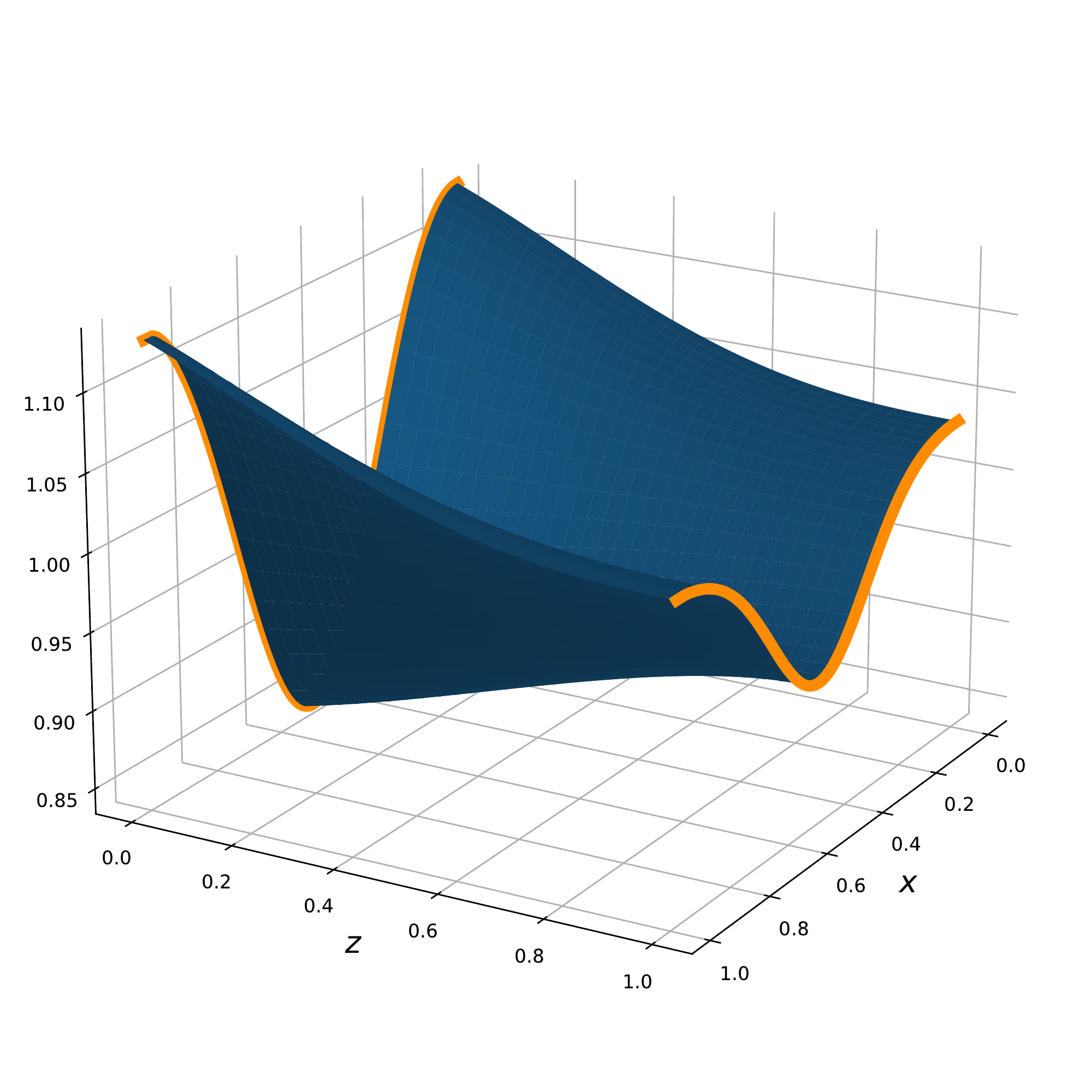}
	\caption{A holographic visualization of the cross-over in response functions
		between $G>\mu$ (left) and $G<\mu$ (right). Plotted is the bulk electric field $F_{tz}$ 
		(whose boundary value is dual to the charge density) in the presence of 
		a 1D spatially varying chemical potential
		$\mu(x)=\bar{\mu}+A\cos(Gx)$ as a function of position and the AdS
		radial direction $z$. For $G>\mu$ ($G=4\mu$) one sees the lattice
		amplitude decrease smoothly as one moves from the AdS boundary to
		the horizon at $z=1$. For $G<\mu$ ($G=0.05\mu$) on the other hand, 
		one sees that
		the RG flow is much slower and ``halts'' around $z \sim 0.5$. This is due
		to the exponential suppression of the coupling between different
		momentum modes at the AdS$_2$ IR fixed point. The results are for 
		a RN black hole at $T/\mu=0.15$.
		}
	\label{fig:rg}
\end{figure}

As is clear from the mathematical expressions this is not a sharp
transition, but a smooth crossover. Nevertheless there is a clear
transition between dominant physics regimes (AdS$_2$ vs hydrodynamics) that
can be made visible through the holographic dynamics.  A finite momentum
conductivity is better viewed as the response when  the system is placed in a fixed spatially
oscillating but static electric field background. The spatial oscillation
imprints a lattice structure in the finite density system. The conventional RG
perspective is that this lattice is irrelevant in the RG. This is the
physics behind the power-law dependence on temperature in
Eq.~\eqref{eq:8}. The AdS$_2$ fixed points of the holographic metals
that we study, either RN or GR, are so-called semi-local quantum
liquids \cite{iqbalSemilocalQuantumLiquids2012}, however. This means that while for $T<\mu$ the two-point correlation
function displays power-law behavior between two time-like separated
points, it is exponentially suppressed between two space-like
separated points. This exponential suppression is so strong that two
points separated spatially by a distance $|x| \gtrsim \frac{1}{\mu}$ have
no causal contact \cite{iqbalSemilocalQuantumLiquids2012}. In momentum space this implies that the coupling
between modes with $ k
\lesssim \mu$ is exponentially small. This decoupling means that
for modes $k \lesssim \mu$ or equivalently a spatially oscillating but
static electric field with $G \leq \mu$ the RG-flow becomes strongly
suppressed once $T$ decreases below $\mu$. One can think of it as that
the $d$-dimensional RG-flow at $T=\mu$ decomposes into individual
RG-flows for each momentum mode. Recalling that in
holography the radial direction encodes the RG-flow, we can visualize
this. In Fig.~\ref{fig:rg} we plot the charge/current density as a function of location
for a modulated chemical potential. For a
lattice momentum $G \gg \mu$ 
the lattice irrelevancy towards the IR is uninterrupted. However for an
oscillating chemical potential with periodicity $G\ll \mu$, the RG
flow ``halts'' around the AdS radius value $r \sim \mu$ corresponding
to $T\sim \mu$. 
For such values of $G\ll \mu$ the lattice thus remains
quite strong in the IR and certainly much stronger than one would
naively expect. The way to understand this is that precisely in this
regime it is the proximity of the
hydrodynamic pole that dominates the response rather than the RG
scaling suppression. Ultimately the RG wisdom does holds for any lattice perturbation and even
for $G\ll \mu$ the lattice will eventually turn irrelevant in the IR (Sec 3.4 in
\cite{donosThermoelectricPropertiesInhomogeneous2015}), and scaling again becomes the pre-eminent physical effect but this only happens at the lowest of temperatures.

For Umklapp hydrodynamics this is relevant because it implies that the
regime where the hydrodynamics results capture the  physics is appreciable.  Below we shall verify that near an AdS$_2$ fixed point Umklapp hydrodynamics is the better way of understanding the physics  for $G<\mu$, whereas AdS$_2$ Hartnoll-Hofman scaling is the better way for $G>\mu$. For the sake of clarity, we emphasize that strictly speaking at a mathematical level both can be, and often are, valid simultaneously as is evidenced by \eqref{eq:9}. However, the physical response is generically dominated by one or the other, and relying on only one of them is not sufficient.

\begin{figure}[t!]
\centering
\includegraphics[width=0.7\textwidth]{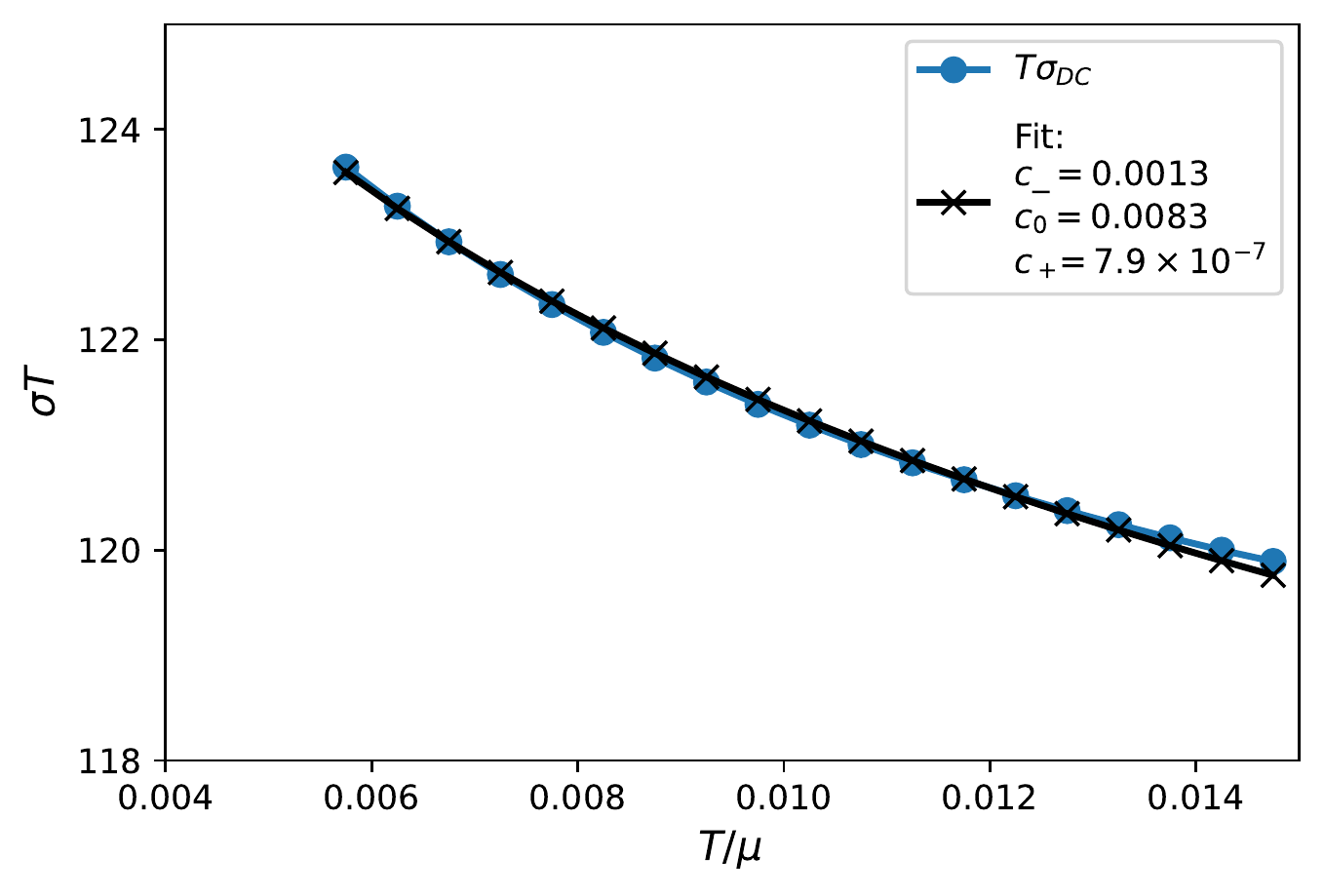}
\caption{The DC conductivity of the GR metal in a weak lattice potential $A=0.05$ is not perfectly inversely linear-in-T. This is due to subleading scaling contributions computable from the AdS$_2$ fixed point. Shown is a three parameter fit of the DC conductivity for $c_-, c_0, c_+$ to the functional form $\sigma_{DC}=1/\left(c_- T^{2\nu_{G}-1} + c_0T^{2\nu^{0}_G-1} + c_+ T^{2\nu^{+}_G-1}\right)$ at low temperature with $\nu_G, \nu_G^0, \nu_G^+$ given by Eq.~\eqref{eq:8} and Eq.~\eqref{eq:extraexponents}, with $k = G$, for ${G}/\mu=0.1$. The values of the exponents $\nu_G, \nu_G^0, \nu_G^+$ at this wave vector are $1.00015,  1.0198$ and $2.53$, respectively. Therefore according to this fit, one expects the exponent $\nu_G$ to be the dominating one only at temperatures $T/\mu < \mathcal{O}(10^{-50})$.
}
\label{fig:triple-HH-scaling-explanation}
\end{figure}

There is a second reason why hydrodynamics is the more appropriate perspective for $G\ll \mu$. A more precise analysis of the momentum-dependent density correlator in an AdS$_2$ metal shows that it has multiple characteristic scaling contributions \cite{anantuaPauliExclusionPrinciple2013}
\begin{align}
	 \operatorname{Im} G_{J^t J^t}^{\mathrm{homogeneous}}(\omega=0,k) \sim c_-T^{2\nu_k}+c_0T^{2\nu_k^0}+c_+T^{2\nu_k^+}
\end{align}
with the additional scaling exponents 
\begin{align}
\label{eq:extraexponents}
\nu_k^0 & =\frac{1+\hat{\eta}}{2}\sqrt{1+ 4\bar{k}^2} \non
		\nu_k^+ & = \frac{1 + \hat \eta}{2 \sqrt{2 + \hat \eta}}\sqrt{10 + \hat{\eta}  + 4 (2 + \hat{\eta}) \bar{k}^2 + 8 \sqrt{1 + (2+\hat{\eta})\bar{k}^2}}
\end{align}
For $k=G\ll \mu$ as one needs for Umklapp between $\operatorname{Im} G_{J^t J^t}^{\mathrm{homogeneous}}(0,k^\prime)$ for $k^\prime =0, \pm G$, all these three exponents take values that are very close to each other. For such small differences in the exponents there is observationally  no clean scaling regime. For low lattice strengths $A$ this is the reason that the observed weak lattice DC conductivities in Fig.~\ref{fig:lin-in-temp-res-plus-saturation} do not scale exactly inversely-linear-in-T as noted in the Introduction. Through Umklapp, the lattice DC conductivity is related to the homogeneous density correlator (which we will review in more details in the next section)
\begin{equation}
		\sigma_{\mathrm{DC,lattice}} \sim \left( \lim_{\omega \to 0} \dfrac{\operatorname{Im} G_{J^t J^t}^{\mathrm{homogeneous}}(\omega,k)}{\omega} \right)^{-1} \sim \dfrac{1}{c_-T^{2\nu_k-1}+c_0T^{2\nu_k^0-1}+c_+T^{2\nu_k^+-1}}~.    
\end{equation}
Fig.~\ref{fig:triple-HH-scaling-explanation} shows that the deviation from linearity is exactly due to the contribution of the additional exponents.

\section{DC vs Optical conductivities in
	explicit lattice (holographic) strange metals from Umklapp}
\label{sec:weaklattice-conductivities}

Having argued that hydrodynamics should dominate the response in holographic
strange metals, we now exploit our ability to do computational experiments to confirm that Umklapp hydrodynamics applies when such holographic strange metals are placed in an explicit periodic
lattice with a small amplitude $A$. Then we shall describe the surprising phenomenological conclusions for electrical DC and optical electrical conductivity. 

\begin{figure}[t!]
	\centering
	\includegraphics[width=\textwidth]{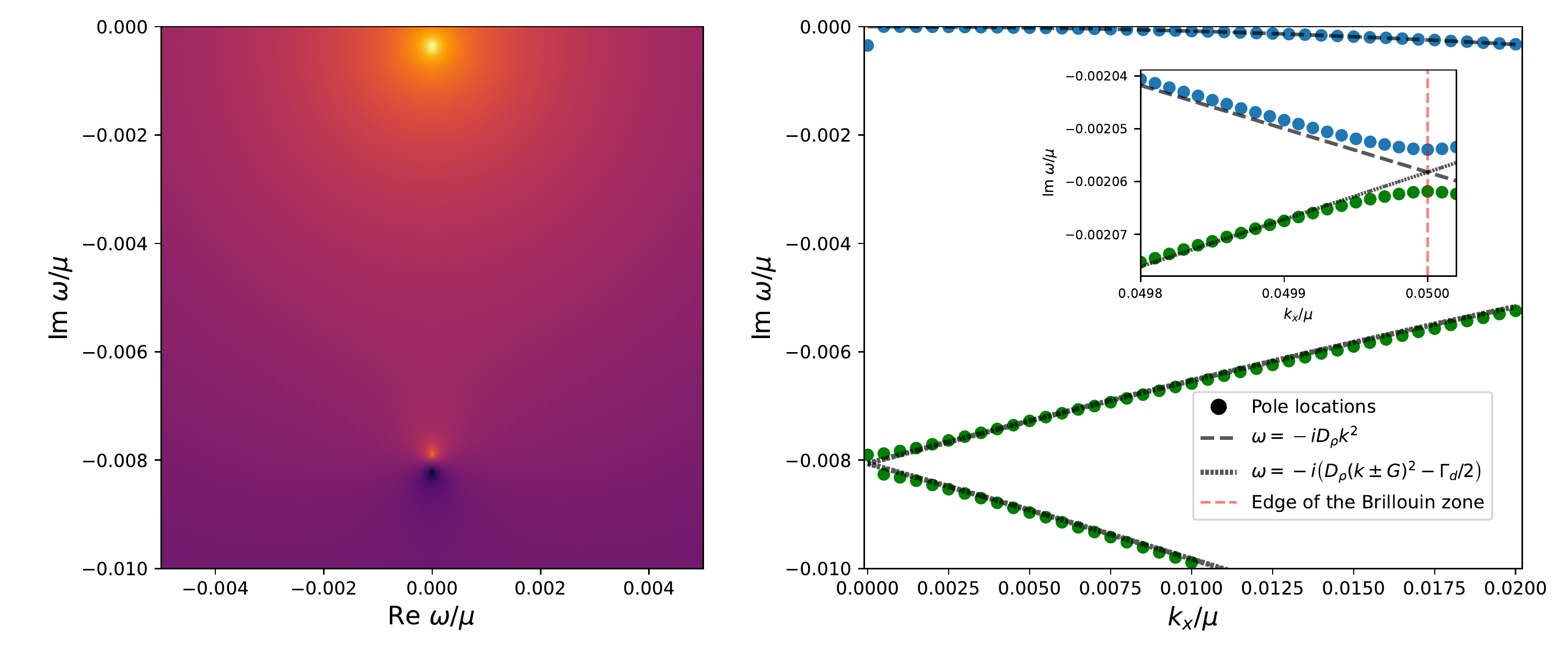}
	\caption{Umklapp hydrodynamics. The left panel shows the presence of both the Drude (upper) pole and the Umklapped charge diffusion (lower) pole and associated zero in the complex frequency plane at $k_x = 0$. The right panel shows the motion of both poles as a function of
		longitudinal momentum $k_x$. The Umklapp is confirmed by matching this motion to the diffusion coefficients of the un-Umklapped hydrodynamics computed in Eqs.~\eqref{eq:9}. The inset shows the level splitting near the Brillouin zone boundary at $k = G/2$. The results are computed in the GR black-hole lattice at $T/\mu = 0.1, G/\mu = 0.1$ with a 1D ionic lattice potential $\mu(x) =
		\mu\left(1+0.05\cos(Gx)\right)$.
		The deviation at low $k$ finds its origin in the next order level splitting in Umklapp which our formula does not account for, similar to the level splitting near the Brillouin zone.
		}
		\label{fig:3}
\end{figure}

\begin{figure}[t!]
		\centering
		\includegraphics[width=\textwidth]{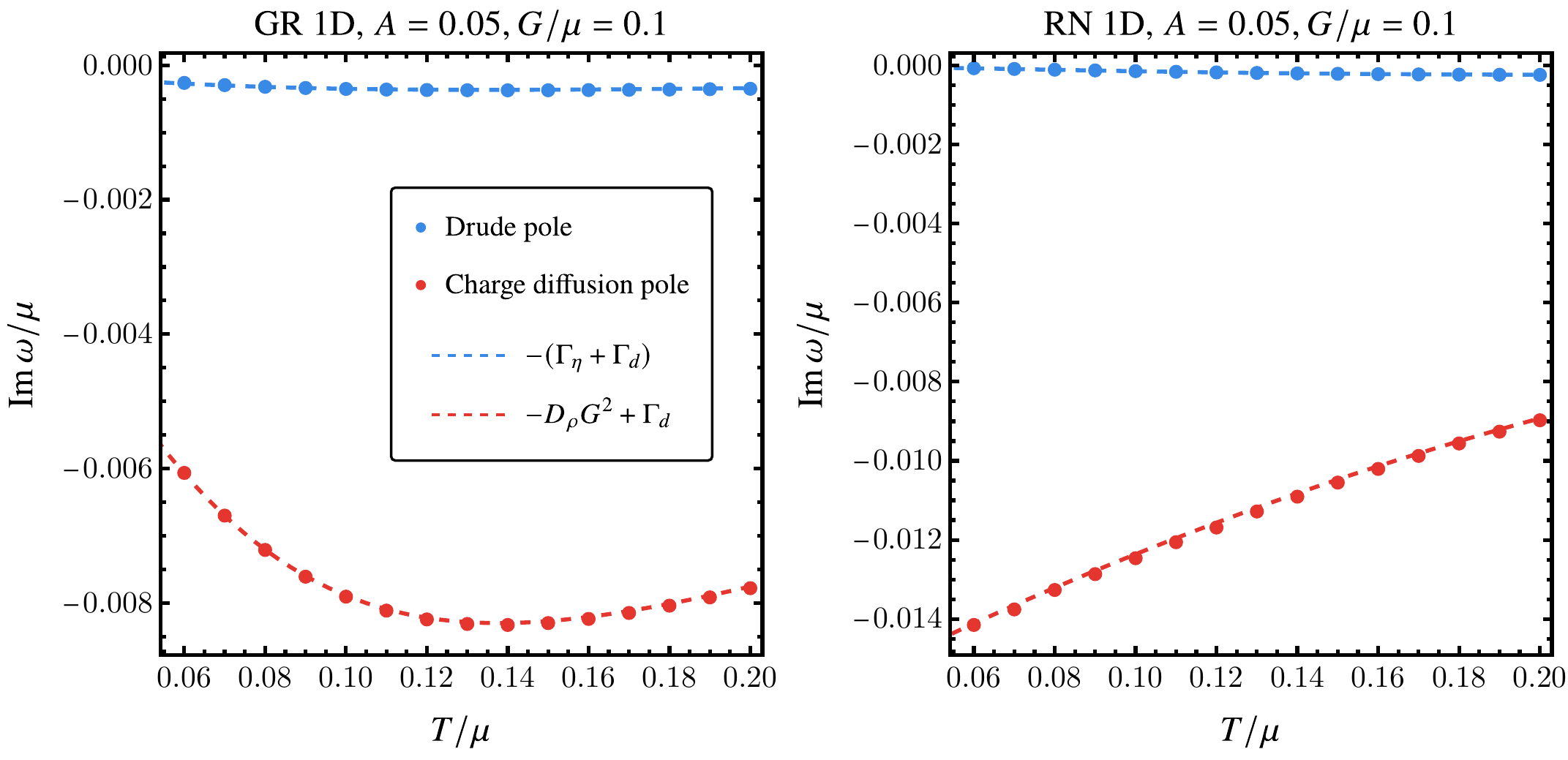}
		\caption{The motion of the poles (points) as one increases temperature compared to hydrodynamics (dashed). As the temperature is increased further the Drude pole eventually collides with the Umklapped charge diffusion pole and gains a real part. At low temperatures where a perturbative Umklapp analysis is valid the behavior of the poles can be understood from the un-Umklapped hydrodynamic analysis. Results are in the GR and RN 1D lattices with $G/\mu = 0.1$, $k_x = 0$ and potential strength $A = 0.05$.}
		\label{fig:5}
\end{figure}

To verify the applicability of Umklapp hydrodynamics in AdS$_2$ metals, we can study the location of the poles in linear response functions. Fig.~\ref{fig:3} shows the poles in the optical conductivity $\sigma(\omega)$ in a GR strange metal in a 1D ionic lattice background $\mu(x) = \bar{\mu}(1 + A\cos(Gx))$. There are multiple poles on the negative imaginary axis and two poles with real part at the location $\omega=\pm v_sG$. The latter are the ones already noted by \cite{horowitzFurtherEvidenceLatticeInduced2012,lingHolographicLatticeEinsteinMaxwellDilaton2013,donosThermoelectricPropertiesInhomogeneous2015} and identified as Umklapped sound modes \cite{donosThermoelectricPropertiesInhomogeneous2015}. That Umklapp is at work is confirmed by tracing the behavior of the poles as a function of temperature. Compare the behavior of the two poles on the negative imaginary axis closest to the origin to the analytically computed values Eqs.~\eqref{eq:polesHydro} 
, we see that the match is very good; see Fig.~\ref{fig:5}.
Moreover, if one also studies the response functions at finite momentum $k$, then one observes the characteristic Umklapp level repulsion at the edge of Brillouin zone
$k=G/2$ (Fig.~\ref{fig:3}).

\subsection{Low temperatures: Drude transport}

We have claimed Umklapp Hydrodynamics explains the remarkable finding summarized in Fig.~\ref{fig:lin-in-temp-res-plus-saturation} that the DC conductivity of a strange metal in a weak lattice remains linear-in-temperature while the mechanism governing the AC-response appears to change. We can now show this.

The DC conductivity from Umklapp Hydrodynamics to lowest order in the lattice strength $A$ equals
\begin{align}
		\sigma_{DC} =\frac{Z_{\mathrm{eff}}}{\Gamma+\frac{\omega_0^2}{\Omega}} + \sigma_Q=\underbrace{\frac{\omega_p^2}{\Gamma_\eta+\Gamma_d}}_{\mathcal O(A^{-2})} + \underbrace{\sigma_{\text{offset}} + \sigma_Q}_{\mathcal O(1)} + \, \mathcal{O}(A^2)
\end{align}
where, in the last equality, the first term is the leading order and the offset term $\sigma_{\text{offset}}$ comes from the higher order terms in Eqs.~\eqref{eq:hydroToCDW}. The first contribution in the DC from the sound part of the Laurent expansion \eqref{eq:LaurentSeries} only comes at order $\mathcal O(A^2)$ and is therefore negligible here.  These expressions already suggest that two physical
mechanisms are at play in the DC result. At first sight this may appear contradictory to the conventional explanation of weak lattice DC conductivity in terms of Drude momentum relaxation $\sigma=\frac{\omega_p^2}{\Gamma_{\text{mom.rel.}}}$. The momentum relaxation rate $\Gamma_{\text{mom.rel.}}$ can be computed in the memory matrix formalism \cite{forsterHydrodynamicFluctuationsBroken2019,hartnollLocallyCriticalResistivities2012} to equal
\begin{align}
		\Gamma_{\text{mom.rel.}} = \frac{g^2G^2}{(\bar{\eps}+\bar{p})} \lim_{\omega \to 0} \dfrac{\text{Im} \langle {\cal{O}}{\cal O}\rangle (\omega, k = G)}{\omega} 
\end{align}
where ${\cal O}(G)$ is the operator that breaks translation invariance with coupling $g$. In the case of an ionic lattice with a cosine potential as we consider, there are two operators ${\cal O}(G)=J^t$, one inserted at wavevector $G$ and one at $-G$ each with coupling strength $g = \bar{\mu} A/2$. Therefore the memory matrix momentum relaxation rate for the ionic lattice is
\begin{equation}
		\label{eq:mem-matrix}
		\Gamma_{\text{mom.rel.}} = \frac{\bar{\mu}^2 A^2 G^2}{2(\bar{\eps}+\bar{p})} \lim_{\omega \to 0} \dfrac{\text{Im} G_{J^t J^t} (\omega, k = G)}{\omega}~.
\end{equation}
Inserting its correlation function computed in a homogeneous background into \eqref{eq:mem-matrix} one in fact finds the exact same answer as computed by Umklapp hydrodynamics $\Gamma_{\text{mom.rel.}}=\Gamma_\eta+\Gamma_d$ (see Appendix \ref{app:memory-matrix} for a derivation of this result).
Theoretically this can be understood through the observation that there are two possible dissipative channels in hydrodynamics. There is sound attenuation controlled by the shear viscosity $\eta$ (and bulk viscosity $\zeta$) and there is charge diffusion controlled by the microscopic conductivity $\sigma_Q$. 
Both are at the same order in the lattice strength $\Gamma_{d,\eta}\sim A^2$. This is the expansion parameter in the memory matrix computation and explains why they both show up. 

The phenomenologically important characteristic is the temperature scaling of the DC resistivity.
Implicitly the lattice scaling implies a scaling with temperature as the effective lattice strength should become irrelevant in the deep IR. This must be encoded explicitly in the scaling of both $\Gamma_\eta$ and $\Gamma_d$, and not in the UV-strength $A$. However, there is a priori no requirement that both $\Gamma_d$ and $\Gamma_\eta$ will scale the {\em same} as a function of $T$.  Generically they ought not. However, in holographic strange metals without a ground state entropy they do. For these systems at low temperatures
\begin{eqnarray}
		\Gamma_{\eta}~~ \sim &\eta(T) \sim s &\sim~ T^{(d-\theta)/z} \non 
		\Gamma_d ~~\sim &{\displaystyle \frac{T^2}{\sigma_Q(T)}\left(T\frac{\partial s}{\partial T}\right)^2}~  &\sim~ \left(\frac{d-\theta}{z}\right)^2T^{(d-\theta)/z}
\end{eqnarray} 
The derivation requires a mild assumption about the low temperature equation of state and is given in Appendix \ref{app:gamma-vs-s-scaling}. Thus for the GR strange metal $\Gamma_{\eta}\sim T$ and $\Gamma_d \sim T$, whereas for the RN metal which has a ground state entropy $\Gamma_\eta \sim T^0$ but the first non-vanishing order for $\Gamma_d$ is $\Gamma_d \sim T^2$. 
Over the range of validity, usually one of them will dominate, though it is conceivable that one dissipative momentum relaxation process switches dominance with the other. If this coincides with a change in scaling this would show up as a change of temperature scaling of the DC resistivity. 

Two observations follow. The first is that despite the numerical results supporting the inference from disordered translational symmetry breaking that the momentum relaxation rate scales as the entropy, this is not true for the contribution from $\Gamma_d$.

The more important observation here and in the following is that which term dominates does not matter. In holographic strange metals the momentum-relaxation rate is set at a deeper level by the non-trivial locally quantum critical IR fixed point. As pointed out by Hartnoll-Hofman and briefly reviewed in the previous Section \ref{sec:appl-of-hydro}, in the regime where Eq.~\eqref{eq:mem-matrix} holds, the frequency scaling enforced by local quantum criticality also sets the temperature scaling of the DC result. For the RN strange metal it is only $\Gamma_\eta$ that is responsible for this, whereas in the Gubser-Rocha strange metal both obey the appropriate scaling. Since $\Gamma_\eta$ also scales as $G^2$, whereas $\Gamma_d$ does not, one can tune the GR response to be dominated by $\Gamma_d$ for $G \ll 2\mu$, and $\Gamma_\eta$ to dominate for $G \gg 2\mu$. This coincides with the applicability of hydrodynamics as we discussed in the previous section, confirming a correlation with a physically observable change (see also section \ref{sec:app-hydro-subsection} below).
This very difference between $\Gamma_\eta \sim G^2$ and $\Gamma_d \sim G^0$ actually causes the order of importance to be opposite in disordered systems. Because disorder can be viewed as an average over an infinite set of lattices, in the decay rate in a disorder system $
\Gamma_{\text{disorder}} \sim \int\!\! G^{d-1} \text{d}G (\Gamma_d + \Gamma_{\eta})$ the $\Gamma_\eta$ term will generically dominate the integral \cite{davisonHolographicDualityResistivity2014}. Since $\Gamma_{\eta} \sim \eta \sim s$, this explains why in disordered systems entropy does directly control the dissipation time scale in contrast to a lattice with a fixed lattice momentum $G_L$ as we explained above.

Independent of the dissipative mechanism, both leading in $A$ momentum-relaxation rates $\Gamma_\eta$ and $\Gamma_d$ become vanishing small at low temperatures suggesting Drude transport. This is readily confirmed in the AC conductivity. Its real part displays a characteristic Drude peak. Mathematically, however, the peak is not exactly a (half-)Lorentzian, but follows from the two-pole expression Eq.~\eqref{eq:12b}.

\subsection{Intermediate temperatures: a mid IR-peak in the
	optical response}
\label{sec:sub:intermediateT}
	
We have just argued that the DC resistivity can remain the same while the physical regime controlling dissipation changes, because it is set at a deeper level by the underlying AdS$_2$ fixed point. Though we have just noted this fact by analyzing the analytic expressions, it is in fact dramatically made clear at an intermediate higher temperature, as we already summarized in the Introduction.

\begin{figure}[t!]
	\centering
\includegraphics[width=0.49\textwidth]{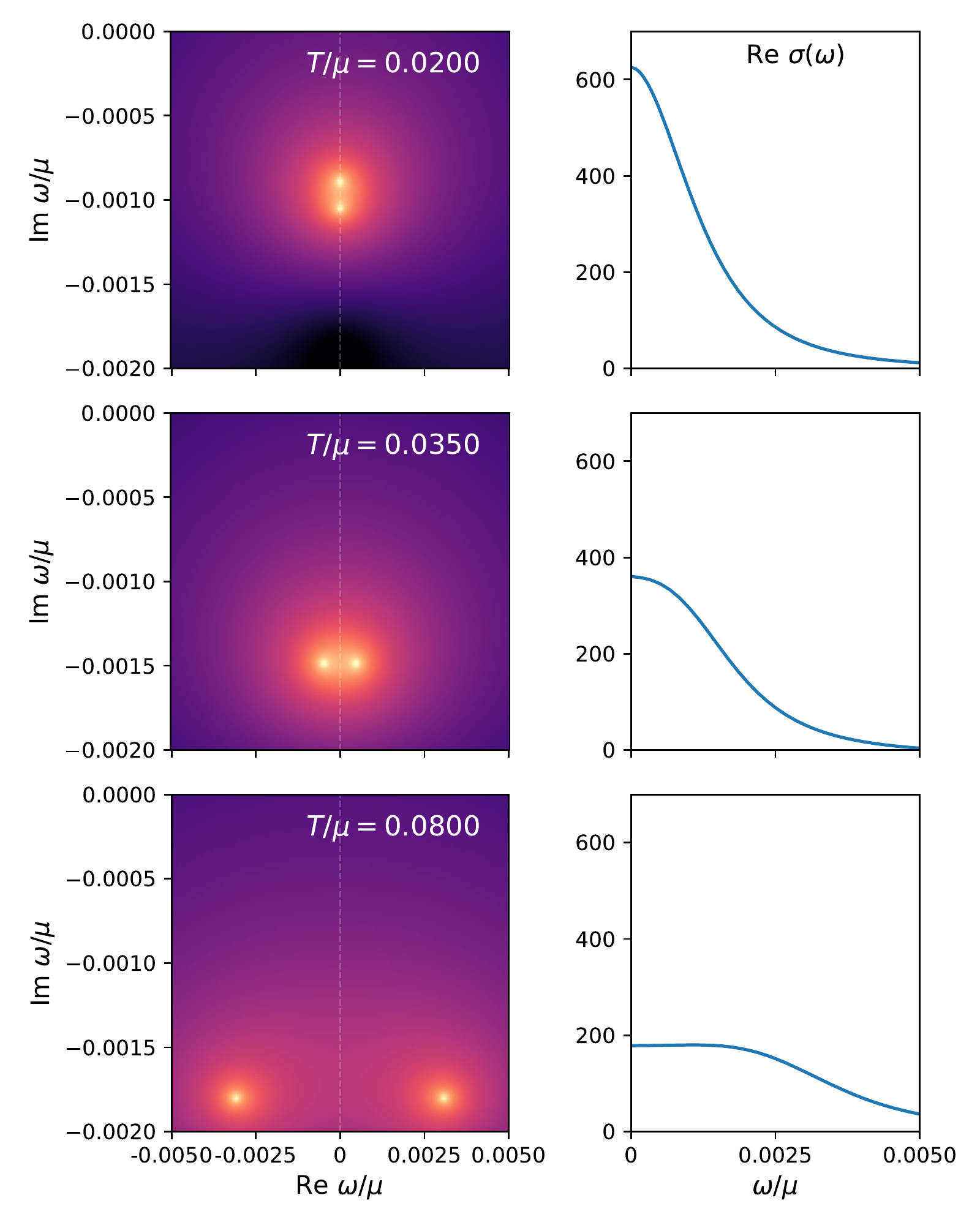}
\raisebox{5.5em}{\includegraphics[width=0.49\textwidth]{RhoDC_LowA_WithInset.pdf}}
	\caption{Emergence of mid-IR peak in the optical conductivity $\sigma(\omega)$ from pole collision. At low
		temperatures the Umklapp has negligible consequences as the response
	is strongly dominated by the conventional Drude pole. At intermediate temperatures the Umklapp pole causes an additional
	broadening. When the temperature increases to
	the point where the poles
	collide and gain a real part the peak still looks Drude to the eye
	even though it arises from two poles symmetrically arranged on both
	sides of the real axis. At even higher temperatures these two poles
	move so far apart that the peaks separate and a mid IR peak at
	finite $\omega$ appears in the optical response. For this figure the parameters are $A = 0.15, G = 0.08\mu$, the same as in Fig.~\ref{fig:lin-in-temp-res-plus-saturation}.
	}
	\label{fig:4}
\end{figure}

In the regime of interest the conductivity computed from Umklapp hydrodynamics is controlled by two poles. In the parametrization
\begin{align}
\label{eq:two-pole-conductivity-formula}
		\sigma(\omega) = \sigma_Q + Z\frac{\Omega-i\omega}{(\Gamma-i\omega)(\Omega-i\omega)+\omega_0^2}
\end{align}
these are the Drude and Umklapp charge diffusion poles at
\begin{align}
		\omega_{\text{Drude}} &= \frac{-i}{2}(\Gamma+\Omega)+\frac{i}{2}\sqrt{(\Gamma-\Omega)^2-4\omega_0^2} = -i(\Gamma_\eta+\Gamma_d)+ \mathcal O(A^4)~, \non 
		\omega_{\text{Um.Ch.Diff.}} & = \frac{-i}{2}(\Gamma+\Omega)-\frac{i}{2}\sqrt{(\Gamma-\Omega)^2-4\omega_0^2} = -i(D_\rho G^2-\Gamma_d)+ \mathcal O(A^4)~.
\end{align}
At low temperatures, the second pole (let alone the two already ignored Umklapped sound poles)  has a small effect. 
Increasing the temperature changes this fundamentally, however. Both poles move as one increases the temperature. However, they do not move in unison. When the argument under the square root $(\Gamma-\Omega)^2-4\omega_0^2$ becomes negative, the poles collide. 
For temperatures higher than the pole-collision temperature, the poles can now acquire a real part and move off the imaginary axis symmetrically; see Fig.~\ref{fig:4}. 
Initially this ``microscopic pole collision'' has little effect on the optical conductivity. In a formal sense it slightly broadens the peak around $\omega=0$ and without an insight into the complex frequency response it is essentially indistinguishable from a conventional Lorentzian Drude peak. However, as one increases temperature further and the poles move further away from the imaginary frequency axis, the peak will split into two, symmetrically arranged around $\omega=0$. For the positive half-line $\omega>0$ one would thus see a peak emerge in the near IR whereas the DC value at $\omega=0$ continues to decrease.

This collision point is controlled by a combination of temperature, lattice strength and lattice periodicity. Already at moderate lattice strengths, this emergence of the mid-IR peak in the AC conductivity happens at temperatures $T<T_{\text{strange}}$ where the DC response is still set by the critical scaling behavior of the underlying AdS$_2$ strange metal. In other words, despite the qualitatively drastic change in the AC-vs-T conductivity, the DC-vs-T response is unaffected.

What is striking is that this emergence of mid-IR peak in the optical response as temperature increases while the DC-resistivity stays linear in $T$ is  precisely what is observed in high $T_c$ cuprates and other strange metals as explained in the introduction.
Given
the earlier hypothesis reviewed there that transport in the high $T_c$-cuprates is hydrodynamical, it is conceivable that this is the explanation of this observed experimental finding.

The mechanism we just explained is tantalizing given  its minimalistic nature. It is in fact {\em ubiquitous} for {\em any} hydrodynamical fluid exposed to a microscopic Umklapp potential where the effective potential strength is rising more rapidly than the momentum diffusivity.  Notice that it does {\em not} apply to a Fermi liquid in metallic background potentials. The onset of equilibration is set by the quasiparticle collision time, but typically a substantial fraction of the centre of mass momentum is absorbed by the Umklapp impeding the total momentum conservation required for hydrodynamics including the mechanism in the above.

\subsection{Intermediate lattice strength: towards an incoherent metal}
\label{polecollargeA}

\begin{figure}
		\centering
		\includegraphics[width=0.8\textwidth]{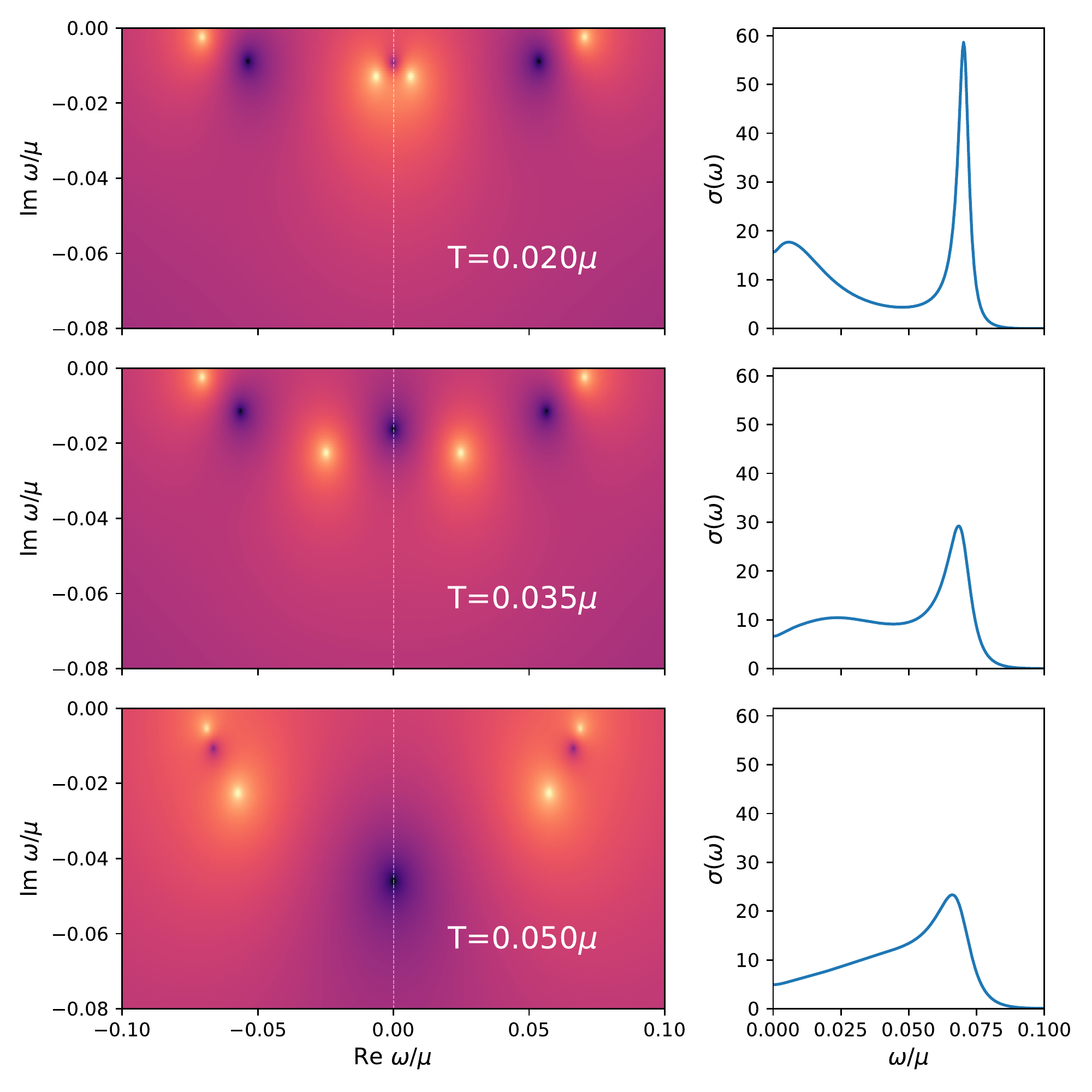}
		\caption{Optical conductivity (right) and the quasinormal mode spectra (left) for intermediate lattice strength GR lattices for $A=1, G/\mu = 0.1\mu$ at three different temperatures. Compared to small $A$ the pole collision (see section \ref{sec:sub:intermediateT}) has already happened even at lowest $T/\mu=0.02$. As one increases $T$ the Umklapped sound poles which stay almost fixed at Re $\omega=\pm c_s G = \pm \frac{1}{\sqrt{2}}G$ (and others not shown) become more important and their dominance in the AC conductivity signals the transition to an incoherent metal regime.}
		\label{fig:evolution-intermediatelattice}
\end{figure}

Our computational experiments on holographic strange metals can also provide us insight in what happens at larger lattice strengths beyond the applicability of perturbative Umklapp hydrodynamics. This is best quantified by tracking the behavior of the complex frequency poles in the AC conductivities. In Fig.~\ref{fig:evolution-intermediatelattice} we show  typical quasinormal mode spectrum computed for lattice strength $A =0.15$. At low temperatures one finds that these are still dominated by the non-linear continuation of the same two-pole structure as we identified for small $A$, i.e. the Drude and Umklapp charge diffusion poles identified in Umklapp hydrodynamics. 

What is notable, is that the pole collision has already happened at a lower temperature than for perturbatively small $A$. Qualitatively this is easy to understand in terms of the RG wisdom that the lattice becomes irrelevant in the IR. If one starts with a stronger $A$ in the UV, one is at a relatively stronger strength at a temperature $T$ or vice versa one is at a comparable strength at a lower temperature $T$. This may seem like semantics, but crucially the DC conductivity linear-in-T scaling remains set by the local quantum critical IR fixed point, which is less affected by an increase in $A$. As a result we can again observe in the AC conductivity a transition in the dissipative mechanism as one increases $T$ during which the resistivity stays essentially linear (Fig.~\ref{fig:lin-in-temp-res-plus-saturation} in the Introduction). The transition in this case is that from the mid-IR-peak regime to an incoherent metal. The latter means that the low frequency AC response is no longer well described by the ``two-coupled-relaxational-current" formula. Other poles now also influence the AC response, especially the two Umklapped sound modes. They feature prominently in the AC response; see Fig.~\ref{fig:evolution-intermediatelattice}.

\begin{figure}
		\centering
		\includegraphics[width=\textwidth]{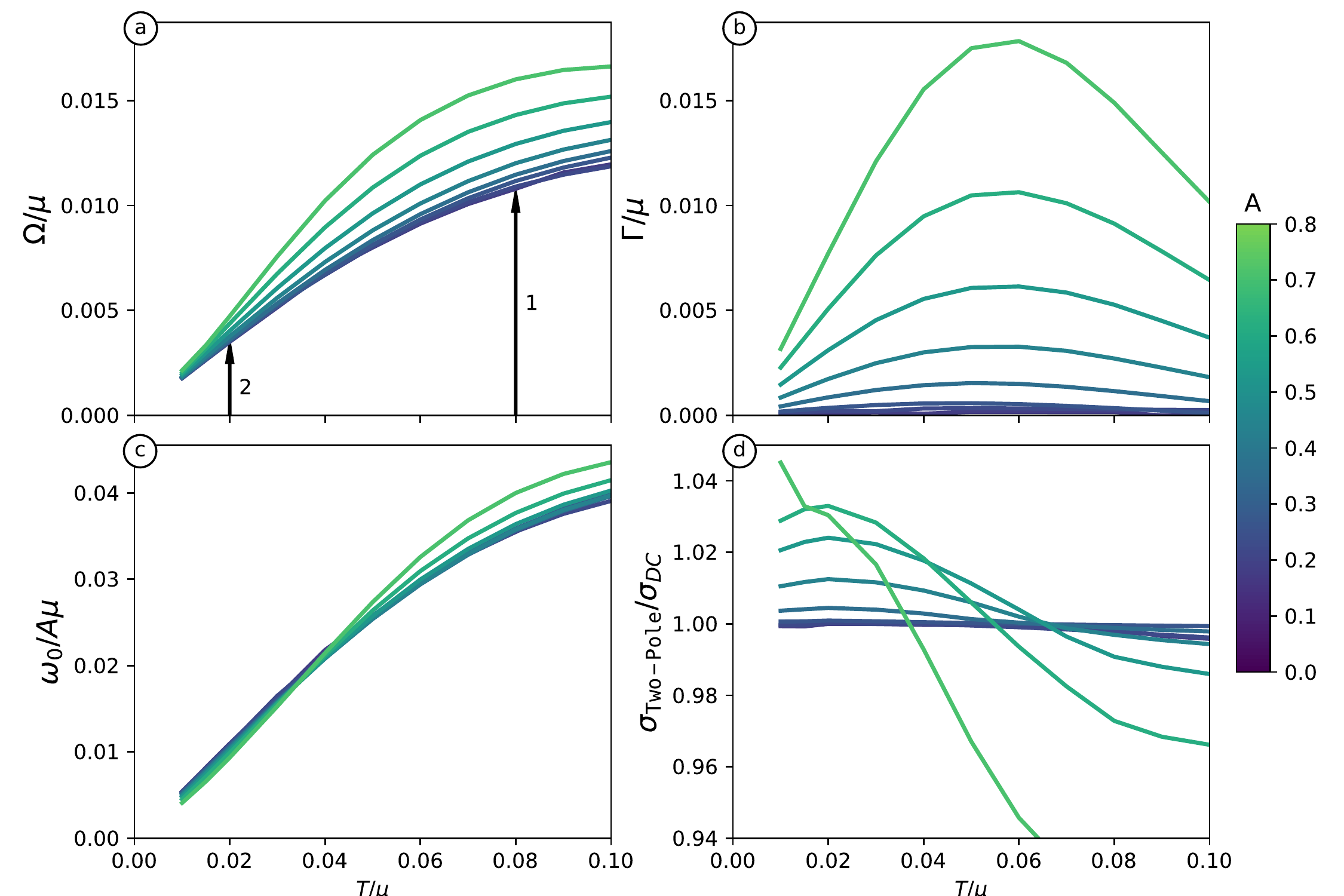}
		\caption{(a-c) The evolution of the phenomenological parameters $\Gamma/\mu, \Omega/\mu, \omega_0/A\mu$ as present in the ``two-coupled-relaxational-current'' expression Eq.~\ref{eq:two-pole-conductivity-formula} as a function of $A$ and $T/\mu$ at $G/\mu=0.12$ in the 1D Gubser-Rocha model. These parameters are extracted from a four-pole fit to the optical conductivity that includes the two lowest-order Umklapped sound peaks which reside at $\text{Re }\omega \approx \pm c_s G$. Both $\Omega/\mu$ and $\omega_0/A\mu$ show little $A$-dependence, whereas $\Gamma/\mu$ depends strongly non-linearly on $A$. In (a), the arrows labelled $1$ and $2$ point to the temperatures at which the pole collision happens at $A=0.1$ and $A=0.2$, respectively. For the stronger lattices, the pole collision has already happened at lower temperatures than we have access to in our numerics. (d) Comparison of $\sigma_{\text{Two-Pole}}$, the conductivity reconstructed from only the ``two-coupled-relaxational-current'' part of the spectrum in figures to $\sigma_{DC}$, the observed DC conductivity. At larger values of $A$, it becomes clear that one must include more information, such as the Umklapped sound modes, in order to accurately reconstruct the DC conductivity at all temperatures.}
		\label{fig:sigmaDC-intermediatelattice}
\end{figure}

Though the AC conductivity really shows the emergence of the incoherent metal regime at larger $T$ and the ``two-coupled-relaxational-current'' expressions fails, for most of the temperature range the DC limit $\omega\rightarrow 0$ is still well described by its asymptotic expression 
\begin{align}
\label{eq:dccond-intermediatelattice}
		\sigma_{DC} = \sigma_0 + \frac{Z}{\Gamma+\frac{\omega_0^2}{\Omega}}
\end{align}
With careful fitting of the optical conductivity as well as the complex location of the four poles, one can fit the parameters $Z, \sigma_0, \Gamma, \Omega, \omega_0^2$ as well as the parameters of the two first Umklapped sound poles as a function of $A$ and $T$. For the full 4-pole ansatz, see Section \ref{sec:4polefit}.
In Fig.~ \ref{fig:sigmaDC-intermediatelattice} we show how the three parameters in the denominator $\Omega, \Gamma$ and $\omega_0$ evolve as function of temperature for intermediate $0.1<A<0.8$.
One sees how these explain the observed DC conductivity quite well.
Given that the DC conductivity is so well captured by Eq. (\ref{eq:dccond-intermediatelattice}), one concludes that for these potentials the DC conductivity is still limited by the momentum life time.

\subsection{On the applicability of Umklapp hydrodynamics}
\label{sec:app-hydro-subsection}

\begin{figure}[!ht]
		\centering
		\includegraphics[width=0.7\textwidth]{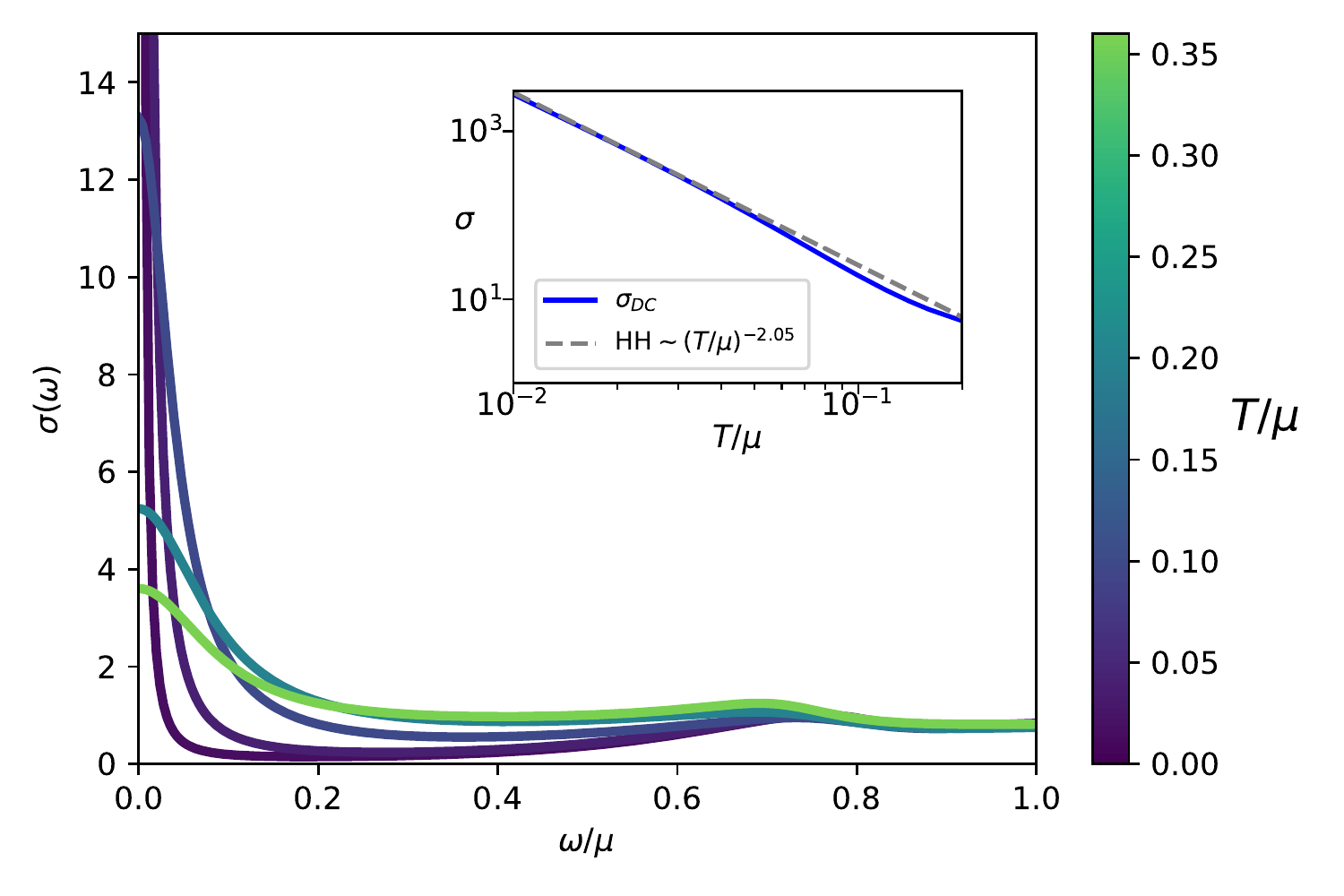}
		\caption{The AC conductivity of the GR model at $G=1.0\mu$ and lattice strength $A=1.0$ for a large temperature range. The low $\omega$-response is of the Drude form for all values and no transition to a mid-IR-peak or an incoherent metal is seen in contrast to lattice momenta $G < \mu$. The small rise at $\omega/\mu=\frac{1}{\sqrt{2}}\frac{G}{\mu} =\frac{1}{\sqrt{2}}$ is the Umklapped sound mode which now has barely noticeable height at low temperatures. The inset shows that the DC conductivity obeys leading order Hartnoll-Hofman scaling at low temperature, which is expected to go as $(T/\mu)^{-2.05}$ at low temperatures.}
		\label{fig:largeG-ACconductivity}
\end{figure}

We end this section with a brief check on our earlier argument in Section \ref{sec:appl-of-hydro} that Umklapp hydrodynamics is the relevant perspective to understand strange metal transport in a weak/intermediate lattice  for $G\lesssim \mu$ rather than Hartnoll-Hofman scaling. The intuitive argument is that momentum dependent conductivities are strongly power-law suppressed as a function of $T$ for $G\gtrsim \mu$ as the RG flow is not ``halted''. Umklapping conductivities that have such marginal weight  should have negligible observable effect. Fig.~\ref{fig:largeG-ACconductivity} shows that this insight is essentially correct. For a lattice with $G=1.0 \mu$, $T/\mu \lesssim 0.35$ and $A = 1.0$ the AC conductivity is Drude-like
, and no transitions to a mid-IR-peak or incoherent metal are seen. An illustration that formally Umklapp hydrodynamics still applies is that one can still notice the now very highly suppressed Umklapped sound peak. Even so, for $G\gtrsim \mu$ the better perspective is Hartnoll-Hofman scaling. Since $G/\mu$ is large here, the various exponents in the resistivity described in Section \ref{sec:appl-of-hydro} are not close and the lowest exponent $\nu_G$ of Eq.~\eqref{eq:8} alone is enough to describe the DC conductivity at low temperatures.

\section{Observations at strong lattice potentials: Planckian dissipation and incoherent metals}
\label{sec:saturation}

\begin{figure}
		\centering
		\includegraphics[width=0.9\textwidth]{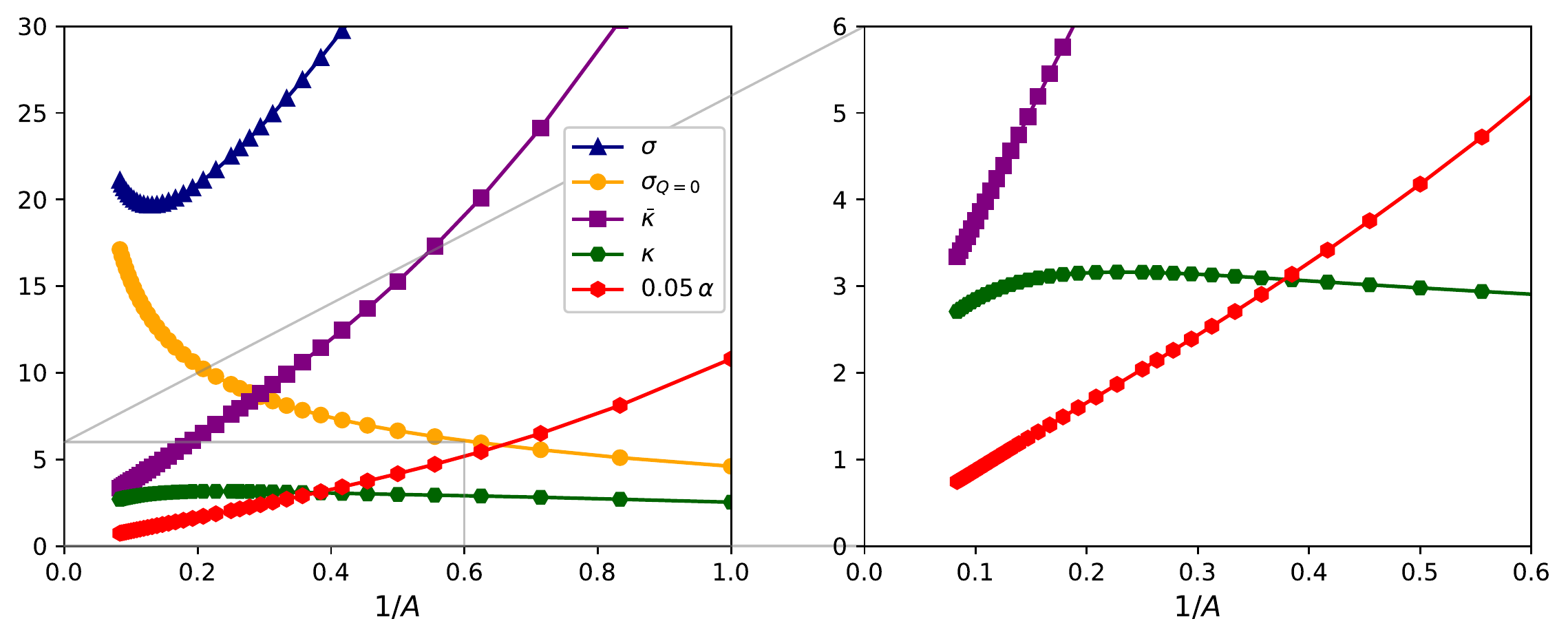}
		\caption{(Left panel) Absence of exact saturation of the conductivities as a function of lattice strength at fixed temperature in the 2D GR model is made quite clear when they are plotted as a function of $1/A$ instead of $A$. The electrical conductivity $\sigma$ reaches a minimum and then starts to grow again at larger $A$, whereas the thermo-electric $\alpha$ and heat conductivity $\bar{\kappa}$ scale as $1/A$ rather than saturate. (Right panel) The open boundary heat conductivity $\kappa$  at first instances does appear to be independent of the lattice strength $A$ for most of the computed values. However, at the largest $A$ it does show  a downturn, asymptoting to $\bar{\kappa}$ which vanishes as $1/A\to 0$. In this large $A$ regime, these asymptotes $\kappa \rightarrow \bar{\kappa}$ and $\sigma_{Q=0}\rightarrow \sigma$ indicate the increased dominance of the diffusive channel. These results are for the 2D GR lattice with $T=0.06\mu, G=0.1\sqrt{2}\mu$.
		}
		\label{fig:saturation-all}
\end{figure}

\subsection{The remarkable ubiquity of Planckian dissipation}

We now switch to analyzing our numerical results at large lattice potentials $A>1$. As we reviewed in Section \ref{homogeneousprop}, for small lattice potentials $A<1$, Planckian dissipation is unlikely to be universal as it will depend on the details of how translational symmetry is broken \cite{blakeUniversalDiffusionIncoherent2016,erdmengerSwaveSuperconductivityAnisotropic2015}. At finite density one must be in a regime where translation is broken strongly and long time transport is controlled by another dissipative mechanism than translational symmetry breaking.

\begin{figure}
		\centering
		\includegraphics[width=0.95\textwidth]{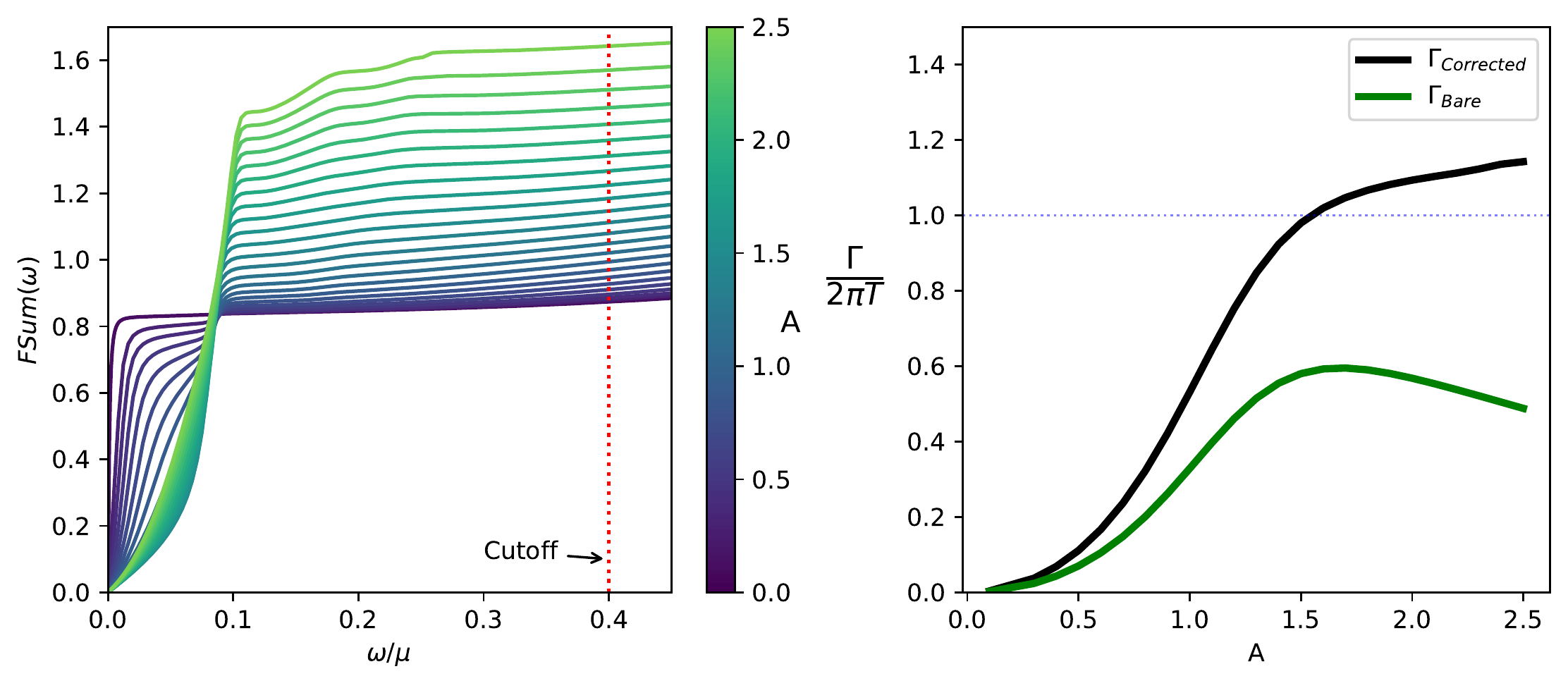}
		\caption{
		At large lattice potential one can construct a naively defined relaxation rate $\Gamma_{\text{bare}}^{-1} = \sigma_{\text{DC}}/\omega_p^2$. Strictly speaking this is only valid for weak lattices. Persisting in the analysis nevertheless, the change in $\Gamma$ is not just given by the change in $\sigma_{\text{DC}}$.
		The integrated optical conductivity $\mathrm{FSum}(\Lambda) =\int_0^{\Lambda}\!\sigma(\omega)\mathrm{d}\omega$ shows that the spectral weight $\omega_p^2$ also increases with $A$. We can account for this effect by normalising the Drude weight to this integrated spectral weight. {The resulting corrected relaxation rate $\Gamma_{\text{corrected}}^{-1} \equiv \sigma_{\text{DC}}/\mathrm{FSum}(\Lambda)$ does appear to show a saturating behavior compared to the bare rate $\Gamma_{\text{bare}}^{-1} = \sigma_{\text{DC}}/\omega_p^2$.} Furthermore, this rate is remarkably close to the Planckian value of $2\pi T/\mu$.  From inspection a cut-off value $\Lambda/\mu =0.4$ is sufficient to account for all the spectral weight in any Drude or Umklapped sound peaks. 
		Tantalizing as these results may be, a correct analysis at large lattice strengths eschews the use of a momentum relaxation rate altogether as it is no longer the unique longest timescale. The results above are taken in the 1D GR model with $T=0.06\mu, G=0.12\mu$. 
		}
		\label{fig:saturation-corrected}
\end{figure}

Performing this numerical experiment where we increase the lattice strength, one sees not only a beautiful sharper linear-in-$T$ resistivity, but also a saturating behavior in that the resistivity appears to become independent of the lattice strength $A$, highlighted in the Introduction (Fig.~\ref{fig:lin-in-temp-res-plus-saturation-part2}).
Though the thermo-electric and heat conductivity also appear to saturate, they do not. Replotting the results as a function of the inverse lattice strength $1/A$ rather than $A$, one sees that they asymptote to zero as $1/A$; see Fig.~\ref{fig:saturation-all}.
One also notes that the electrical conductivity does not saturate but turns over when inspected this precisely.
Treating the numerical results as a purely experimental finding, a naive Drude analysis does suggest that the dissipative process saturates --- even though this does not apply for strong momentum relaxation.
Increasing the lattice potential $A$ has two effects, it changes the strength and possibly mechanism of dissipation, but it can also shift degrees of freedom from lower to higher energy and vice versa. In simple Drude language where one postulates $\sigma_{DC}=\omega_p^2/\Gamma$, increasing the lattice strength cannot only affect $\Gamma$, but also the Drude weight $\omega_p^2$. Again, the Drude formula doesn't necessarily apply at large $A$, of course. Nevertheless, to focus on the dissipation we must also account for possible shifts in the weight. Because the total weight of the optical conductivity is protected and conserved, a more appropriate measure of the dissipation is to normalize the measured DC conductivity by the total weight $\int_0^{\Lambda}\!\text{d}\omega \sigma(\omega)$ and study the resultant rate $\Gamma^{-1}_{\text{corrected}} = \sigma_{\text{DC}}/\int_0^{\Lambda}\!\text{d}\omega \sigma(\omega)$. Fig.~\ref{fig:saturation-corrected} shows both the bare naive Drude rate $\Gamma^{-1}_{\text{bare}}=\sigma_{\text{DC}}/\omega_p^2$ and the corrected rate. Indeed in terms of the naive Drude rate even at the largest $A$ the saturating behavior in the conductivity is not exact. {However, when corrected for a possible spectral shift, the postulated relaxation rate does start to saturate. Not only does this relaxation rate appear to start to saturate, as Fig.~\ref{fig:saturation-corrected} shows, it does so at a value that is numerically close to the Planckian dissipation rate $\Gamma_{\text{corrected}} \simeq 2\pi /\tau_{\hbar} = 2\pi T$.}
A naive Drude weak momentum relaxation analysis applied in the strong lattice regime may therefore inadvertently lead one to conclude to have detected Planckian dissipation. However, to understand whether Planckian dissipation is really occurring, we must resort to a different theoretical framework.

\subsection{An incoherent metal explained with microscopic scrambling}
\label{sec:incoherent-diffusivities}

How to understand transport in a system where translation invariance is badly broken was discussed in detail by Hartnoll \cite{hartnollTheoryUniversalIncoherent2015}, and its connection with Planckian dissipation was set out in a series of papers \cite{blakeUniversalChargeDiffusion2016,blakeUniversalDiffusionIncoherent2016,blakeDiffusionChaosAdS22017,niuDiffusionButterflyVelocity2017,blakeThermalDiffusivityChaos2017} in the context of systems with strong translational disorder. The essence is that in this regime only energy and charge are the conserved currents that survive at long distances.
For this section we shall
not just focus on the electrical conductivity but on the full thermo-electric transport matrix
\begin{align}
		\begin{pmatrix} \vec{J} \\ \vec{j}_{Q} \end{pmatrix} = \begin{pmatrix} \sigma & \alpha T \\
	 \alpha  &  \bar{\kappa} \end{pmatrix} \begin{pmatrix} \vec{E} 
		\\
		\frac{1}{T} \vec{\nabla}T \end{pmatrix}
\end{align}
with ${j}^i_Q = \frac{1}{T}(T^{0i}-\mu J^i)$.
Here $\bar{\kappa} = \kappa + \frac{T \alpha^2}{\sigma}$ is the heat conductivity in the absence of electric field, and $\kappa$ is the heat conductivity in the absence of electric current (open boundary heat conductivity).
Fig.~\ref{fig:lin-in-temp-res-plus-saturation-part2} shows the result for all conductivities for increasing lattice strength into the incoherent regime, both in the Gubser-Rocha ($s_{\text{GR}} \sim T+\ldots$) and in the Reissner-Nordstr\"om AdS$_2$ metal ($s_{\text{RN}} \sim c_0 + c_1 T + \ldots$). 
The conductivities are rescaled such that their dominant power-law scaling with $T$ is scaled out. In detail one observes also that the thermo-electric and the heat conductivity conform sharper to the conjectured appropriate temperature scaling as $A$ increases, culminating again in a saturating behavior for large $A$. 

It is tempting to view this scaling of the thermo-electric conductivities as validating that the system is dominated by a single common relaxation time that scales like the entropy at low temperatures, even though it does not apply here as $A$ is large. Single relaxation time Drude theory would suggest that $\sigma = \omega_p^2/\Gamma$, $\alpha = \frac{s}{n}\sigma$, and $\frac{\bar{\kappa}}{T} = \frac{s^2}{n^2}\sigma$. If $\Gamma \sim s(T)$ as naively guessed above, it is consistent with the above observations. As we will now explain, and confirmed with counterexamples in studies of strong translational disorder, this single relaxation time description is not correct.

\begin{figure}
		\centering
		\includegraphics[width=0.95\textwidth]{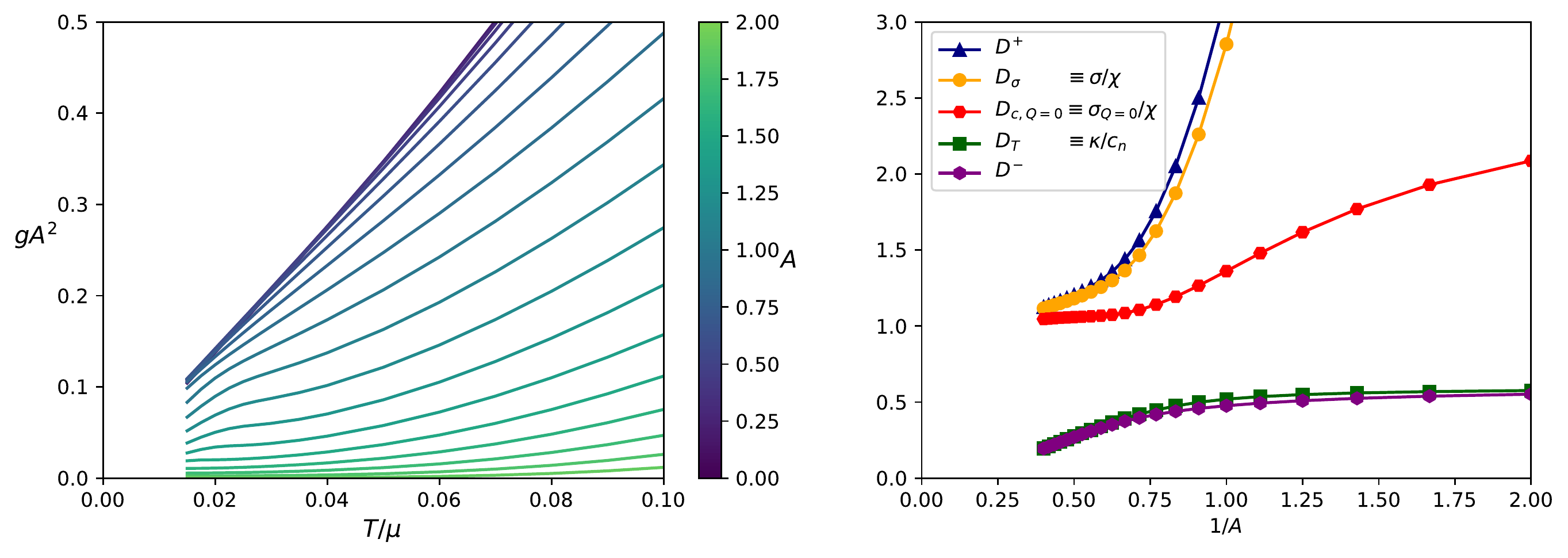}
		\caption{Left: The cross-coupling between the heat and electrical conductivity in the strongly coherent regime is governed by the combination  $g=\frac{T\sigma}{c_{n}}\left(\frac{\alpha}{\sigma}-\left(\frac{\partial s}{\partial n} \right)\right)^2$ respectively at low temperatures. Clearly $g$ decreases linear in temperature at low $T$, but it also decreases with stronger lattice potential $A$. Right: As a consequence the diffusivities at low $T$ in a strong lattice become independent. Shown are the empirical combinations $D_\sigma \equiv \frac{\sigma}{\chi}, D_T=\frac{\kappa}{c_n}, D_{\sigma_{Q=0}}=\frac{\sigma_{Q=0}}{\chi}$ as a function of $1/A$ for fixed $T/\mu=0.05$.
		}
		\label{fig:incoherent-crosscoupling}
\end{figure}

To extract possible relaxation rates in an incoherent metal with strong translational symmetry breaking, one posits constitutive relations for the two remaining currents and does a hydrodynamic analysis.
One finds that the DC conductivities are the zero frequency limit of the dynamics of two independent diffusive modes with diffusion constants $D_+$ and $D_-$. These are
\begin{align}
		D_++D_- &= \frac{\kappa}{c_n} + \frac{\sigma}{\chi} +\frac{T\sigma}{c_{n}}\left(\frac{\alpha}{\sigma}-\left(\frac{\partial s}{\partial n}\right)_T\right)^2 \non
		D_+D_- & =\frac{\kappa}{c_n}  \frac{\sigma}{\chi}
\end{align}
Here $c_{n}=T\left(\frac{\partial s}{\partial T}\right)_n$ is the specific heat at fixed charge density, $\chi = \left(\frac{\partial n}{\partial \mu} \right)_T$ is the isothermal charge compressibility, and the conductivities $\sigma, \kappa$ are both the transport coefficients as well as the DC values.
One recognizes a charge diffusion and a heat/energy diffusion mode (the remnant of sound in absence of a nearly conserved momentum), cross coupled through the combination $g \equiv \frac{T\sigma}{c_{n}}\left(\frac{\alpha}{\sigma}-\left(\frac{\partial s}{\partial n}\right)_T\right)^2$. 
If we are to make the case that a single dissipative mechanism dominates, this cross-coupling is important, as in its absence, charge and energy diffusion are clearly independent. 
Fig.~\ref{fig:incoherent-crosscoupling} shows what the strength of this coupling is numerically. As was shown in \cite{blakeThermalDiffusivityChaos2017}, this coupling behaves as $g/\sigma \sim T^{(z+d - \theta)/z}$ if the scaling of the homogeneous non-trivial IR fixed point remains valid in the presence of strong translational symmetry breaking. For the GR metal this means $g \sim T$. 
Compared to $\sigma/\chi \sim T^{-1}$ it is therefore small and can be  treated perturbatively in the low temperature limit.

Solving for $\sigma, \kappa$
in the limit where the terms in the cross coupling $\frac{T\alpha^2}{c_n\sigma}\sim T$, $\frac{T\alpha}{c_n}\left(\frac{\partial s}{\partial n}\right) \sim T$ and $\frac{T \chi \sigma}{c_n}\left(\frac{\partial s}{\partial n}\right)^2_T\sim T$
are small compared to $\sigma \sim T^{-1}$, one finds\footnote{Note that the coupling term $\frac{T \chi}{c_n}\left(\frac{\partial s}{\partial n}\right)_T = \frac{n T}{(\eps + P)} - \dfrac{c_s^2 \mu}{\alpha_n s}$ contains the same thermodynamic factor as $\Gamma_d$. If the temperature scaling in the strong lattice is the same as in the homogeneous system, this coupling scales as $\dfrac{n T}{(\eps + P)} - \frac{c_s^2 \mu}{\alpha_n s}\sim T$ since $\alpha_n \sim T^{-2}$ as was shown in Appendix \ref{app:gamma-vs-s-scaling}. Numerics confirms that this is the case.}
\begin{equation}
		\begin{aligned}
				\frac{\sigma}{\chi} & = D_+ \left[1 + \frac{T}{c_n \chi} \dfrac{\left(\alpha - D_+ \chi \left( \frac{\partial s}{\partial n} \right)_T \right)^2}{D_+ (D_- - D_+)} \right]\\
				\frac{\kappa}{c_n} & = D_- \left[1 - \frac{T}{c_n \chi} \dfrac{\left(\alpha - D_+ \chi \left( \frac{\partial s}{\partial n} \right)_T \right)^2}{D_+ (D_- - D_+)} \right]\\
		\end{aligned}
\end{equation}
To lowest order in the temperature the electrical and heat conductivity are therefore determined by independent diffusion constants; see Fig.~\ref{fig:incoherent-crosscoupling}. The electrical conductivity is determined by $D_+\sim T^{-1}$ and the heat conductivity by $D_- \sim T$. There is therefore no simultaneous explanation for both conductivities in terms of  universal Planckian dissipation. In holographic models with strong translational disorder there are systems where both conductivities are set by Planckian dissipation \cite{blakeUniversalChargeDiffusion2016,blakeUniversalDiffusionIncoherent2016}. This happens when the charge susceptibility is relevant. For irrelevant or marginal charge susceptibility, the electrical conductivity is set by a different dissipative mechanism. The Gubser-Rocha model with strong disorder belongs to this class \cite{niuDiffusionButterflyVelocity2017}, and so does our strong ionic lattice model with $\chi \sim T^0$.

Despite the existence of two independent dissipative mechanisms,  the heat conductivity can be explained from Planckian dissipation. Very strongly coupled systems are similar to weakly coupled dilute classical gases in that their macroscopic transport can be understood from microscopic processes. For weakly coupled dilute gases this is through the Boltzmann equation summing microscopic scattering; for ultrastrongly coupled systems this is through parameters of microscopic scrambling as measured through the out-of-time-ordered correlation function $C(t,x) = \langle W(t,x)V(0)W(t,x)V(0)\rangle_T \sim e^{\lambda (t-x/v_B)}$.\footnote{This ``ballistic`` OTOC expression applies to large $N$ systems such as holographic and SYK systems. The more generic answer is ``diffusive'' $C(t,x) \sim e^{\lambda (t - x^2/v_Bt) }$.} In holographic systems this connection manifests itself in that the OTOC is equivalent to computing the hydrodynamic response function (of longitudinal sound) at imaginary $\omega$ and $k$  \cite{grozdanovBlackHoleScrambling2018}. The Lyapunov exponent $\lambda$ and the butterfly velocity $v_B$ can then be read off from a skipped pole in the hydrodynamic dispersion relation \cite{grozdanovBlackHoleScrambling2018}. One finds that in holographic systems $\lambda$ saturates the Maldacena-Shenker-Stanford unitarity bound $\lambda \leq 2\pi/T$. The butterfly velocity is more sensitive to the theory. On general grounds it scales near (translationally invariant) quantum critical IR fixed points as $v_B^2 \sim T^{2-2/z}$. The fact that both macroscopic transport and the scrambling parameters $\lambda, v_B$ are encoded in the hydrodynamic response means that they are not unrelated. In particular the thermal diffusivity $D_T = \dfrac{\kappa}{c_n} = E \dfrac{v_B^2}{\lambda}$  with $E=\frac{1}{2}$ for AdS$_2$ $z\rightarrow \infty$ metals in strong disorder \cite{blakeUniversalDiffusionIncoherent2016,blakeDiffusionChaosAdS22017,blakeThermalDiffusivityChaos2017}. Since the natural units of diffusivity are $v^2\tau$, this is interpreted as Planckian dissipation with $\tau = \frac{1}{\lambda} = \frac{1}{2\pi T}$. The RN metal is a special case. As explained in \cite{blakeDiffusionChaosAdS22017}, there the butterfly velocity is controlled by a dangerously irrelevant operator instead of universal scaling. A careful computation reveals that for the RN strange metal $v_B \sim \sqrt{T}$. Combined with Planckian dissipation $\tau = \frac{1}{\lambda}=\frac{1}{2\pi T}$, this explains the observed RN thermal diffusivity $D_- = \frac{\kappa}{c_n} = T^0\sim v_B^2\tau$.

This result is established and confirmed in the many studies cited above on connecting scrambling to hydrodynamics for vanishing, weak momentum relaxation or ``homogeneous'' momentum relaxation
\cite{lingHolographicButterflyEffect2017,ahnThermalDiffusivityButterfly2018,lucasOperatorSizeFinite2019,langleyQuantumCriticalDiffusion2021,liThermalDiffusionQuantum2019,inkofQuantumCriticalScaling2020,baggioliUniversalBoundsTransport2020,jeongBoundDiffusionConstants2021a,jeongQuasinormalModesDyonic2022a} and \cite{patelQuantumChaosCritical2017,patelQuantumButterflyEffect2017,wermanQuantumChaosElectronphonon2017}.\footnote{Presumably for all models with so called maximal quantum chaos \cite{blakeSystemsMaximalQuantum2021}. There are counterexamples \cite{lucasChargeDiffusionButterfly2016,guEnergyDiffusionButterfly2017,davisonSlowRelaxationDiffusion2019,wuClassicalModelSubPlanckian2021}; see \cite{wuUniversalityAdS2Diffusion2021} for a good discussion}
We postulate that the same applies in the explicit strong lattice systems studied here. This need not be, for computing the butterfly velocity $v_B$ in a non-translationally invariant system is not straightforward (the Lyapunov exponent on the other hand is universally $\lambda=2\pi/T$ \cite{maldacenaBoundChaos2016}). At the same time the scaling we observe for strong lattice potentials is the same as that which is observed for strong translational disorder. This is strong evidence in favor of the argument that the same should apply here.

Within the framework of incoherent metals there is no universal explanation of the observed inverse-in-T scaling of the conductivity for the Gubser-Rocha metal. Its tantalizing behavior $\sigma \sim s(T)^{-1}$ or rather $\sigma \sim \frac{1}{\Gamma_\hbar}$ on the other hand {\em does} suggest that some type of universality is at work. This is confirmed by the RN results. The obvious conjecture is that $D_- = (v^{\text{charged}})^2/\lambda$ where the velocity $v^{\text{charged}}$ relevant for diffusion of charged objects {\em differs} from the butterfly velocity for neutral objects. Some evidence that this can be the case is presented in \cite{ageevWhenThingsStop2019,sorokhaibamPhaseTransitionChaos2020,colangeloChaosOverlineQSystem2020,chenManybodyQuantumDynamics2020}. If $v^{\text{charged}}$ were independent of temperature, this would explain the observed incoherent metal phenomenology in the large lattice GR and RN metals in terms of a single Planckian relaxation time, but differing scrambling velocities. We leave this for future research.

\subsection{Saturating behavior and Planckian dissipation}

The diffusivities in the incoherent regime should be insensitive to the details of translational symmetry breaking. This is what allows them to expose universal dissipative physics. This resulting explanation of universality in terms of microscopic scrambling also makes physical sense: the onset of chaos is controlled by the short-range interactions and is not expected to be influenced significantly by a background lattice.  The data we present is obviously dependent on the lattice strength $A$. For most values of $A$ we are therefore not in the universal regime. However, as $A$ increases to the largest value we can observe in our numerical data, there is a saturating behavior in the electrical conductivity that together with its sharper single power behavior argues strongly that we are close to this universal incoherent limit. Such saturating behavior in the incoherent electrical conductivity at large lattice strength was already noted in \cite{cheslerConformalFieldTheories2014}. That study focused on the regime where the dimensionless combinations  $\frac{\mu}{G} \rightarrow 0, \frac{A\mu}{G}$ fixed and large. Here we focus on the regime where both $\frac{\mu}{G}$ and $\frac{A\mu}{G}$ are fixed and large with the latter parametrically larger. 

\begin{figure}
		\centering
		\includegraphics[width=0.9\textwidth]{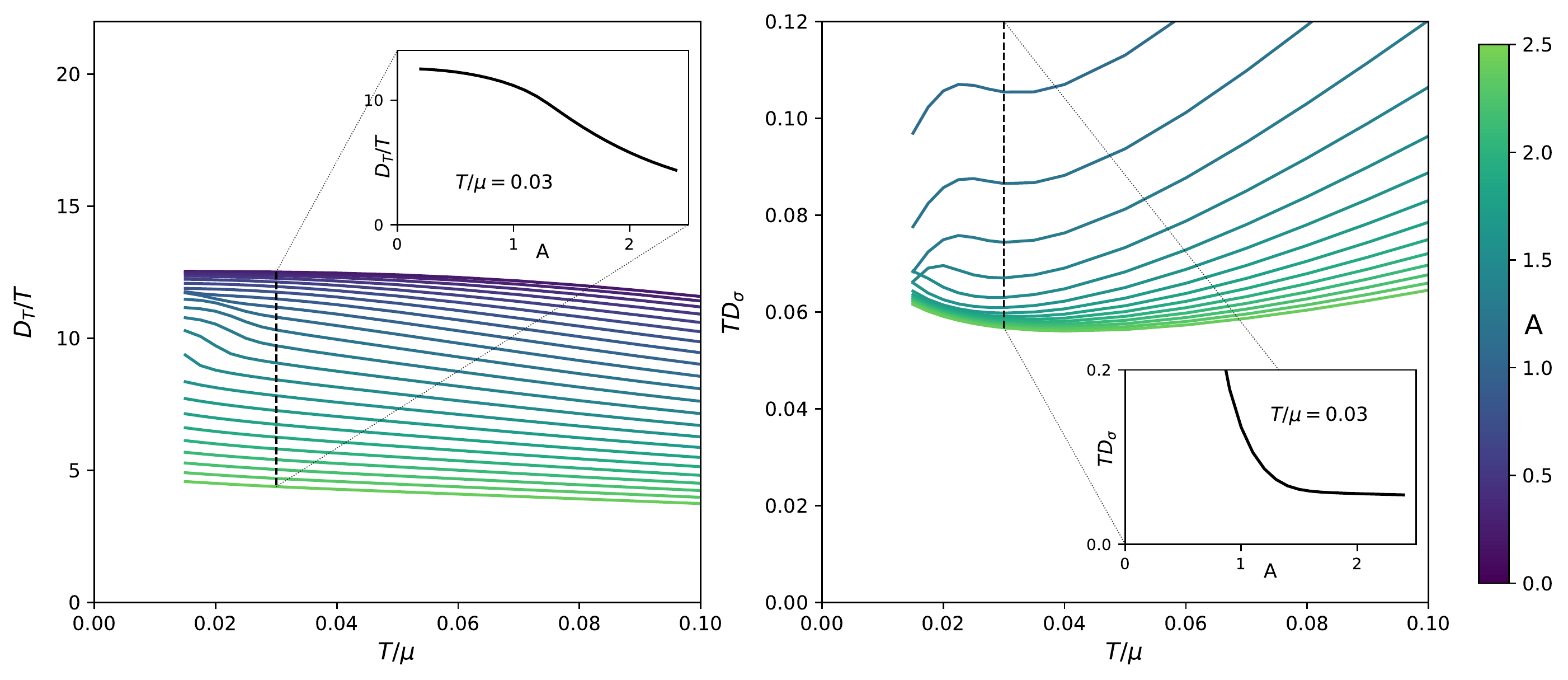}
		\caption{Electrical and thermal diffusivities as a function of $T$ for various $A$. The $T$-dependence shows how they become more single-power like at larger $A$. The $A$-crosssection shows the saturation for large $A$ for the charge diffusivity, but an increasing $A$ dependence for the thermal diffusivity. These results are in the GR lattice at $G=0.1\mu$.}
		\label{fig:diff-saturation}
\end{figure}

We can use our numerical results to directly check these assertions. Rather than observing the conductivities we do so for the diffusivities 
\begin{align}
		D_\sigma = \frac{\sigma}{\chi},~D_T=\frac{\kappa}{c_n},~D_{\sigma_{Q=0}}=\frac{\sigma_{Q=0}}{\chi}
\end{align}
We have introduced here a charge-without-heat diffusivity $D_{\sigma_{Q=0}}=\frac{\sigma_{Q=0}}{\chi}$ as this is the appropriate counterpart to the heat-without charge open boundary thermal diffusivity $D_T \equiv \kappa/c_n$. Fig.~\ref{fig:incoherent-crosscoupling} shows indeed how the charge diffusivities $D_\sigma$, $D_{\sigma_{Q=0}}$ not only both saturate, but also become approximately equal. The latter shows indeed that we have entered the incoherent regime. A more detailed depiction of the saturation is given in Fig.~\ref{fig:diff-saturation}.

We have already shown in the Introduction that the crossover into the incoherent sector can also be seen in the conductivities directly (Fig.~\ref{fig:incoherentransportcoef}). 
The open boundary thermal conductivity $\kappa$ starts to comprise more than 80\% of the full heat conductivity. A stronger statement extrapolated from the incoherent metal considerations is that the open boundary heat conductivity $\kappa$ is rather insensitive to momentum relaxation for any translational symmetry breaking potential irrespective of its strength \cite{blakeThermalDiffusivityChaos2017}.  
According to  Fig. (\ref{fig:saturation-all}) this is indeed the case in the perturbative small $A$ case. Upon pushing the potential to extremely large values we do observe that some changes in $\kappa$ start to arise. This is fully in the incoherent regime, where we can equate $\kappa \equiv c_nD_T$ with one of the physical diffusion constants $\kappa =c_nD_-$. This diffusion constant also changes from $A$-independent to slight decaying behavior, explaining the change in behavior in $\kappa$.
We conclude that at least for $D_T$ our computations confirm the universal nature of the diffusion constants.

\section{Discussion: is it relevant for condensed matter physics?} 
\label{outlook}

We started this paper with just presenting the data as these rolled out of the computer. As such these are highly suggestive. We focus in on a holographic strange metal that fulfills minimal conditions that appear to be imposed by experiment: local quantum criticality ($z \rightarrow \infty$) and a Fermi-liquid like thermodynamics in the form of a Sommerfeld entropy ($s \sim T$). We then invoke a lattice potential that may become very strong, again a minimal requirement suggested by experiment. For a wavevector of the potential that is not too large (smaller than the inverse local length) we find a resistivity that is to good approximation linear in temperature for a large range of potential strength. Ramping up the potential the slope of the linear resistivity saturates at a value that is consistent with a Planckian ($\tau_{\hbar}$) current life time. Although the dynamical range in temperature and potential strength is limited in our computations,  we can track the temperature evolution of the optical conductivity in the regime where the saturation is setting in. This temperature evolution is also suggestive with regard to experiment: at low temperature we find a simple Drude response that turns into an incoherent mid IR peak, and this gross change does not imprint on the DC resistivity that stays linear. Taken together, this shines an unusual light on three problems of principle in strange metal transport: (a) Why is the resistivity linear in temperature down to the lowest temperatures? (b) Why is the empirically extracted current relaxation time so close to the Planckian rate $\tau_{\hbar}$?
(c) Why does the cross-over from good metal (Drude optical conductivity) to bad metal (the mid IR peak response) not affect the DC resistivity at all? 

The question remains whether the resemblances between numerical observations from this holographic toy model and the complicated reality of the copper oxide electron systems are just a coincidence or whether 
they reveal a truly universal principle governing transport that supersedes all the differences between them. To get a better understanding, we focused in on both the small- and large 
lattice potential regimes. We showed that in the perturbative small potential regime the transport behavior can be completely reconstructed on basis of the thermodynamics and transport properties of the unbroken homogeneous system. This is based on hydrodynamical flow behavior in the presence of a weak periodic potential and we discovered a generic principle governing linear response: next to the usual shear drag, a mode coupling emerges with the Umklapped charge diffusion mode. As we increase temperature the coupling between two relaxation modes can account for a second new phenomenon: the two poles can collide and this explains the emerging mid-IR peak in the AC conductivity. Even though the temperature dependence of the DC-resistivity is formally set by the same thermodynamic quantities, the underlying non-trivial IR fixed point constrains these in such a way that the DC resistivity temperature scaling can be independent of the dynamical change in the AC conductivity.

The large lattice potential regime on the other hand is where the resistivity slope saturates. Our numerics indicate that this happens in the ``incoherent metal'' regime where momentum conservation does not play any role. Accordingly, the temperature dependence of the resistivity should be inversely proportional to the {\em charge} diffusivity.
This charge diffusivity in the incoherent regime $D_-\sim T^{-1}$ should not be compared with the hydrodynamical charge diffusivity for weak or zero momentum relaxation which scales as $D_{\rho} \sim T$. The {\em thermal} diffusivity $D_T \sim T$ on the other hand is essentially insensitive to the strength of the lattice potential. It scales similarly for both small and large potential, though only at large potential can it be explained in terms of microscopic chaos anchored in a saturated Lyapunov bound $\Gamma =\lambda = 2\pi T$ having a Planckian magnitude. Although this is presently not well understood this is consistent with the analytical findings in a homogeneous holographic strange metal with momentum relaxation (Q-lattice).

Although this does shed light on various aspects we do not claim a complete understanding of our numerical results. The above suggests that there are quite different forms of physics  at work pending the strength of the potential. Nevertheless, we do find that the evolution of the transport quantities is of a strikingly smooth kind.  Another striking aspect is the contrast between the GR and RN results in Fig.~\ref{fig:lin-in-temp-res-plus-saturation-part2}: the differences in temperature dependencies appear to be entirely linked to the different temperature dependence of the entropy. The above analysis, where we can expose the different origins in the weak and large lattice potential regime, does make clear that this connection with entropy is almost certainly a coincidence, though we cannot exclude that some yet to be identified greater universality may be at work linking the dissipative properties in the convective and diffusive regimes together where entropy may play a crucial role.

To use this to explain the experimental observations, the critical holographic input is in the form of the current being controlled by ``generalized" hydrodynamics (including the incoherent metal) that in turn requires (a) an existence of hydrodynamics up to microscopic length scales shorter than the lattice spacing, (b) thermodynamical behavior
that is anchored in a non-trivial IR fixed point, and (c) a saturation of the chaos bound (with a charge dependent butterfly velocity $v_B^{\text{charge}}$). 

In fact, the most critical question is whether experimental strange metal transport is governed by hydrodynamics, and not by the usual quasiparticle transport. In this regard our finding that hydrodynamics provides a most natural explanation for the temperature evolution of the DC and AC charge response is encouraging: it is an elementary mechanism that offers a minimal and simple explanation for this otherwise mysterious affair.\footnote{Note that the focus here is only on the mysterious behavior of the longitudinal linear-in-$T$ resistivity. We do not address other experimental cuprate strange metal conundrums, such as the concurrent $T^2$-scaling in the Hall angle.} However, to prove it one would like to mobilize the mesoscopic transport devices of the kind that have proven successful in this regard observing hydrodynamical flow behavior in graphene (e.g., \cite{sulpizioVisualizingPoiseuilleFlow2019}). 

The next issue is, are the hydrodynamical modes surviving down to length scales of order of the microscopic lattice spacing $1/ G$? We found this to be a special property of the local quantum critical holographic metals, but is this also at work in the cuprate strange metals? This is far from obvious. Besides the Umklapped charge diffusion mode, we also saw the sharp and prominent Umklapped sound peak in the optical conductivity when the potential becomes sizable. This relates directly to a first discrepancy between our results for the optical conductivity and the experimental results in the cuprates. We find that for the strongest potentials that our numerics can handle,  the optical response rather abruptly switches off at frequencies above the Umklapped sound peak ( Fig.~\ref{fig:evolution-intermediatelattice}). In experiment no sound peak is seen, and a  power law (branch cut) tail is found instead, extending all the way up to $\mu \simeq 1$ eV \cite{marelQuantumCriticalBehaviour2003,michonPlanckianOpticalConductivity2022,vanheumenStrangeMetalElectrodynamics2022}. Our holographic results do not shed any light on this matter, although one could imagine that perhaps an Umklapped {\em overdamped} sound channel could conspire to give rise to such a quasi-critical behavior. But the issue is whether the charge diffusion hydro-mode that is responsible for the mid IR peak in holography may survive up to large momenta in the experimental systems. Different from sound, this mode is non-convective and perhaps less sensitive to translational symmetry breaking. Presently we have no answer to this question. It could be interesting to study the optical conductivity of the cuprate metals experimentally at high temperatures.  The data in so far available are sketchy and it would be interesting to find out what a systematical and high precision study would reveal regarding for instance the way in which  the mid IR peak depends on temperature. Alternatively the sound contribution to the density-density response can been measured directly by EELS \cite{mitranoAnomalousDensityFluctuations2018,husainCrossoverChargeFluctuations2019}, with the caveat that sound is promoted to a plasmon in the presence of dynamical electromagnetism.This may be hard, because the plasmon is damped stronger in strange metallic states than ordinary Fermi liquids \cite{Gran:2017jht,Mauri:2018pzq,Romero-Bermudez:2018etn,Andrade:2019bky}. The results are at this moment inconclusive, and need to still be found consistent with the AC optical conductivity.

Perhaps the most delicate issue relates to the connection with microscopic chaos. The connection with Planckian dissipation requires a saturation of the Maldacena-Shenker-Stanford bound on the Lyapunov exponent of the OTOC $\lambda \leq 2\pi T$. It appears that a necessary condition for this to happen is in the form of {\em dense many body entanglement}. One may argue that this is the secret of the experimental strange metals: these are born from strongly interacting {\em fermion} systems at a finite density and it may well be that the concomitant sign problem enforces dense entanglement in the non-Fermi-liquids \cite{zaanenLecturesQuantumSupreme2021a}. But this may not be a sufficient condition.  The chaos bound is known to saturate in  matrix large $N$ systems at strong coupling with a holographic dual as well as the disorder averaged SYK models.  These systems are characterized by dense matrix interactions. 

However, the Hubbard models that are the community standard as microscopic point of departure for the cuprate electrons are characterized by local interactions and the associated Hamiltonians correspond with rather sparse matrices. As with regard to the transport properties, the present benchmark is in the form of finite temperature quantum Monte Carlo computations for the resistivity \cite{huang2019strange}. The temperatures that can be reached are still quite high ($ \simeq 1000$ K) but arguably approaching the linear resistivity regime. However, the outcomes are quite different from what we find. 

The Hubbard models are of course in their own way toy models, capturing the largest scales in the problem but ignoring a lot of other physics. Could it be that long range interactions arising e.g. from Coulomb interactions and/or phonon mediated interactions are crucial to support the rapid scrambling near the Lyapunov bound given their non-local nature \cite{Xu:2022vko}? Could there be a direct relation to SYK physics associated with the observation of spin glass physics  \cite{Tsuji:2018mth,Christos:2022lma} , with the obvious difficulty that this has only been observed in the spin striped 214 system?

At the least, holography inspires to ask quite unusual questions to experiment: it suggests a physics that is tantalizingly  different from the usual Fermi-liquid quasiparticle physics. Eventually, it should be possible by targeted experimentation to reach a verdict. This is not easy: the cuprates have been subjected to unprecedented experimental scrutiny over the last 35 years  but on basis of the available information it is still impossible to decide the issue. 

An example of this law-of-Murphy that insightful results may be the hardest to obtain experimentally is the thermal transport. So much is clear that the thermal conductivity $\kappa$ of the GR metal acquires a universal temperature dependence that is up to very high lattice potentials independent of the potential strength. Numerically we observe that $\kappa \sim T^2$. But this in gross contrast with the thermal conductivity in a Fermi liquid, where $D_T \sim \tau_c$ where $\tau_c \sim 1/T^2$ is the quasiparticle collision time  such that $\kappa \sim 1/T$~\cite{abrikosovTheoryFermiLiquid1959}.  There is a large difference of the order  $T^3$  in the temperature dependence of the thermal conductivity between the holographic metal and a Fermi liquid! 

This should be the smoking gun but why can this not be used? The reason is that at the high temperatures where the strange metal is realized ($ > 100$ K) the thermal transport is rather completely dominated by the {\em phonons}. The phonon heat conduction short circuits the  heat transport and it is virtually impossible to extract the electronic contributions. 
The same problem is there for a measurement a charge transport without heat $\sigma_{Q=0}$. Aside from the experimental hurdle of zeroing out heat transport cleanly, the definition of $\sigma_{Q=0}=\sigma-T\alpha^2/\bar{\kappa}$ implicitly refers to the {\em electronic} component of   the heat transport only.

Finally, there is one thermo-electrical transport coefficient that is readily available experimentally: the Seebeck coefficient enumerating the thermopower. This is given by $s = \alpha / \sigma$.  According to  Fig.~\ref{fig:lin-in-temp-res-plus-saturation-part2}, $\alpha \sim T^0$ and $\sigma \sim 1/T$, and we predict $s \sim T$: although for different reason this is the same temperature dependence generic for a Fermi-liquid (the Mott formula), this is indeed the scaling that has been  observed in cuprate strange metals , e.g. \cite{collignonThermopowerPhaseDiagram2021,gourgoutSeebeckCoefficientCuprate2022}.

\acknowledgments

\noindent
We thank R. Davison, B. Gout\'{e}raux, S. Grozdanov, S. Hartnoll and A. Krikun for discussions. This research was supported in part by the Dutch Research Council (NWO) project 680-91-116 ({\em Planckian Dissipation and Quantum Thermalisation: From Black Hole Answers to Strange Metal Questions.}), the FOM/NWO program 167 ({\em Strange Metals}), and by the Dutch Research Council/Ministry of Education. K.G. acknowledges funding from the European Union’s Horizon 2020 research and innovation programme under the Marie Sk\l odowska-Curie grant agreement No 101024967. The numerical computations were carried out on the Dutch national Cartesius and Snellius national
supercomputing facilities with the support of the SURF Cooperative as well as on the ALICE-cluster of Leiden
University. We are grateful for their help.

\appendix
\newcommand{\dd}{\text{d}}
\section{AdS RN and GR black holes}
\label{app:set-up}
We will be interested in perturbations of both Reissner-Nordstr\"{o}m and Gubser-Rocha black holes.

\subsection{Reissner-Nordstr\"{o}m}

The RN black holes start from the Einstein-Maxwell action
\begin{equation}
	S = \int \dd^4 x \, \sqrt{-g}\left[ \dfrac{L^2}{2 \kappa^2} \left(R - 2 \Lambda \right) - \dfrac{L^2}{4 e^2} F_{\mu\nu}F^{\mu\nu}\right]~,
\end{equation}
with $2 \kappa^2 = e^2 = L^2 = 1$ and $\Lambda = - 3$. The equations of motion are
\begin{equation}
\label{eq:RNEOMs}
	\begin{aligned}
		R_{\mu\nu} - \Lambda g_{\mu\nu} & = \dfrac{1}{2} \left[ F_{\mu \rho} F_\nu^\rho - \dfrac{1}{4} g_{\mu\nu} F_{\rho\sigma} F^{\rho\sigma} \right]~,\\
		\nabla_\mu F^{\mu\nu} & = 0~.
	\end{aligned}
\end{equation}
These equations admit an electrically charged black hole solution, the AdS-Reissner-Nordström (RN) solution in asymptotically AdS\textsubscript{4} space-time, for which the metric and gauge field are given by\footnote{\label{fn:DG1}Sometimes, it is more convenient to make a change of variable $z \rightarrow 1-(1-r)^2$ \cite{donosMinimallyPackedPhases2016}.}
\begin{equation}
\begin{aligned}
	\dd s^2 & = g_{\mu\nu} \dd x^\mu \dd x^\nu = \dfrac{1}{z^2} \left[ - f(z) \dd t^2 + \dfrac{\dd z^2}{f(z)} + \dd x^2 + \dd y^2 \right]~,\\
	A &= A_t(z) \dd t ~,
	\end{aligned}
\end{equation}
where
\begin{equation}
f(z) = \left(1-z\right)\left(1 + z + z^2 - \frac{\mu^2 z^3}{4}\right),\quad A_t(z)= \mu(1-z).
\end{equation}
 The radial coordinate $z$ can be scaled such that the horizon is located at $z_h = 1$ and the boundary of the space-time is at $z = 0$. The temperature of the black hole can be computed by considering the surface gravity of the horizon, and is given by
\begin{equation}
T_{\mathrm{RN}} = \left|\frac{f'(z_h)}{4\pi}\right| =  \frac{12 - \mu^2}{16\pi}
\end{equation}

\subsection{Einstein-Maxwell-Dilaton}

For the dilatonic black holes, we depart from the Einstein-Maxwell-Dilaton action\cite{gubserPeculiarPropertiesCharged2010,lingHolographicLatticeEinsteinMaxwellDilaton2013}
\begin{equation}
\label{eq:EMDAnsatz}
S =\frac{1}{2\kappa^2} \int \dd^4 x \sqrt{-g}\left[R - \frac{Z(\phi)}{4}F_{\mu\nu}F^{\mu\nu} - \frac{1}{2}\left(\partial_\mu\phi\right)^2 + V(\phi)\right].
\end{equation}
The potentials $Z,V$ are given by 
\begin{equation}
Z(\phi) = \exp(\phi/\sqrt{3}), \qquad V(\phi) = \frac{6}{L^2}\cosh(\phi/\sqrt{3}).
\end{equation}
The cosmological constant is given by $\Lambda = -V(0)/2 = -3$. Setting $2\kappa^2 = L^2 = 1$, the equations of motion for this system are given by
\begin{equation}
		\label{eq:EMDEOMs}
		\begin{aligned}
				R_{\mu\nu} & = \dfrac{Z(\phi)}{2} \left[ F_{\mu}{}^{\rho} F_{\nu \rho} - \dfrac{1}{4} g_{\mu\nu} F^2 \right] + \dfrac{1}{2} \partial_\mu \phi \partial_\nu \phi + \dfrac{1}{2} g_{\mu\nu} V(\phi)~,\\
				\nabla_{\mu} \left[ Z(\phi) F^{\mu\nu} \right] & = 0~,\\
				\Box \phi & = V^\prime(\phi) + \dfrac{Z^\prime(\phi)}{4} F^2~,
		\end{aligned}
\end{equation}
where we used that on-shell $R = -2 V(\phi) + \dfrac{1}{2} (\partial \phi)^2$. This setup also has an analytic solution which is given by a metric, gauge field and non-trivial scalar $\phi$ in the form of
\begin{equation}
\begin{aligned}
ds^2 & = g_{\mu\nu} \dd x^\mu \dd x^\nu = \frac{1}{z^2}\left[ -h(z)\dd t^2 + \frac{1}{h(z)}\dd z^2 + g(z)(\dd x^2 + \dd y ^2)\right]\\
A &= \frac{\sqrt{3Q z_h (1+Q z_h)}}{z_h} \frac{(1-z/z_h)}{1+Qz} \dd t\\
\phi &= \frac{\sqrt{3}}{2}\log\left(1 + Q z\right)
\end{aligned}
\end{equation}
where
\begin{equation}
\begin{aligned}
h(z)& = \frac{(1-z/z_h)}{g(z)}\left[1+(1+3Q z_h) \frac{z}{z_h}+ \left( 1 + 3 Q z_h (1+Q z_h ) \right)\left(\frac{z}{z_h}\right)^2\right],\\
g(z)&= (1+Qz)^{3/2}.
\end{aligned}
\end{equation}
The parameter $Q$ encodes the charge of the black hole. The chemical potential is given by $\mu = \sqrt{3 Q z_h (1 + Q z_h)}/z_h$. The near-horizon form of the potentials in equation~\eqref{eq:EMDAnsatz} corresponds to a scaling behavior of $z,-\theta \rightarrow \infty$, also identified by $\gamma, -\delta = 1/\sqrt{3}$ in \cite{charmousisEffectiveHolographicTheories2010}. The temperature here is given by
\begin{equation}
T_{\mathrm{GR}} = \left|\frac{h'(1)}{4\pi}\right| =  \frac{3\sqrt{1 + Q z_h}}{4\pi z_h}.
\end{equation}

\subsection{Lattice Backgrounds}
\label{app:numerics}

The translational symmetry of the black hole backgrounds is broken by applying a periodic ionic lattice in the boundary field theory through the modulation of the chemical potential \cite{horowitzOpticalConductivityHolographic2012}. In the gravitational theory, this corresponds to imposing a boundary condition on the gauge field 
\begin{equation}
A_t(x,y,z = 0) = \bar{\mu}\left(1 + A_x \cos(G_x x) + A_y \cos(G_y y)\right).
\end{equation}
$A_{x,y}$ parameterize the strength of the lattice, while $G_{x,y}$ are the reciprocal lattice dimensions, respectively. Our computational domain in $(x,y)$ is chosen to always contain a whole number of lattice periods, i.e. $x \sim x + 2\pi n_x/G_x$, $y \sim y + 2\pi n_y/G_y$ where $n_x, n_y \in \mathbb{Z}$. Throughout this work, we take $G_x = G_y \equiv G$ and $A_x = A_y \equiv A_0/2$ in a 2D lattice and $A_x \equiv A_0, A_y = 0$ for a 1D lattice.

This breaking of translational symmetry influences the solutions dramatically. The additional curvature generated by the periodic lattice means that in principle all the off-diagonal components of the metric as well as all components of the gauge field will become non-trivial.

For RN, the ansatz for the fields is adapted from reference \cite{donosMinimallyPackedPhases2016}:
\begin{equation}
\begin{aligned}
ds^2 &= \frac{1}{z^2}\left(-Q_{tt} f(z)\eta_t^2 + Q_{xx}\eta_x^2 + Q_{yy}\eta_y^2 +\frac{Q_{zz}}{f(z)}\eta_z^2\right),\\
\eta_t &= \dd t,\\
\eta_x &= \dd x + Q_{xy} \dd y + Q_{xz}\dd z,\\
\eta_y &= \dd y + Q_{yz} \dd z,\\
\eta_z &= \dd z,\\
A &= \mu(1-z) A_t\dd t 
\end{aligned}
\end{equation}

Our EMD ansatz looks similar and is given by 
\begin{equation}
\begin{aligned}
ds^2 &= \frac{1}{z^2}\left(-Q_{tt} h(z)\eta_t^2+ g(z)\left( Q_{xx}\eta_x^2 +Q_{yy}\eta_y^2\right) +\frac{Q_{zz}}{h(z)}\eta_z^2 \right),\\
\eta_t &= \dd t,\\
\eta_x &= \dd x + Q_{xy} \dd y + Q_{xz}\dd z,\\
\eta_y &= \dd y + Q_{yz} \dd z,\\
\eta_z &= \dd z,\\
A &=\frac{\mu(1-z)}{1+Qz}A_t \dd t, \qquad \phi = \frac{3}{2}\log\left(1 + \varphi Q z \right).
\end{aligned}
\end{equation}
For both types of solutions, we are interested in stationary solutions, and therefore all functions $F = \left\{Q_{ij}, A_{i}, \varphi\right\}$ are functions of $(x,y,z)$, each periodic in $(x,y)$ with a periodicity of $L_{x,y} = 2\pi n_{x,y}/G_{x,y}$. The equations of motion in equation~\eqref{eq:RNEOMs} and \eqref{eq:EMDEOMs} form very complicated systems of non-linear partial differential equations in three dimensions, which in general cannot be solved analytically. For numerical convenience, the DeTurck trick and another gauge fixing term for the gauge field can be used to turn this set of equations into an elliptic boundary value problem \cite{headrickNewApproachStatic2010,adamNumericalApproachFinding2012,donosMinimallyPackedPhases2016}. The UV boundary conditions on the radial coordinate come from imposing an asymptotically AdS solution while imposing that the dilaton be a marginal operator with no source, as was highlighted in \cite{chagnetProperQuantizationGubserRocha2022}. The horizon boundary conditions arise from requiring regularity at the horizon, which means that in a series expansion in powers of $(z-1)$ we can relate each $\partial_z F_i$ to all functions and their tangential derivatives at the horizon,\footnote{If the change of coordinates in footnote~\ref{fn:DG1} is used, this simplifies to $\partial_r F_i = 0\,\forall\,i$, as only even powers of $r$ will appear in the near-horizon expansion. This comes at the cost of accuracy near the horizon.} together with the condition that  $Q_{tt}\bigr|_{z=1} = Q_{zz}\bigr|_{z=1} $.\footnote{This ensures a constant temperature across the (corrugated) horizon.}
\\

The boundary value problems are solved using a self-developed software
package in \texttt{C}, using the PETSc
library \cite{balayEfficientManagementParallelism1997,balayPETScTAOUsers2021}. A Newton line-search
algorithm employing second- and third order finite difference schemes
on rectangular grids is used to find solution to the non-linear
problem. The computational grids are either uniformly spaced or have
the radial coordinate run over the Chebyshev-Lobatto nodes for
increased accuracy near the boundaries of the problem. Typical grid
sizes for the simulations run between $N_x\times N_y \times N_z = 40
\times 40 \times 60$ to $80 \times 80 \times 120$. For convergence
checks, the vanishing of the norm of the DeTurck vector provides a
good measure \cite{adamNumericalApproachFinding2012}. Due to the large number of
degrees of freedom involved ($O(10^7)$ for the largest lattices) most
of the numerical work was done using the ALICE cluster at Leiden
University and the Dutch national
Cartesius and Snellius supercomputers with the support of SURF
Cooperative. The code is publicly available \cite{floris_balm_2022_7284816,floris_balm_2022_7285050}.

\subsection{DC Conductivity}

The DC conductivity is computed by solving a Stokes flow problem on the black hole horizon \cite{donosThermoelectricDCConductivities2014,banksThermoelectricDCConductivities2015,donosDCConductivityMagnetised2016,donosHolographicDCConductivity2017}. Using a set of time-independent perturbations, one can show that the bulk linear response problem of computing (thermo)electric DC conductivities can be reduced to a linearized version of the Navier-Stokes equations for an auxiliary fluid that lives on a static black hole horizon background. The equations take a similar form for both EMD and RN black holes, and can be written as \cite{donosDCConductivityMagnetised2016}
\begin{equation}
\label{eq:StokesDC}
\begin{aligned}
\eta^{(0)}\bigl(-2\nabla^j \nabla_{(i}v_{j)} + &3 v^j\nabla_j\phi^{(0)}\nabla_i{\phi}^{(0)}\bigr)- d\chi_{ij}^{(0)}Q^j - F^{(0)}_{ij}J^j = \\
& \rho^{(0)}\left(E_i + \nabla_j w\right) + Ts^{(0)}\left(\zeta_i - \nabla_i \frac{p}{4\pi T}\right) \\
\partial_i Q^i = 0,& \qquad\qquad \partial_i J^i  = 0.
\end{aligned}
\end{equation}
The superscript $(0)$ indicates that these are background quantities evaluated at the horizon.  These are the values we extract from the numerical solutions to the background lattices described above.\footnote{For the RN black holes, one should take $\phi = 0, Z(\phi) = 1, V(\phi) = 6$.} The Stokes equations~\eqref{eq:StokesDC} is then a set of four equations for the four unknown functions $v_x,v_y, w, p$. The currents $Q,J$ and transport coefficients $\rho^{(0)}, \eta^{(0)}, s^{(0)}, \chi^{(0)}$ can be written in terms of these four functions, the background horizon quantities and the induced metric on the horizon $h^{(0)}_{ij}$\cite{donosDCConductivityMagnetised2016}. $E_i, \zeta_i$ source the electric field and thermal gradient, and are taken to be constant over the unit cell. The thermo-electric DC conductivities are then extracted by evaluating
\begin{equation}
\label{eq:ConductivityMatrix}
\begin{pmatrix}
\vec{J}\\
\vec{Q}\end{pmatrix} = 
\begin{pmatrix}
\sigma & \alpha T\\
\bar{\alpha} T & \bar{\kappa}T
\end{pmatrix}
\begin{pmatrix}
\vec{E}\\
\vec{\zeta}\end{pmatrix},
\end{equation}
where $\vec{J},\vec{Q}$ are here the spatial averages of the solutions when evaluating equation~\eqref{eq:StokesDC}. These averaged values do not renormalize when lifted to the boundary to be interpreted in the field theory. As a result, the thermo-electric conductivity matrix defined in equation~\eqref{eq:ConductivityMatrix} is that of the field theory.

In previous works, e.g. \cite{donosHolographicDCConductivity2017}, these equations have been used to study simpler systems that do not fully break spatial translational symmetry or break it in a homogeneous way. That simplification allows for a largely analytic treatment of these equations. The systems we are interested in do not permit us such luxuries and therefore we have to solve this coupled linear PDE in two periodic dimensions numerically. For this, we developed a package in Python which can solve coupled (non-)linear partial differential equations for backgrounds as well as perturbations. This package is rather flexible, in that it can make use of both (pseudo)spectral and arbitrary-order finite difference methods to solve the equations. This package will be made available publicly at a later date.

\section{Semi-local criticality and an induced IR length scale}
\label{app:causal}

Semi-local quantum liquids can be defined by a ``self-energy'' that is
either a power-law in frequency $\Sigma \sim \omega^{2\nu(k)}$ or
exponential $\Sigma \sim  \exp(-\frac{k^{z/(z-1)}}{\omega^{z-1}})$ with $z$ the
dynamical critical exponent. In the limit
$z\rightarrow \infty$ the latter reduces to the former
\cite{faulknerHolographicNonFermiLiquid2011}. Both $\omega$ and $k$ are dimensionless frequencies
and momenta in units of the chemical potential $\mu$. As emphasized in
\cite{iqbalSemilocalQuantumLiquids2012} the spatial structure of such semi-local quantum
liquids is that the spread of local perturbations decays very rapidly
and is bounded by an emergent length scale $\xi \propto \mu^{-1}$.

Though the emergence of this semi-local physics is poorly understood
from a conventional point of view, its emergence bound is surprisingly
clear from a dual holographic perspective. It is a direct consequence
of the existence of a maximal distance, $x_{\rm max}\sim 1/\xi_\mu$
that two light-rays emitted from near the AdS black hole horizon can
spread \cite{iqbalSemilocalQuantumLiquids2012}. It implies that a local perturbation in the IR can only
originate from/influence a finite spatial region (in the UV
variables).

This supplementary section shows how this maximal distance arises.
A light-ray parametrized by $X^{\mu}(\tau)$ follows a null geodesic, i.e. 
\begin{equation} \label{constraint}
	 g_{\mu\nu}\dot{X}^{\mu} \dot{X}^{\nu}=0\,.
\end{equation}
Consider a generic $z=\infty$ metric 
\begin{eqnarray} \label{eq:ds}
ds^2 &=&  \left(\frac{r}{R}\right)^{-\frac{2 \eta }{d}}\left[ -\mu ^2 R^2 r^2 dt^2 + \frac{R_2^2}{r^2} dr^2 + \mu ^2 R^2 dx_i dx^i \right] 
\end{eqnarray}
The parameter $\eta = -\frac{\theta}{z}$ is the remnant of the
hyperscaling violation exponent $\theta$ in the limit $z\rightarrow
\infty$ and $R_2 = R/\sqrt{6}$ is the emergent AdS${}_2$ radius. For a geodesic emanating from the horizon we can use
the radial $r$ coordinate as the affine parameter $\tau$, and the
physical gauge $X^r=r$. Then solving
the geodesic equation $D_{\tau}\dot{X}^{\mu}=0$ subject to the null
length constraint, one finds for the motion in the transverse directions
\begin{align} \label{eqxi2}
X_{i,\pm} (r) &= x^{(0)}_i \pm \frac{v_i R_2}{\mu  R v} \tan ^{-1}\left(\frac{r v}{\sqrt{v_t^2-r^2 v^2}}\right)\,.
\end{align}
with $v^2=\sum_i v_i^2$.
Two light-rays starting from the same point $x_i^{(0)}$ one pointing to
the left and one to the right therefore
arrive at the boundary ($r=\infty$) a distance
$2\frac{v_i}{v}\frac{R_2}{\mu R}\frac{1}{\tan(i)}$ apart.

We are now interested in the intersection of two lightcones $x_{i,+}$ and $x_{i,-}$, which can be found from
\begin{equation} \label{r0sol}
x_{i,+}(r_0; x^{(0)} = 0) = x_{i,-} (r_0; x^{(0)} )\implies r_0 = \frac{v_t}{v} \sin\left( \frac{\mu  R v x_i^{(0)} }{2 v_i  R_2}\right)\,.
\end{equation}
After combining \eqref{eqxi2} with \eqref{r0sol}, we find that the maximal allowed distance is
\begin{equation} \label{critdistance}
x^{\rm (max)}_{1} = \frac{R_2}{R \mu} \pi \cos\theta \,, \qquad x^{\rm (max)}_2 = \frac{R_2}{R \mu} \pi \sin\theta \,.
\end{equation}
where we have chosen the parametrization for the initial velocity components along $\left( x_1,x_2\right)$ as $v_1 = v \, \cos\theta\,,\quad v_2 = v\, \sin\theta\,,$ where $\theta\in [0,\pi/2]$ is the initial angle, measured with respect to the $x_1$-axis. 

The relative initial distance between the two geodesics $\Delta s$ reads
\begin{equation}\label{eq:deltas}
\Delta s = \sqrt{x^{\rm (max)}_{1}{}^2 + x^{\rm (max)}_{2}{}^2} = \frac{R_2}{R \mu} \pi\,,
\end{equation}
which is universal and does not depend on the initial conditions. It coincides with the result presented in \cite{iqbalSemilocalQuantumLiquids2012}. 

\begin{figure}
	\begin{center}
	\includegraphics[scale=1]{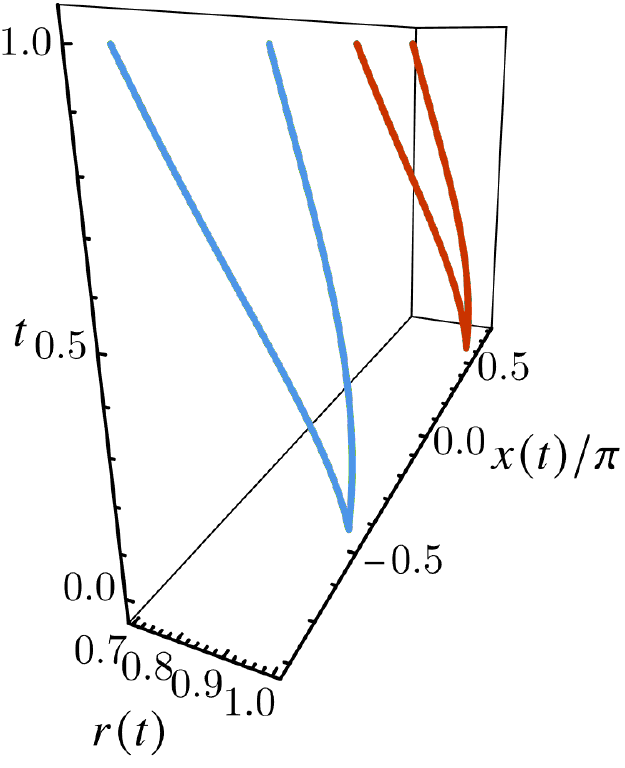}
	\includegraphics{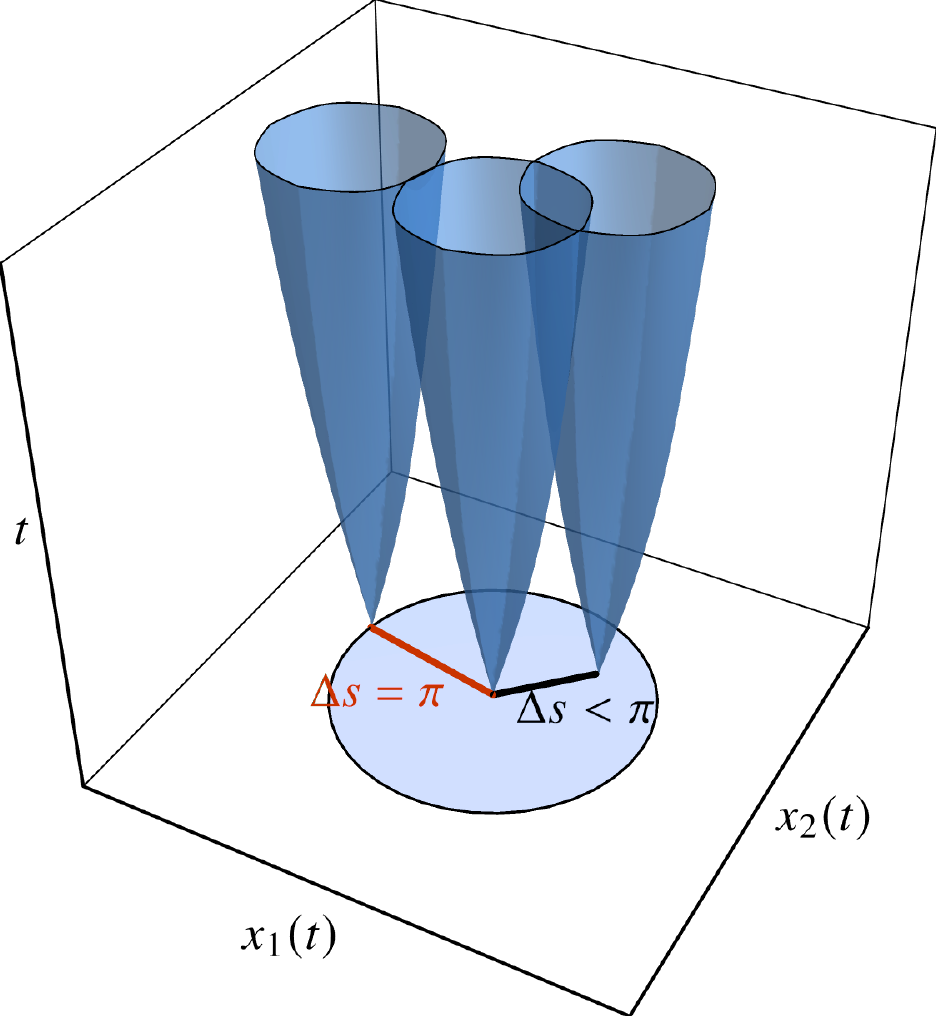}
	\caption{Left figure: Causal structure for two light-rays separated at a relative distance $x_0=\pi$. The $x$-axis corresponds to the $x_1(t)$, while the $y$-axis to the $r(t)$ coordinate. The external time has been taken as the dialing parameter, along the $z$-axis in the figure. For the present purposes, we have considered no motion along the $x_2$ direction, and we have also set $R_2 = R = \mu = 1$. Right figure: Causal structures for three light-cones as functions of $\left(x_1(t),x_2(t)\right)$. The $z$-axis corresponds to time, for which we have set $t=1$ as the time that the geodesics reach the boundary. Those geodesics that start at any point within the disk of radius $\pi$ will be causally connected, while disconnected if otherwise. \label{fig:causal} }
\end{center}
\end{figure}

In figure \ref{fig:causal}, we plot the causal structure for two light-rays separated by a certain initial distance. For an initial separation larger than the critical distance, \eqref{critdistance}, both light-rays are not causally connected anymore. To illustrate this, we have chosen as a dialing parameter the external time $t$. After some computations, we get
\begin{equation}
t(x_i) = -\frac{R_2 v}{\mu  R v_t} \cot \left[\frac{\mu  R v}{v_i R_2} \left(x_i-x_i^{(0)}\right)\right]\,.
\end{equation}
which is plotted in the second figure in \ref{fig:causal}. From here, we highlight that any geodesic that starts at an initial relative distance $\Delta s\leq \pi$, will be causally connected, whereas if $\Delta s> \pi$, it will be causally disconnected. 

Based on \eqref{eq:deltas} and on the fact that the 2-point correlation function $G\sim 1/\xi_{\rm m}\sim \pi$, we conclude that the maximal correlation distance in Planckian dissipation is related to the existence of this maximal causality distance in geodesic. 

\section{Four pole fitting formula}\label{sec:4polefit}
The full 4-pole fitting formula that can fit the four poles nearest to the origin in the complex frequency plane is given by the following nine-parameter formula
\begin{equation}
\begin{aligned}
		\sigma(\omega) = \sigma_0 + &Z\frac{\Omega - i\omega}{(\Omega - i\omega)(\Gamma - i\omega) +\omega_0^2} + \\ &\left(\frac{Z_{s,1}+iZ_{s,2}}{\omega - (\omega_{s,1}+i\omega_{s,2})} + \text{time-reversed}\right)~.
		\end{aligned}
\end{equation}
The weights and positions of the sound poles is constrained by time reversal symmetry, which dictates that 
\begin{equation}
		\sigma^*(-\omega^*) = \sigma(\omega)~.
\end{equation}

\section{Memory matrix formalism}
\label{app:memory-matrix}

The correlation functions of the homogeneous GR and RN fluids are well described by the standard hydrodynamics of relativistic conformal fluids with U(1) charge (see \cite{kovtunLecturesHydrodynamicFluctuations2012}). To compute \eqref{eq:mem-matrix}, we simply need the correlator $G_{J^t J^t}$ which is given by
\begin{equation}
		G_{J^t J^t}(\omega, k) = \dfrac{\sigma_Q k^2}{D_\rho k^2 - i \omega} - \dfrac{k^2 \omega_p^2}{\omega^2 + i D_\pi k^2 \omega - c_s^2 k^2}~.
\end{equation}
This form quite readily shows how this dynamical response has both a convective part (sound) and a dissipative part. At low frequencies, this correlator can be expanded as
\begin{equation}
		G_{J^t J^t}(\omega, k) = \dfrac{\omega_p^2}{c_s^2} +\dfrac{\sigma_Q}{D_\rho} + i \omega \left[ \dfrac{\sigma_Q}{D_\rho k^2} + \omega_p^2 \dfrac{D_\pi}{c_s^4} \right] + \mathcal O(\omega^2)~.
\end{equation}
The leading term is entirely real and will not contribute to the imaginary part. Therefore, we can eventually obtain \eqref{eq:mem-matrix} as
\begin{equation}
		\Gamma_{\mathrm{mom.rel.}} = \dfrac{\bar{\mu}^2 A^2}{2 (\bar{\eps} + \bar{p}) D_\rho \alpha_n} + \dfrac{\bar{\mu}^2 A^2 D_\pi \bar{n}^2}{2 c_s^4 (\bar{\eps} + \bar{p})^2} G^2 = \Gamma_d + \Gamma_\eta~,
\end{equation}
where we recognize the quantities $\Gamma_d,\Gamma_\eta$ introduced in \eqref{eq:gamma-definitions}.

\section{Scaling of hydrodynamical relaxation rates}
\label{app:gamma-vs-s-scaling}

Consider an equation of state $P(T,\mu)/\mu^3 = a_0 + a_1 (T/\mu)^{\hat{\eta} + 1}$ where $\hat{\eta} = (d-\theta)/z$ is the generic effective dimension in the presence of a dynamical critical exponent $z$ and hyperscaling violation exponent $\theta$. This equation of state will be a valid approximation for the low-temperature regime of the holographic Einstein-Maxwell-Dilaton systems, such as RN and GR. Then, the entropy and charge density one obtains from this pressure are $s/\mu^2 = (\hat{\eta}+1) a_1 (T/\mu)^{\hat{\eta}}$ and $n/\mu^2 = 3 a_0 - (\hat{\eta}+2) a_1 (T/\mu)^{\hat{\eta} + 1}$.  We will now look at the momentum relaxation rate \eqref{eq:mem-matrix} for a relativistic charged fluid such that the viscosity saturates the minimal viscosity bound $\eta = s/(4\pi)$ and we will take the EMD T-scaling $\sigma_Q = \hat{\sigma}_Q (T/\mu)^{\hat{\eta}+2}$ \cite{davisonIncoherentTransportClean2015}. From the integrated first law $\eps + P = sT + \mu n$, we find $\eps = 2 P$ for this choice such that we still have a conformal system and therefore $\zeta = 0$.

The relaxation rate has two contributions, one $G$-dependent and one $G$-independent, reminiscent of our result \eqref{eq:gamma-definitions}, which we will by analogy name $\Gamma_\eta$ and $\Gamma_d$. In the general non-conformal case we have now introduced, these therefore take the form
\begin{equation}
		\begin{aligned}
				\Gamma_\eta/\mu & = A^2 (G/\mu)^2 \dfrac{\pi a_1 (\hat{\eta}+1)}{6 a_0} \left( \dfrac{T}{\mu} \right)^{\hat{\eta}}\dfrac{\left( 1 - \frac{a_1(\hat{\eta}-2)}{3 a_0} (T/\mu)^{\hat{\eta}+1} \right)^2}{\left(1+\frac{a_1}{a_0} (T/\mu)^{\hat{\eta}+1}\right)^3}~,\\
				\Gamma_d/\mu & = A^2 \dfrac{a_1^2 (\hat{\eta}+1)^2}{6 a_0 \hat{\sigma}_Q} \left( \dfrac{T}{\mu} \right)^{\hat{\eta}} \dfrac{\left( \hat{\eta} + \frac{a_1 (\hat{\eta}-2)}{3 a_0} (T/\mu)^{\hat{\eta} + 1} \right)^2}{\left( 1 + \frac{a_1}{a_0} (T/\mu)^{\hat{\eta} + 1} \right)^3}~.
		\end{aligned}
\end{equation}
The leading order of $\Gamma_\eta$ can therefore be obtained as
\begin{equation}
		\Gamma_\eta/\mu \sim A^2 (G/\mu)^2 \dfrac{\pi a_1 (\hat{\eta}+1)}{6 a_0} \left( \dfrac{T}{\mu} \right)^{\hat{\eta}} \sim  A^2 (G/\mu)^2 \dfrac{\pi}{2} \dfrac{s}{n}~.
\end{equation}
This shear drag contribution is therefore entirely determined by the entropy at low temperature. The other contribution, $\Gamma_d$, is slightly less straightforward. When $\hat{\eta} > 0$, a similar behavior arises
\begin{equation}
		\label{eq:GammaDFunT}
		\Gamma_d/\mu \sim A^2 \dfrac{a_1^2 (\hat{\eta}+1)^2 \hat{\eta}^2}{6 a_0 \hat \sigma_Q} \left( \dfrac{T}{\mu} \right)^{\hat{\eta}} \sim \dfrac{A^2}{2 n/\mu^2 }\left( T \dfrac{\partial s}{\partial T} \right)^2 \sigma_Q^{-1} \left( \dfrac{T}{\mu} \right)^2~.
\end{equation}
Therefore in this general case, which encompasses the GR case $\hat{\eta} = 1$, $\Gamma_d$ and $\Gamma_\eta$ have the same temperature dependence although $\Gamma_d$ is more sensible to the susceptibilities like the specific heat $T \frac{\partial s}{\partial T}$ and the hydrodynamic transport coefficient $\sigma_Q$.
A counterexample of this general rule however arises when $\hat{\eta} = 0$, as it is for the RN black hole for instance, where the leading order of $\Gamma_d$ vanishes and instead one must expand to second order to have
\begin{equation}
		\Gamma_d/\mu \overset{\hat{\eta} = 0}{\sim} A^2 \dfrac{a_1^4}{3 a_0^3 \hat{\sigma_Q}} \left( \dfrac{T}{\mu} \right)^2~.
\end{equation}

Finally, we can explain how this $(T/\mu)^2$ factor in \eqref{eq:GammaDFunT} arises naturally from the $\alpha_n$ factor introduced in \eqref{eq:5}. To do so, consider the quantity $D_\rho \Gamma_d = A^2 \frac{\bar{\mu}^2}{2 (\bar{\eps} + \bar{p}) \alpha_n}$. We will relax here our assumptions about the equation of state and only assume some Sommerfeld entropy $s = \gamma(\bar \mu) \bar T$ and $n = n_0(\bar \mu)$ at low temperature. The scaling of $D_\rho \Gamma_d$ is therefore entirely determined by that of $\alpha_n^{-1}$ which can be determined using
\begin{equation}
		\begin{aligned}
				\left( \dfrac{\partial \bar n}{\partial \bar T}\right)_{\bar \mu} & = \left( \dfrac{\partial \bar s}{\partial \bar \mu}\right)_{\bar T} \sim \gamma^\prime(\bar \mu) \bar T~,\\
				\left( \dfrac{\partial \bar n}{\partial \bar \mu}\right)_{\bar T} & \sim n_0^\prime(\bar \mu)~,\\
				\left( \dfrac{\partial \bar \eps}{\partial \bar T}\right)_{\bar \mu} & = \bar T \left( \dfrac{\partial \bar s}{\partial \bar T}\right)_{\bar \mu} + \bar \mu \left( \dfrac{\partial \bar n}{\partial \bar T}\right)_{\bar \mu} \sim \left( \gamma(\bar \mu) + \bar \mu \gamma^\prime(\bar \mu) \right) \bar T~,\\
				\left( \dfrac{\partial \bar \eps}{\partial \bar \mu}\right)_{\bar T} & = \bar T \left( \dfrac{\partial \bar s}{\partial \bar \mu}\right)_{\bar T} + \bar \mu \left( \dfrac{\partial \bar n}{\partial \bar \mu}\right)_{\bar T} \sim \bar \mu n_0^\prime(\bar \mu) + \gamma^\prime(\bar \mu) \bar T^2~.
		\end{aligned}
\end{equation}
Then, we can plug these relations into Eqs.~\eqref{eq:5} and obtain
\begin{equation}
		\alpha_n \sim \dfrac{\bar T^2 \left( \gamma(\bar \mu) + 2 \bar \mu \gamma^\prime(\bar \mu) \right) + \bar \mu n_0^\prime(\bar \mu)}{\bar T \left[ n_0^\prime(\bar \mu) \left( \gamma(\bar \mu) + \bar \mu \gamma^\prime(\bar \mu) \right) \bar T - \gamma^\prime(\bar \mu) \bar T \left(  \bar \mu n_0^\prime(\bar \mu) + \gamma^\prime(\bar \mu) \bar T^2 \right) \right]} \sim \dfrac{\bar \mu}{\bar T^2 \gamma(\bar \mu)}~.
\end{equation}
Therefore, given Sommerfeld entropy, we naturally get that $D_\rho \Gamma_d \sim T^{2}$. Provided then that $D_\rho \sim T$, which is the case for the GR holographic metal, you recover the scaling $\Gamma_d \sim T$.

\section{Lorentz oscillator decoupling}
\label{app:lorentz}

Consider a system of modes coupled to one another in the following way
\begin{equation}
		\label{eq:LorentzOscillator}
		\begin{aligned}
				\partial_t J_1 + \Gamma_1 J_1 + \gamma_1 J_2 & = E_1~,\\
				\partial_t J_2 + \Gamma_2 J_2 - \gamma_2 J_1 & = E_2~,
		\end{aligned}
\end{equation}
where $\Gamma_{1,2}$ are relaxation rates for the currents $J_{1,2}$, $E_{1,2}$ are explicit sourcing and $\gamma_{1,2}$ couple the two modes to one another. In matrix notation $\partial_t J_a + M_{ab} J_b = E_a$, this leads to the following evolution matrix
\begin{equation}
		M_{\mathrm{LO}} = \begin{pmatrix}
				\Gamma_1     & \gamma_1 \\
				-\gamma_2    & \Gamma_2 \\
		\end{pmatrix}~.
\end{equation}
We can then solve this dynamic system and obtain, in frequency space,
\begin{equation}
		\begin{aligned}
				J_1(\omega) & = \dfrac{\left( \Gamma_2 - i \omega \right) E_1}{(\Gamma_1 - i\omega)(\Gamma_2 - i \omega) + \gamma_1 \gamma_2}~,\\
				J_2(\omega) & = \dfrac{\gamma_2 E_1}{(\Gamma_1 - i\omega)(\Gamma_2 - i \omega) + \gamma_1 \gamma_2}~,\\
		\end{aligned}
\end{equation}
where we have set $E_2 = 0$ as we are only interested in externally sourcing one of the currents. Critically, we will be interested in a total current $J$ which overlaps with both $J_1$ and $J_2$ through
\begin{equation}
		J = \sigma_0 E_1 + a J_1 + b J_2
\end{equation}
where $\sigma_0$ is some explicit contribution by the external sourcing. Therefore, the conductivity associated to this current is
\begin{equation}
		\sigma = J/E_1 = \sigma_0 + \dfrac{a \left( \Gamma_2 - i \omega \right) + b \gamma_2}{(\Gamma_1 - i\omega)(\Gamma_2 - i \omega) + \gamma_1 \gamma_2}~.
\end{equation}
This form is very reminiscent of \eqref{eq:12b} with
\begin{equation}
		\begin{gathered}
				Z_{\mathrm{eff}} = a~, \quad \Omega = \Gamma_2 + \frac{b}{a} \gamma_2\\
				\Gamma = \Gamma_1 - \frac{b}{a} \gamma_2~, \quad \omega_0^2 = \gamma_1 \gamma_2  - \frac{b}{a} \gamma_2 \left[  \Gamma_1 - \Gamma_2 - \frac{b}{a} \gamma_2 \right] ~.
		\end{gathered}
\end{equation}

Let us now compare to the matrix $M$ \eqref{eq:dynMatrixHydro} describing the dynamical hydrodynamic system in the small lattice expansion. From this system of $4$ coupled fields, it is possible to decouple two by taking the large speed of sound limit $c_s \to \infty$\footnote{To take this limit carefully, one needs to rescale the momentum modes $\delta \eps^{(C)},\delta \pi_x^{(C)}$ by a factor of $c_s^2$ beforehand.} which formally just encodes the assumption that the sound poles live far from the two poles close to origin. While this is a relatively simple limit to illustrate the qualitative behavior of the isolated two pole sector, we must emphasize that this limit will not reproduce quantitatively the mapping \eqref{eq:hydroToCDW} exactly, and that is because there are higher order effects of the coupling to the sound sector which should be more carefully disentangled. It will be however a helpful illustration of the dynamics of the low frequency sector. 
The two currents remaining $J_{1,2}$ are then the momentum current density $\delta \pi_x^{(0)}$ and the parity-odd charge density $\delta n^{(S)}$.

The decoupled system then takes the form
\begin{equation}
		M = \begin{pmatrix}
				0 & A G \bar{\mu}/2\\
				-\frac{A G \bar{\mu}}{(\bar{\eps} + \bar{p})\alpha_n} & D_\rho G^2
		\end{pmatrix} = \begin{pmatrix}
				0 & A G \bar{\mu}/2\\
				-\dfrac{2 D_\rho G \Gamma_d}{A \bar{\mu}}  & D_\rho G^2
		\end{pmatrix}~,
\end{equation}

while the total current of interest is $J = \sigma_Q E_x + \omega_p^2 \delta \pi_x^{(0)} - \frac{\bar{\mu}A}{2} \omega_p^2 D_\rho G \delta n^{(S)}$. Thus, we deduce from this that $Z_{\mathrm{eff}} = \omega_p^2$ while the effective momentum relaxation rates and effective couplings are
\begin{equation}
		\label{eq:hydroToCDW2by2}
		\begin{aligned}
				\Omega & = D_\rho G^2 \left[1 - D_\rho \Gamma_d \right]~,\\
				\Gamma & = (D_\rho G)^2 \Gamma_d~,\\
				\omega_0^2 & = D_\rho G^2 \Gamma_d \left[ 1 - (D_\rho G)^2 + D_\rho^3 G^2 \Gamma_d \right]~.
		\end{aligned}
\end{equation}
As expected, there is a discrepancy between Eqs.~\eqref{eq:hydroToCDW} and Eqs.~\eqref{eq:hydroToCDW2by2} which just highlights that the limit $c_s \to \infty$ should be refined. However, this correctly predicts the leading order in $A$ of every coefficient and gives a very close, qualitative estimate of the corrections at the next order.
\newpage
\bibliographystyle{custom}
\bibliography{UniversalStrangeMetal}

\end{document}